\documentstyle[11pt]{article}
\begin{document}

  \makeatletter
  \@addtoreset{equation}{section}
  \makeatother
  \renewcommand{\theequation}{\thesection.\arabic{equation}}
  \baselineskip 15pt

  \newtheorem{defofentang}{Definition}[section]
  \newtheorem{factorizability}{Theorem}[section]
  \newtheorem{rank}[factorizability]{Theorem}
  \newtheorem{onedimensional}[factorizability]{Theorem}
  \newtheorem{correlation}[factorizability]{Theorem}
  \newtheorem{defofentangidentical}{Definition}[section]
  \newtheorem{factorizabilityidentical}{Theorem}[section]
  \newtheorem{defofentangidentical2}[defofentangidentical]{Definition}
  \newtheorem{factorizabilityidentical2}[factorizabilityidentical]{Theorem}
  \newtheorem{defofentangidentical3}[defofentangidentical]{Theorem}
  \newtheorem{defofentangN}{Definition}[section]
  \newtheorem{onedimensionalN}{Theorem}[section]
  \newtheorem{factorizabilityN}[onedimensionalN]{Theorem}
  \newtheorem{rankN}[onedimensionalN]{Theorem}
  \newtheorem{corrbetween}[onedimensionalN]{Theorem}
  \newtheorem{mutualentang}[defofentangN]{Definition}
  \newtheorem{mutualfactor}[onedimensionalN]{Theorem}
  \newtheorem{sviluppodipi}{Theorem}[section]
  \newtheorem{giancarlo}[sviluppodipi]{Theorem}
  \newtheorem{definizionedient}{Definition}[section]
  \newtheorem{defentanggroupfermion}{Definition}[section]
  \newtheorem{defofentangNidentical}[defentanggroupfermion]{Definition}
  \newtheorem{relevant}[sviluppodipi]{Theorem}
  \newtheorem{teodianti}[sviluppodipi]{Theorem}

 \newtheorem{localfact1}[sviluppodipi]{Theorem}
  \newtheorem{localfact2}[sviluppodipi]{Theorem}
  \newtheorem{factorizabilityNidentical}[sviluppodipi]{Theorem}
  \newtheorem{setcompletobosoni}{Definition}[section]
  \newtheorem{relevantboson}{Theorem}[section]
  \newtheorem{relevantboson2}[relevantboson]{Theorem}
  \newtheorem{mixed}{Theorem}[section]

\title{\bf Entanglement and Properties of Composite Quantum Systems: a
 Conceptual and Mathematical Analysis
\footnote{Work supported in part by Istituto Nazionale di Fisica Nucleare,
 Sezione di Trieste, Italy}}
\author{Giancarlo Ghirardi\footnote{e-mail: ghirardi@ts.infn.it}\\
 {\small Department of Theoretical Physics of the University of Trieste,}\\
 {\small Istituto Nazionale di Fisica Nucleare, Sezione di Trieste, and}\\
 {\small International Centre for Theoretical Physics, Trieste, Italy,}\\
 \\ Luca Marinatto\footnote{e-mail: marinatto@ts.infn.it}\\
 {\small Department of Theoretical Physics of the University of Trieste,
 and}\\
 {\small Istituto Nazionale di Fisica Nucleare, Sezione di Trieste, Italy,}\\
 and \\
 \\ Tullio Weber\footnote{e-mail: weber@ts.infn.it}\\
 {\small Department of Theoretical Physics of the University of Trieste,
 and}\\
 {\small Istituto Nazionale di Fisica Nucleare, Sezione di Trieste, Italy.}}

 \date{}

 \maketitle

\begin{abstract}
Various topics concerning the entanglement of composite quantum systems are
 considered with particular emphasis concerning the strict relations of such
 a problem with the one of attributing objective properties to the
 constituents.
Most of the paper deals with composite systems in pure states. After a
 detailed discussion and a precise formal analysis of the case of systems of
 distinguishable particles, the problems of entanglement and the one of the
 properties of subsystems of systems of identical particles are thoroughly
 discussed. This part is the most interesting and new and it
 focuses in all details various subtle questions which have never been
 adequately discussed in the literature. Some inappropriate assertions which
 appeared in recent papers are analyzed.
The relations of the main subject of the paper with the nonlocal aspects of
 quantum mechanics, as well as with the possibility of deriving Bell's
 inequality are also considered.  \\


  PACS: 0.3.65.Bz

\end{abstract}


\section{Introduction}

One of the crucial points of any theory aiming to account for natural
 phenomena concerns the possibility of identifying the properties objectively
 possessed by individual physical systems and/or by their constituents.
Such a problem acquires a completely different status within different
 theoretical schemes, typically in the classical and quantum cases.
First of all, quantum mechanics, if the completeness assumption is made,
requires a radical change of attitude about
 the problem of attributing objective properties to physical systems due to
 its fundamentally probabilistic character.
Secondly, and even more important for our analysis, it gives rise to specific
 and puzzling situations concerning the properties of the constituents of a
 composite system due to its peculiar feature, Entanglement - the direct
 English translation of the original German form {\em Verschrankung} used by
 Schr\"odinger~\cite{ref1} - which Schr\"odinger himself considered
 {\em ``the characteristic trait of Quantum Mechanics, the one that enforces
 its entire departure from classical line of thoughts''}.

This paper is devoted to analyze  quantum entanglement, to characterize it in
 a clear and rigorous way, to derive various new theorems allowing to identify
 its occurrence and to point out some misleading and/or erroneous arguments
 about it which can be found in the literature.
Particular emphasis is given to the conceptual and formal changes which are
 necessary to deal in a logically correct way with the problem of entanglement
 of systems involving identical constituents.

The paper is divided in four parts and is organized as follows: in Part I the
 possible probabilistic features of various (classical and quantum) theories
 are discussed with the purpose of illustrating the interplay between the
 epistemic and nonepistemic aspects of the  description of natural processes
 and of characterizing the different types of ``states'' which one has to take
 into account according to the information he has about the system.
The problem of the attribution of objective properties to individual physical
 systems in the classical and quantum cases is also discussed.
The rest of the paper is entirely devoted to quantum systems.
In Parts II and III the most extended and relevant, we investigate the
 implications of entanglement of composite quantum systems concerning
 properties, by confining our considerations to systems in pure states, or,
 equivalently, to homogeneous quantum ensembles.
Part II deals with the case of distinguishable constituents, while Part III
 is entirely devoted to present a detailed and original analysis of the case
 involving identical constituents.
In Part IV we take briefly into account the non-pure cases, or, equivalently,
 the non-homogeneous quantum ensembles, and we discuss  some relevant
questions connected with quantum nonlocality
 and Bell's inequality.


 \vspace{1cm}

\begin{center}
  {\LARGE\bf  Part I: Probabilities and Properties}\\
\end{center}


 \section{Physical theories and their probabilistic features}

Any theoretical scheme aiming to account for natural processes describes the
 state of individual physical systems and of the physically observable
 quantities by appropriate mathematical entities. The scheme must contemplate
 rules mirroring the crucial steps of the unfolding of a process: the
 preparation of the system, its evolution and the recipes by which one can
 make predictions concerning the outcomes of prospective measurement processes
 on the system.
All the just mentioned stages can exhibit deterministic or probabilistic
 aspects.
In the case in which one has to resort (for various reasons which we will
 analyze in what follows) to a probabilistic description, one is naturally
 led to raise the conceptually relevant question of the precise status
 assigned to probabilities within the scheme, in particular whether they
 have an epistemic or nonepistemic character.
Obviously, answering  such a question requires a specific analysis of the
 logical and formal structure of the theory. In fact it is quite easy to
 exhibit physically equivalent theoretical schemes (one of the best known
 examples being Bohmian Mechanics and Standard Quantum Mechanics) whose
 probabilities have a completely different conceptual status.

Let us therefore start by discussing the notion of \textit{state of an
 individual physical system} within the hypothetical theory under
 consideration.
The crucial point, from a conceptual point of view, consists in identifying
 which is the most accurate characterization that the theory allows concerning
 the situation of an individual physical system.
By taking this attitude, we are disregarding (for the moment) the unavoidable
 difficulties one meets in actually preparing a system in such a way that
 its physical situation corresponds to the just mentioned most accurate
 specification allowed by the formalism, and/or in knowing precisely its
 situation at a given time.
For our present purposes we assume that such a ``most accurate preparation''
 or ``most exhaustive knowledge'' is, in principle, possible.
Such a characterization is expressed by mathematical entities which we will
 denote as the {\bf States} (with capital S) of the theory.
As the reader certainly knows, the problem we are facing is strictly related
 to the so called assumption of completeness of the theoretical framework:
 such an assumption amounts simply to accept that no specification more
 precise than the one given by the {\bf States} is possible.

One can immediately exhibit some elementary examples of what we have in mind.
For instance, the {\bf States} of a system of {\it N} point particles within
Newtonian mechanics are the points $P$ of the $6N$-dimensional phase space
 of the system.
Similarly, non-relativistic quantum mechanics with the completeness assumption
 asserts that the {\bf States} of a system of $N$ spinless particles are the
 state vectors of the associated Hilbert space, i.e.,  the square integrable
 functions $\Psi (\mathbf{r}_{1},...,\mathbf{r}_{N})$ of the $3N$ coordinates
 of the particles.

As already remarked it can very well happen that one is not able to prepare
 a system in a precise {\bf State} or to have a precise knowledge of it.
This impossibility may derive from practical limitations but  it can also
 occur for reasons inherent to the theory itself. When we do not know the
 {\bf State}, but we still have some control on the preparation or some
 meaningful information about the physical situation, we will speak of the
 {\bf state} (with lower case s) of the physical system\footnote{Typically
 a state can be identified with a probability measure on the ensemble of
 States.}.
Once more simple cases can be mentioned: for a mechanical classical system
 we can (in practice) know at most its {\bf state} and never its {\bf State},
 this fact being due to the practical impossibility of identifying with
 infinite precision the point in phase space characterizing the precise
 physical situation of the system.
Within Bohmian Mechanics, in spite of the fact that the most accurate
 specification of an individual physical system is given by the combined
 assignment of its wave function $\Psi ({\bf{r}}_{1},...,{\bf{r}}_{N})$ plus
 the positions $({\bf{r}}_{1},...,{\bf{r}}_{N})$ of its particles, it is
usually
 assumed that while we can prepare a system in any chosen state vector,
 there is no possibility of controlling or knowing the positions of the
 particles in more detail than is conveyed by the quantum position
 probabilities\footnote{For a detailed discussion about the conceptual
 status of this assumption see the detailed analysis of ref.\cite{refDGZ}}.
We remark that, as is well known, if one could control the positions
 (which are often called -- absurdly, according to Bell -- the hidden variables
 of the theory) one would be able to falsify quantum mechanics as well as achieve 
 superluminal signaling.

Having clarified this point, let us come back to our general theoretical
 scheme and let us confine, for the moment, our considerations to the case
 in which our physical system is in a precise {\bf State}.
As already stated, a satisfactory theory must contain some recipe (usually
 an  evolution equation) allowing to deduce from the knowlegde of the
 {\bf State} $S(0)$ at the initial time, some meaningful information about
 the physical situation at later times.
Once more, the evolution may be deterministic or stochastic; in the first
 case it is a mapping of the set of the {\bf States} into (or onto) itself,
 in the second case it is a mapping from the set of {\bf States} to the set
 of {\bf states}.
Since we are interested in discussing the cases of standard quantum mechanics
 and of classical mechanics, we will assume that the evolution is perfectly
 deterministic and reversible, i.e. it is an injective and surjective mapping
 of the set of the {\bf States} onto itself.

At this point we are led to analyze the last and essential feature of the
 theory, i.e., its allowing to make predictions about the outcomes of
 prospective measurements of physical observables. The question should be
 clear: we assume that we know the {\bf State} $S(t)$ (i.e. the most accurate
 specification which the theory makes legitimate concerning an individual
 physical system) at a given time $t,$ and we are interested in what the
 theory tells us about the outcomes of measurement procedures concerning all
 conceivable observable quantities at the considered time.
Once more the predictions of the theory can have a deterministic or a
 probabilistic character, the two paradigmatic cases being classical and
 quantum mechanics.

In classical mechanics the observable quantities are functions of the
 {\bf State}, i.e. of the point in phase space associated to the precise
 physical situation we are considering.
Let us denote as $F({\bf{r}}_{i},{\bf{p}}_{i})$ a generic observable.
Such a quantity takes a precise value at time $t,$ which is simply given by
 $F({\bf{r}}_{i}(t), {\bf{p}}_{i}(t))$.
Accordingly, classical mechanics is a deterministic theory at its fundamental
 level, i.e. when analyzed in terms of its {\bf States}.
Obviously, probabilities can enter into play also within such a theory; as
 already mentioned this happens when we are not dealing with the {\bf States}
 but with the {\bf states} of the theory.
However, the need to pass from the {\bf States} to the {\bf states}
 corresponds, for the considered case, to a lack of information about the
 system with respect to the one that the theory considers in principle
 possible: this allows to conclude that the probabilities of classical
 mechanics (and in particular the practically unavoidable ones of classical
 statistical mechanics) have an epistemic status, i.e., they are due to our
 ignorance about the physical system under consideration.

The situation in quantum mechanics is quite different. First of all, even
 when we deal with the {\bf States} (i.e. with the so called pure cases in
 which we know precisely the state vector of the system) the theory attaches
 (in general) probabilities different from 0 and 1 to the outcomes of
 measurements concerning almost all physical observables $F,$ which, as is
well known, are represented by self-adjoint operators $\widehat{F}$.
This means that, when the completeness assumption is made, quantum
 probabilities have a nonepistemic status.
However, it is useful to remark that for any given state vector (a {\bf State}
 in our language) there is always one (actually infinitely many) self-adjoint
 operator such that the considered state vector is an eigenstate of it
 belonging to an appropriate eigenvalue.
For such an observable the theory attaches probability 1 to the outcome
 corresponding to the eigenvalue in a measurement of the related observable,
 so that one can predict with certainty the outcome.
Even more: the formalism tells us that for a system in a pure state there
 are complete sets of commuting observables such that the state vector is a
 simultaneous eigenstate of all of them\footnote {To be more precise, given
 any complete set of commuting observables such that all its members commute
 with the projection operator on the one-dimensional manifold spanned by the
 state vector, the  state vector itself is a common eigenvector of the
 considered complete set of observables.
Accordingly, in the case under consideration there are various complete sets
 of observables such that the theory attaches probability 1 to a corresponding
 specific set of eigenvalues for each set.}.

When, within a quantum scheme, one passes from the consideration of the
 {\bf States} to that of the {\bf states}, the situation becomes richer and
 deserves further comments.
In fact, two conceptually different situations can occur, depending on the
 information one has about the system (or the inhomogeneous ensemble)
 associated to a {\bf state}.

In the first case one knows the precise probabilities $p_{i}$ of the system
 being in the {\bf State} $\vert \phi _{i}\rangle,$ or, equivalently, the
 precise fractions $p_{i}$ of the members of the ensemble which are in the
 considered state $\vert \phi_{i}\rangle$.
Then an interplay between epistemic and nonepistemic probabilities occurs:
 if we are interested in a specific observable $G$, in order to evaluate the
 probability of getting the outcome $g_{k}$ in a measurement we have to argue
 as follows.
There is an epistemic probability $p_{i}$ that my system (or the individual
 which is picked up from the ensemble) is described by the pure state
 $\vert \phi_{i}\rangle,$ and such a state attaches (in general) a
 nonepistemic probability, let us say $\pi_{i,k}$ to the outcome $g_{k}.$
It has to be remarked that if one takes such an attitude (i.e. he knows that
 the system can be only in one of the state vectors $\vert \phi_{i}\rangle$
 but he does not know precisely which one) it remains true that for each such
 state there are precise observables (different for different values of the
 index $i$) which have probability one of giving appropriate outcomes.
This remark is relevant for the problem of identifying the properties which
 can be considered as objectively possessed by individual physical systems.
As it is well known, the consideration of the statistical operator $\rho$
 (a trace class, trace one, semipositive definite operator) is the
 mathematically appropriate entity to deal with the {\bf states} of the system
 (or of the ensemble).
For the case considered above, it has the expression\footnote {It goes without
 saying that also in the case of a {\bf State} one can resort to the
 statistical operator formalism. In the considered case the statistical
 operator turns out to be the projection operator on the one-dimensional
 manifold spanned by the state vector and, as such, it satisfies the condition
 $\rho^{2}= \rho$.} $\rho =\sum_{r} p_{r}\vert \phi_{r}\rangle \langle
 \phi_{r}\vert $.

In the second case one knows the statistical operator but one is ignorant
 about the precise composition of the ensemble associated to the {\bf state}
 under consideration.
We stress the conceptual relevance of making the just mentioned distinction
  between the two above cases.
It derives from the precise formal fact that, while in classical mechanics
 non-pure states (i.e. {\bf states} in our language) are in one-to-one
 correspondence with statistical ensembles, in quantum mechanics this is by
 no means true; actually the correspondence between statistical ensembles and
 statistical operators is infinitely many to one.
To give just an elementary example, we can mention the case of an ensemble
 which is the union of two subensembles which are pure cases associated
 to the orthogonal states $\vert \varphi_{1}\rangle$ and $\vert \varphi_{2}
 \rangle,$ with weights $p_{1}$ and $p_{2}$, respectively, and suppose
 $p_{1}>p_{2}$.
The statistical operator can be written
\begin{eqnarray}
\label{b}
  \rho & = & p_{1}\vert \varphi _{1}\rangle \langle \varphi _{1}\vert +p_{2}
  \vert \varphi_{2}\rangle \langle \varphi_{2} \vert \nonumber\\
  & \equiv & p_{2}\left[ \vert \varphi_{1}\rangle \langle \varphi _{1}\vert +
  \vert \varphi_{2}\rangle \langle \varphi_{2}\vert \right] +
  \left( p_{1}-p_{2}\right) \vert \varphi_{1}\rangle \langle \varphi_{1}\vert.
\end{eqnarray}
One can then notice that in the second expression for $\rho $ the number
 $p_{2}$ multiplies the projection operator on the two dimensional manifold
 spanned by the set $\left\{|\varphi_{1}\rangle,|\varphi_{2}\rangle\right\}.$
Such a projection can be written in terms of any pair of orthogonal vectors
 $|\mu_{1}\rangle,|\mu _{2}\rangle,$ spanning the same manifold, so that
 $\rho $ can also be written as:
\begin{equation}
 \label{c}
 \rho = \left( p_{1}-p_{2}\right) |\varphi_{1}\rangle \langle \varphi_{1}|
 + p_{2}|\mu_{1}\rangle\langle \mu_{1}|+p_{2}|\mu_{2}\rangle \langle \mu_{2}|.
\end{equation}
Equation~(\ref{c}) shows that $\rho$ is also the statistical operator
 describing an ensemble which is the union of three pure subensembles
 associated to the nonorthogonal states $\{|\varphi_{1}\rangle,$
 $|\mu_{1}\rangle,|\mu_{2}\rangle\}$ with the indicated weights.
Since the state vectors $|\mu_{1}\rangle,|\mu_{2}\rangle$ are, in general,
 eigenstates of observables different and incompatible with those having
 $|\varphi_{2}\rangle$ as an eigenvector, the observables that have definite
 values when one member of the ensemble is chosen become ambiguous when we
 specify only the statistical operator and not the actual composition of the
 ensemble.
Obviously, the probabilities that the theory attaches to the outcomes of all
 conceivable measurement processes coincide for all ensembles associated to
 the same {\bf state} $\rho$, but from a conceptual point of view there is a
 subtle difference between the two cases, which should be clear to the reader
 and which we will reconsider in Section 3.

The considerations of this section, when reference is made to the two cases
 of interest for us, i.e., Classical Mechanics and Quantum Mechanics, can be
 summarized as follows: \\ \\
{\it\bf Classical Mechanics}: \vspace{0.6cm}

  \begin{center}
  \begin{tabular}{|c|c|c|}
  \hline
  & & \\
  \textbf{State} &

\begin{tabular}{c}
Point of the phase space  \\
$(\,q_{i}\,,\,p_{i}\,)$
\end{tabular}
 & Determinism \\
  & & \\
  \hline
  & & \\
  \textbf{state} &

\begin{tabular}{c}
 Probability measure on the phase space \\
$\rho\,(\,q_{i}\,,\,p_{i}\,)$
\end{tabular}
&  Epistemic Probabilities \\
  & & \\
  \hline
  \end{tabular}
  \end{center}\vspace{1cm}


  \noindent
  {\it\bf Quantum Mechanics}: \vspace{0.6cm}

  \begin{center}
  \begin{tabular}{|c|c|c|}
  \hline
  & & \\
 \textbf{State} &

\begin{tabular}{c}
State Vector \\
$\vert \psi \rangle$
\end{tabular}

 & Nonepistemic Probabilities \\
  & & \\
  \hline
  & & \\
  \textbf{state} &

\begin{tabular}{c}
 Statistical operator \\
$\rho = \sum_{i}p_{i}\vert \psi_{i} \rangle\langle \psi_{i}\vert $
\end{tabular}

&  Epistemic and Nonepistemic Probabilities \\
  & & \\
  \hline
  \end{tabular}
  \end{center}\vspace{1cm}
As discussed above, within quantum mechanics it is useful to keep in mind
 that for a given {\bf state} we can still have different information about
 the ensemble associated to it, according whether we know the composition
 (weights and state vectors) of the ensemble itself in terms of its pure
 subensembles or we know nothing besides the statistical operator.


\section{Properties of individual physical systems}
\label{sezionetre}

In this Section we tackle the problem of attributing properties to individual
 physical systems. In order to come to the most interesting point of our
 analysis, i.e. to discuss the specific problems which arise in connection
 with this matter in the case of composite quantum systems in entangled
 states, it is appropriate to reconsider briefly the case of Classical
 Mechanics.
Within such a theory, as already stated, all conceivable observables, both
 referring to the whole system as well as to all its subsystems, are functions
 of the positions and momenta of the particles, so that, when one knows the
 {\bf State} of the system, i.e., the phase space point associated to it, one
 knows also the precise values of all physical observables.
We can claim that within Classical Mechanics all properties are objectively
 possessed, in the precise sense that the measurement of any given observable
 simply reveals the pre-existing value possessed by the
 observable~\footnote{Here we have tacitly assumed  that one can perform
 ideally faithful measurements, i.e., measurements which reveal precisely the
 value of the quantities they are devised to measure. In what follows we will
 also make the corresponding idealized assumption for the quantum case, i.e.
 that if the state vector is an eigenstate of an observable, its measurement
 will yield with certainty the associated eigenvalue.}.
It goes without saying that if we lack the complete information about the
 system, then we can make statements only concerning the (epistemic)
 probabilities that it possesses precise properties.
Nevertheless, it remains true that any individual system (and its subsystems)
 has all conceivable properties, in spite of the fact that we can be ignorant
 about them \footnote{Stated differently, claims of the kind ``the energy of
 this particle has this specific value" have truth values, i.e. they are
 definitely either true or false.}.

As everybody knows, the situation is quite different in quantum mechanics
 due to the nonabelian structure of the set of the observables. Accordingly,
 as already discussed, the theory, in general, consents to make only
nonepistemic
 probabilistic predictions about the outcomes of measurement processes even
 when the {\bf State} of the system is known.
However, in such a case, there are always complete sets of commuting
 observables such that the theory attaches probability one to a precise
 outcome in a measurement process of any one of them.
It is then natural to assume (as we will do) that when we can make certain
 (i.e. with probability one) predictions  about the outcomes, the system
 possesses objectively the property, or element of physical reality ``such
 an observable has such a value", independently of our decision to measure it.
Here we have used the expression {\em objective properties} and {\em elements
 of reality} with the same meaning that Einstein \cite{ref2} gave  them in
 the analysis of the EPR paradox$\,$:
\begin{quotation}
  {\sl If, without in any way disturbing a system, we can predict with
 certainty (i.e. with a probability equal to unity) the value of a physical
 quantity, then there exists an element of physical reality corresponding to
 this physical quantity.}
\end{quotation}
When one takes into account the just outlined situation, one can concisely
 express the lesson that quantum mechanics has taught us, by stating that
 within such a theory one cannot consider (even in principle) an individual
 physical system as possessing objectively too many properties.
Some of them can be legitimately considered as {\it actual}, all the other
 have the ontological status of \textit{potentialities}.
At any rate, according to the remarks of the previous Section, a system in a
 pure state always has complete sets of definite and objective properties.

Obviously, when we have not the most accurate knowledge of the physical
 situation of the system under consideration, i.e., when we know only its
 {\bf state}, then, in general, we can at most make epistemic probabilistic
 statements even about the limited set of properties that the system might
 possess.
As discussed in the previous Section we can, in principle, have a different
 knowledge of the state of the system according whether we know the precise
 composition of the ensemble to which it belongs or only the statistical
 operator associated to it.
The difference can be easily appreciated  by considering a particular
 instance of the situation analyzed in the previous section concerning
 different ensembles associated to the same statistical operator.
Suppose we have an unpolarized beam of spin $1/2$ particles and we are
  interested only in their spin properties.
The statistical operator corresponding to it is ($1/2)I$ ($I$ being the
  identity operator).
However, such a state can describe, e.g., an ensemble of particles with
 uniform distribution of their spins over all directions, or an  ensemble
 obtained by putting together an equal number of particles in the
eigenstates associated to the eigenvalue
+1 and -1 of the observable $\sigma_{z}$, respectively.
While in the second case the statement ``each particle of the ensemble has
 surely either the spin {\it up} or {\it down} along the {\it z}-axis" is
 legitimate and true independently of any measurement being actually
 performed, it is certainly illegitimate in the first one.\footnote{Obviously,
 in both cases, if one subjects all particles to a measurement of their
 {\it z}-spin component, he will get almost in $1/2$ of the cases the outcome
 {\it up} and in $1/2$ of the cases the outcome {\it down}. However, here
 we are not making exclusive reference to the outcomes, but to the possibility
 of considering a property as objectively (i.e. independently of our decision
 to perform a measurement) possessed. From this point of view the two cases
 are radically different.}

Up to this point we have confined our attention  to quantum systems
 considered as a whole.
However, as already mentioned, the phenomenon of quantum entanglement makes
 the situation much more puzzling when consideration is given to composite
 quantum systems and one raises the problem of the properties of their
 constituents.
As we will see, in such a case it is very  common to meet situations (most
 of which arise as a consequence of the interactions between the constituents)
 in which the constituents themselves do not possess {\it any property
 whatsoever}.
This is a new feature which compels us to face a quite peculiar state of
 affairs: not only must one  limit drastically the actual properties of
 physical systems (being in any case true that the system as a whole always
 has some properties), but one is forced also to accept that the parts of a
 composite system can have no property at all.
Only the entire system, even if its parts are far apart and noninteracting,
 has some properties, while its parts have only {\it potentialities} and no
 {\it actualities}. In this way  the quantum picture of the universe as an
 ``unbroken whole", or as ``undivided", emerges.

Quantum entanglement has played a central role in the historical development
 of quantum mechanics, in particular since it has compelled the scientific
 community to face the essentially nonlocal features of natural processes.
Nowadays, entangled states have become the essential ingredients of all
 processes involving teleportation and quantum cryptography and constitute
 an important tool for implementing efficient quantum algorithms.
This explains why a great deal of efforts has been spent by theorists during
 the last years in trying to characterize the very nature and properties of
 entanglement, and this is also the reason which motivates our attempt to
 deepen some questions about these matters.
\newpage


\vspace{1cm}

\begin{center}
 {\LARGE\bf  Part II: Entanglement and properties of quantum systems of
 distinguishable particles in pure states}
\end{center}

As already stressed, the problem of attributing properties to the constituents
 of composite systems in entangled states is a rather delicate one.
Here we will discuss this  problem and we will derive a series of significant
 theorems with particular reference to systems of distinguishable particles
 in pure states.


\section{Entanglement of two distinguishable particles}

In this section we study the {\bf Entanglement} between two distinguishable
 particles ${\cal S}_{1}$ and ${\cal S}_{2}$.
Let us suppose that the two particles are parts of a larger quantum system
 ${\cal S}={\cal S}_{1} + {\cal S}_{2}$, whose associated Hilbert space
 ${\cal H}$ is the direct product of the Hilbert spaces of the single
 subsystems, ${\cal H}={\cal H}_{1}\otimes {\cal H}_{2}$.
As already stated, in this Part of the paper we will always assume that the
 composite quantum system ${\cal S}$ is  described by a state vector $\vert
 \psi(1,2) \rangle \in {\cal H}$ or, in a totally equivalent manner, by a
 pure density operator $\rho =\vert \psi(1,2) \rangle\langle \psi(1,2) \vert$.


\subsection{General definition and theorems}
\label{generaldefinitions}

Let us start by characterizing a non-entangled (a separable) composite system
 by making explicit reference to the fact that one of its two constituent
 subsystems possesses complete sets of properties (as we will see this in
 turn implies that the same is true for the other constituent):
\begin{defofentang}
 \label{defofentang}
 The system ${\cal S}_{1}$, subsystem of a composite quantum system
 ${\cal S}={\cal S}_{1} + {\cal S}_{2}$ described by the pure density
 operator $\rho$, is {\bf non-entangled} with subsystem ${\cal S}_{2}$ if
 there exists a projection operator $P^{(1)}$ onto a  one-dimensional manifold
 of ${\cal H}_{1}$ such that:
\begin{equation}{\nonumber}
 \label{separabilitanonidentici}
  Tr^{(1+2)}[\,P^{(1)}\otimes I^{(2)}\rho\,]=1.
\end{equation}
\end{defofentang}
The fact that in the case of non-entangled states it is possible to consider
 each one of the constituents as possessing complete sets of well definite
 physical properties, independently of the existence of the other part,
 follows directly from the following theorem:

\begin{factorizability}
  \label{factorizability}
  If consideration is given to a composite quantum system ${\cal S}=
 {\cal S}_{1} + {\cal S}_{2}$ described by the pure state vector $\vert
 \psi(1,2)\rangle $ (or, equivalently by the pure density operator
 $\rho =\vert \psi(1,2) \rangle\langle \psi(1,2) \vert$) of ${\cal H}=
 {\cal H}_{1}\otimes {\cal H}_{2}$, each of the following three conditions
 is necessary and sufficient in order that subsystem ${\cal S}_{1}$ is
 non-entangled with subsystem ${\cal S}_{2}$:
\begin{enumerate}
 \item  there exists a projection operator $P^{(1)}$ onto a one-dimensional
  manifold of ${\cal H}_{1}$ such that $Tr^{(1+2)}\left[ P^{(1)}\otimes
  I^{(2)}\rho \right] =1;$

 \item  the reduced statistical operator $\rho ^{(1)}=Tr^{(2)}\left[\,\rho\,
 \right] $ of subsystem $S_{1}$ is a  projection operator onto a
 one-dimensional manifold of ${\cal H}_{1}$;

 \item  the state vector $\left| \psi (1,2)\right\rangle $ is factorizable,
 i.e., there exist a state $\vert\phi(1) \rangle \in {\cal H}_{1}$ and a
 state $\vert \xi(2) \rangle \in {\cal H}_{2}$ such that $\vert \psi(1,2)
 \rangle= \vert \phi(1) \rangle\,\otimes \vert \xi(2) \rangle$.
\end{enumerate}
\end{factorizability}

\noindent {\it Proof:}

If subsystem $S_{1}$ is non-entangled with $S_{2}$ then, according to
 Definition~\ref{defofentang} condition 1 is satisfied, i.e.
 $Tr^{(1+2)}\left[ P^{(1)}\otimes I^{(2)}\rho \right] =Tr^{(1)}\left[ P^{(1)}
 \rho ^{(1)}\right]=1$.
Since $P^{(1)}$ projects onto a one-dimensional manifold and $\rho^{(1)}$
 is a statistical operator ( i.e., a trace-class, trace one, semipositive
 definite and hermitian operator), the last equality implies $\rho^{(1)} =
 P^{(1)},$ i.e., $\rho^{(1)}$ is a projection operator onto a one-dimensional
 manifold.

If $\rho^{(1)}$ is a projection operator onto a one-dimensional manifold
 (condition 2) then it is useful to resort to  von Neumann's biorthonormal
 decomposition of the state $|\psi (1,2)\rangle$ in terms of states of
 ${\cal H}_{1}$ and ${\cal H}_{2}$ (if there is any accidental degeneracy
 we can dispose of it as we want):
\begin{equation}
\label{vN}
 |\psi (1,2)\rangle=\sum_{k}\pi_{k}|\phi_{k}(1)\rangle \otimes
 |\xi_{k}(2)\rangle
\end{equation}
where the $\pi_{k}$ are real and positive numbers satisfying
 $\sum_{k}\pi_{k}^{2}=1.$ Equation~(\ref{vN}) implies:
\begin{equation}
 \label{vN2}
 Tr^{(2)}[\,|\psi (1,2)\rangle\langle \psi (1,2)|\,]=\sum_{k}\pi_{k}^{2}\,|
  \phi_{k}(1)\rangle\langle \phi_{k}(1)|.
 \end{equation}
The r.h.s of this equation, due to the orthogonality of the states
 $|\phi_{k}(1)\rangle,$ can coincide with a projection operator onto a
 one-dimensional manifold iff the sum in~(\ref{vN2}) contains only one term,
 let us say the first one, the corresponding coefficient $\pi_{1}$ taking
 the value 1. Accordingly, from~(\ref{vN}) we get:
\begin{equation}
 \label{vN3}
 |\psi (1,2)\rangle=|\phi_{1}(1)\rangle \otimes |\xi_{1}(2)\rangle,
\end{equation}
i.e. $|\psi (1,2)\rangle$ is factorized.

Finally, if $\vert \psi(1,2)\rangle $ is factorized as in~(\ref{vN3}),
 then the one-dimensional projection operator $P^{(1)}=|\phi_{1}(1)\rangle
 \langle \phi_{1}(1)|$ satisfies:
\begin{equation}
 Tr^{(1+2)}\left[ P^{(1)}\otimes I^{(2)}\rho \right] =1,
\end{equation}
and subsystem $S_{1}$ is not entangled with $S_{2}.$$\:\:\:\:\:\:\:\:\:\Box$


\subsection{Entanglement and properties of two distinguishable particles}
\label{entanglementandpropertiesoftwo}

The analysis of the previous subsection allows us to conclude that if a
 quantum system composed of two subsystems is non-entangled, the states of
 subsystems ${\cal S}_{1}$ and ${\cal S}_{2}$ are completely specified, in
 the sense that it is possible to associate to each of them a unique and
 well-defined state vector.
According to our previous discussion, the individual subsystems can therefore
 be thought of as having complete sets of definite and objective properties
 of their own.

We pass now to analyse the case of composite systems of two subsystems in
 entangled states. According to Theorem~\ref{factorizability}, the reduced
 density operator of each subsystem is not a projection operator onto a one
 dimensional manifold. It is then useful to analyse whether there exist
 projection operators on manifolds of dimension greater than or equal to 2 of
 ${\cal H}_{1}$, satisfying condition~(\ref{separabilitanonidentici}).
As shown by the following theorem, there
 is a strict relation between such projection operators and the range
 ${\cal R}[\, \rho^{(1)}\,]$ of the reduced statistical operator $\rho^{(1)}$:
\begin{rank}
 \label{rank}
 A necessary and sufficient condition for the  projection
 operator $P^{(1)}_{{{\cal{M}}_{1}}}$ onto the linear manifold
 ${\cal M}_{1}$ of ${\cal H}_{1}$ to satisfy the two following conditions:

\begin{enumerate}
 \item $Tr^{(1)}[\,P^{(1)}_{{\cal{M}}_{1}} \rho^{(1)}\,]=1$;
 \item there is no projection operator $\tilde{P}^{(1)}$ of ${\cal H}_{1}$
 satisfying the conditions $\tilde{P}^{(1)} < P^{(1)}_{{\cal{M}}_{1}}$
 ($\,$i.e. it projects onto a proper submanifold ${\cal N}_{1}$ of
 ${\cal M}_{1}$) and   $Tr^{(1)}[\, \tilde{P}^{(1)} \rho^{(1)}\,] =1$,
\end{enumerate}

\noindent is that the range ${\cal R}[\, \rho^{(1)}\,]$ of the reduced
 statistical operator $\rho^{(1)}$ coincides with ${\cal M}_{1}$.
\end{rank}

\noindent {\it Proof:}

If ${\cal R}[\, \rho^{(1)}\,] = {{\cal M}}_{1}$, then
 $P^{(1)}_{{{\cal M}}_{1}} \rho^{(1)} = \rho^{(1)}$ implying that
 $Tr^{(1)}[\,P^{(1)}_{{{\cal M}}_{1}} \rho^{(1)}\,] = Tr^{(1)}[\,\rho^{(1)}\,]
 =1$, while any projection operator $\tilde{P}^{(1)}$ on a proper submanifold
 of ${\cal M}_{1}$ satisfies $Tr^{(1)}[\,\tilde{P}^{(1)} \rho^{(1)}\,] < 1$.

Conversely if $Tr^{(1)}[\,P^{(1)}_{{\cal{M}}_{1}} \rho^{(1)}\,]=1$ we consider
 the spectral decomposition of $\rho^{(1)}$, where only eigenvectors
 corresponding to non-zero eigenvalues appear:
\begin{equation}
  \rho^{(1)}=  \sum_{i} p_{i} \vert \varphi_{i}(1)\rangle \langle
  \varphi_{i}(1)\vert\,, \:\:\:\:\:\:\:\:\:\:\sum_{i} p_{i}=1,\:\:\: p_{i}
  \neq 0 \:\:\:\:\: \forall i.
\end{equation}
We then have
\begin{equation}
 Tr^{(1)}[\,P^{(1)}_{{\cal{M}}_{1}} \rho^{(1)}\,]=
 \sum_{i} p_{i}\, Tr^{(1)}[\,P^{(1)}_{{\cal{M}}_{1}}
 \vert \varphi_{i}(1)\rangle \langle\varphi_{i}(1)\vert \,]=
 \sum_{i} p_{i}\, || P^{(1)}_{{\cal{M}}_{1}} \vert \varphi_{i}(1)\rangle||^{2}
 =1
\end{equation}
implying, since $\sum_{i}p_{i}=1$ and $|| P^{(1)}_{{\cal{M}}_{1}} \vert
 \varphi_{i}(1)\rangle ||^{2}\leq 1$, that $P^{(1)}_{{\cal{M}}_{1}} \vert
 \varphi_{i}(1)\rangle= \vert \varphi_{i}(1)\rangle$.

The last equation shows that $P^{(1)}_{{\cal{M}}_{1}}$ leaves invariant
 ${\cal R}[\, \rho^{(1)}\,]$ so that ${{\cal M}}_{1} \supseteq {\cal R}[\,
 \rho^{(1)}\,]$.
If the equality sign holds we have proved the sufficiency. On the contrary if
 ${{\cal M}}_{1} \supset {\cal R}[\, \rho^{(1)}\,]$, then the projection
 operator on the closed submanifold ${\cal R}[\, \rho^{(1)} \,]$ of
 ${{\cal M}}_{1}$ satisfies condition $Tr^{(1)}[\, P^{(1)}_{{\cal R}
 [\rho^{(1)}]} \rho^{(1)}\,]=1$ contrary to the assumptions.
We have therefore proved that ${\cal R}[\, \rho^{(1)} \,]={{\cal M}}_{1}$.
 $\:\:\:\:\:\:\Box$

Let us analyze in detail the consequences of the above theorem by studying
 the following two cases concerning the range of the reduced statistical
 operator $\rho^{(1)}$:

\begin{enumerate}
 \item ${\cal R}[\,\rho^{(1)}\,]={\cal M}_{1} \subset {\cal H}_{1}$,
 \item ${\cal R}[\,\rho^{(1)}\,]={\cal H}_{1}$,
\end{enumerate}

As it has just been shown, in the first case (recall that we are considering
 the case in which  the dimensionality of the manifold ${\cal M}_{1}$ is
 strictly greater than one) the projection operator $P^{(1)}_{{\cal{M}}_{1}}$
 on ${\cal M}_{1}$ is such that $Tr^{(1)}[\,P^{(1)}_{{\cal{M}}_{1}}
 \rho^{(1)}\,]=1$.
Accordingly, given any self-adjoint operator $\Omega ^{(1)}$ of ${\cal H}_{1}$
 which commutes with $P^{(1)}_{{{\cal M}}_{1}},$ if consideration is given to
 the subset ${\cal B}$ (a Borel set) of its spectrum coinciding with the
 spectrum of its restriction $\Omega_{R}=P^{(1)}_{{{\cal M}}_{1}}\Omega^{(1)}
 P^{(1)}_{{{\cal M}}_{1}}$ to ${\cal M}_{1},$ we can state that subsystem
 ${\cal S}_{1}$ has the objective (in general unsharp) property that
 $\Omega^{(1)}$ has a value belonging to ${\cal B}$.
In particular, all operators which have ${\cal M}_{1}$ as an eigenmanifold,
 have a precise objective value.
Summarizing, even though in the considered case we cannot say that subsystem
 ${\cal S}_{1}$ has a complete set of properties by itself (i.e. objectively),
 it still has some sharp or unsharp properties associated to any observable
 which commutes with $P^{(1)}_{{{\cal M}}_{1}}$.

On the contrary, in the second  of the above cases, i.e. the one in which
 ${\cal R}\,[\rho^{(1)}\,]={\cal H}_{1}$, we have to face the puzzling
 implications of entanglement in their full generality. In fact, repetition
 of  the argument we have just developed leads in a straightforward way,
 when ${\cal  M}_{1}$ coincides with ${\cal H}_{1}$, to the conclusion that
 the only projection operator $P^{(1)}$ on ${\cal H}_{1}$ satisfying
 $Tr^{(1)}[P^{(1)}\rho ^{(1)}]=1$ is the identity operator $I^{(1)}$ on the
 Hilbert space ${\cal H}_{1}$.
The physical meaning of this fact should be clear to the reader: it amounts
 to state that subsystem ${\cal S}_{1}$ does not possess objectively any
 sharp or unsharp property, i.e., that there is no self-adjoint operator for
 which one can claim with certainty that the outcome of its measurement will
 belong to any proper subset of its spectrum.
The only certain but trivial statement\footnote{Obviously, the theory attaches
 precise probabilities to the outcome belonging to any chosen Borel subset of
 the spectrum, but no one of such probabilities takes the value 0 or 1.
Stated differently, in the considered case, we cannot speak of any (even
 quite unsharp) property as actual, all conceivable properties having the
 ontological status of potentialities.} which is legitimate is that the
 outcome will belong to the spectrum.
In the above situation we say that subsystem ${\cal S}_{1}$ is {\em totally
 entangled} with subsystem ${\cal S}_{2}$.

In the just considered case of total entanglement we have (appropriately)
 assumed the Hilbert spaces ${\cal H}_{1}$ and ${\cal H}_{2}$ to be
 infinite-dimensional.
If one is interested in finite, let us say $N$-dimensional Hilbert spaces,
 then, besides the fact that the range of $\rho^{(1)}$ is the whole Hilbert
 space ${\cal H}_{1}$, it can also happen that $\rho^{(1)}$ itself is a
 multiple (by the factor $1/N$) of the identity operator.
In such a case, the subsystem ${\cal S}_{1}$ not only does not possess any
 objective property, but it is characterized by the fact that the
 probabilities of giving any outcome in a measurement of any complete set of
 commuting observables are all equal to $1/N$.
In a very precise sense one could state that the system has only
 potentialities and moreover that they are totally indefinite.
Since in many processes involving quantum teleportation, quantum cryptography
 and in the studies about quantum information one often makes reference to
 finite-dimensional Hilbert spaces, and thus one can easily meet the just
 mentioned situation, a specific term has been introduced to deal with this
 state of affairs in which the properties of each subsystem are completely
 indefinite.
Accordingly, the entangled states of two subsystems for which the reduced
 statistical operators are multiples of the identity are usually referred to
 as {\em maximally entangled states} \footnote{In the infinite-dimensional
 case, to meet maximal entanglement one is compelled to enlarge the class of
 states and to resort to states which cannot be associated to bounded, trace
 class, trace one, semipositive definite operators. We will not analyze here
 this case.}.

In order to clarify the two paradigmatic situations we have just analyzed,
 let us consider the simple case of a system composed by an electron and a
 positron, which we will label as particle 1 and particle 2 respectively. \\

\noindent {\em Example 1:} Let us suppose that the $e^{-}e^{+}$ system is
 described by the following state vector (with obvious meaning of the symbols):

\begin{equation}
 \label{esempioprimo}
   \vert \psi(1,2) \rangle = \frac{1}{\sqrt{2}}\,[\, \vert \vec{n}\uparrow
   \rangle_{1}\vert \vec{n}\downarrow \rangle_{2} - \vert \vec{n}\downarrow
   \rangle_{1} \vert \vec{n}\uparrow \rangle_{2} \,]\otimes \vert R \rangle_{1}
   \vert L \rangle_{2},
\end{equation}

\noindent where we have indicated with $\vert R \rangle$ and $\vert L \rangle$
 two orthonormal states, whose coordinate representations are two specific
 square-integrable functions having compact disjoint supports at Right and
 Left, respectively.
The reduced density operator describing the electron (obtained by taking the
 trace on the degrees of freedom of the second particle) acts on the
 infinite dimensional Hilbert space,  ${\cal H}_{1}={\cal{C}}^{2} \otimes
 {\cal{L}}(R^{3})$ and has the following form:

\begin{equation}
 \label{esempiouno}
   \rho^{(1)}= \frac{1}{2}[\,\vert \vec{n}\uparrow \rangle_{11} \langle
   \vec{n}\uparrow \vert + \vert \vec{n}\downarrow\rangle_{11} \langle
   \vec{n}\downarrow\vert \,] \otimes \vert R \rangle_{11} \langle R \vert
   =\frac{1}{2}I \otimes \vert R \rangle_{11} \langle R \vert
\end{equation}

Even though we cannot say anything about the value of the spin along any
 arbitrary direction $\vec n$, we can nevertheless state that the electron is,
 with certainty, inside the bounded right region $R$, and an analogous
 statement, i.e. that it is inside the bounded region $L$, can be made
 concerning the positron.
Therefore the subsystems do not possess a complete set of properties with
 respect to both spin and position, but have at least the element of reality
 of being in definite spatial regions.
This possibility of making claims about some properties is due to the fact
 that the range of the statistical operator of equation~(\ref{esempiouno})
 is a proper submanifold of ${\cal {H}}_{1}$, i.e. the two dimensional
 manifold  spanned by $\vert \vec{n}\uparrow \rangle_{1}\vert R\rangle_{1}$
 and $\vert \vec{n}\downarrow \rangle_{1}\vert R\rangle_{1}$. \\

\noindent {\em Example 2:} In place of state~(\ref{esempioprimo}) we consider
 now the following state vector for the $e^{-}e^{+}$ system:

\begin{equation}
 \label{totallyentangled}
  \vert \psi(1,2) \rangle = \frac{1}{\sqrt{2}}\,[\, \vert \vec{n}\uparrow
  \rangle_{1}\vert \vec{n}\downarrow \rangle_{2} - \vert \vec{n}\downarrow
  \rangle_{1} \vert \vec{n}\uparrow \rangle_{2}\,]\otimes
  \,[\sum_{i}c_{i}\vert\varphi_{i}(1)\rangle\,\vert\theta_{i}(2)\rangle
  \,],\qquad c_{i}\neq 0\:\:\forall i,
\end{equation}

\noindent $\left\{\,\vert\varphi_{i}(1)\rangle \right\}$ and $\left\{\,\vert
 \theta_{i}(2)\rangle\,\right\}$ being two complete orthonormal sets of the
 Hilbert spaces ${\cal{L}}(R^{3})$ associated to the spatial degrees of
 freedom of the constituents.
The reduced density operator for the electron is:

\begin{equation}
\label{hegel}
 \rho^{(1)}=Tr^{(2)}[\,|\psi (1,2)\rangle \langle \psi (1,2)|\,]=
 \frac{1}{2}I^{(1)}\otimes \sum_{i}|c_{i}|^{2}|\varphi_{i}(1)\rangle \langle
 \varphi_{i}(1)|.
\end{equation}

\noindent In eq.~(\ref{hegel}), $I^{(1)}$ is the identity operator in the spin
 space of the electron.
Since the range of $\rho^{(1)}$ is now the whole Hilbert
 space of the first particle, according to the previous discussion we cannot
 attribute any element of reality referring to any conceivable observable of
 the electron: it possesses only potential and no actual properties.

We hope to have succeeded in giving the appropriate emphasis to the remarkable
 peculiarities of {\it the most characteristic trait of quantum mechanics}
 and in having made clear that it compels us to accept that {\it the
 subsystems of a composite system may have no property at all which can be
 considered as objectively possessed}.

Summarizing, with reference to the range ${\cal R}\,[\, \rho^{(1)}\,]$ of the
 reduced statistical operator, we can conclude that:

\begin{itemize}
 \item ${\cal R}\,[\, \rho^{(1)}\,] = \mbox {a one-dimensional manifold}
  \:\:\Rightarrow\:\:$   subsystem ${\cal S}_{1}$ is {\em non-entangled} with
  ${\cal S}_{2} \:\:\Rightarrow\:\:$ it possesses complete sets of objective
  properties, the same holding true for ${\cal S}_{2}$;
 \item ${\cal R}\,[\, \rho^{(1)}\,] = \mbox {a proper submanifold of} \:\:\
  {\cal H}_{1}\:\: \mbox {of dimension greater than 1} \:\:\Rightarrow\:\:$
  subsystem ${\cal S}_{1}$ is {\em partially entangled} with ${\cal S}_{2}
  \:\:\Rightarrow\:\:$ it possesses some  objective properties, however not a
  complete set of them;
 \item ${\cal R}\,[\, \rho^{(1)}\,] ={\cal H}_{1}\:\:\Rightarrow\:\:$
   subsystem ${\cal S}_{1}$ is {\em totally entangled} with ${\cal S}_{2}
   \:\:\Rightarrow\:\:$ it does not possess any objective property whatsoever.
\end{itemize}

\noindent Note that in the second case it may very well happen that while
 ${\cal R}\,[\, \rho^{(1)}\,]$ is a proper submanifold of ${\cal H}_{1}$,
 ${\cal R}\,[\, \rho^{(2)}\,]$  coincides with ${\cal H}_{2}$.
Analogously, in the last case it may happen that ${\cal R}\,[\,\rho^{(2)}\,]$
 is a proper submanifold of ${\cal H}_{2}$.


\subsection {Entanglement and correlations}
\label{quattropuntotre}

Another consequence of the entanglement of a composite quantum system is the
 occurrence of strict correlations between appropriate observables of the
 component subsystems, even when they are far apart and non-interacting.
This is expressed by the following theorem:

\begin{correlation}
 \label{correlation}
 Subsystem ${\cal S}_{1}$ is non-entangled with subsystem ${\cal S}_{2}$
 iff, given the pure state $\vert \psi(1,2) \rangle$ of the composite system,
 the following equation holds for any pair of observables $A(1)$ of
 ${\cal H}_{1}$ and $B(2)$ of ${\cal H}_{2}$ such that $\vert\psi(1,2)\rangle$
 belongs to their domains:
\begin{equation}
 \label{correlazione}
   \langle \psi(1,2)\vert A(1)\otimes B(2) \vert\psi(1,2) \rangle =
   \langle \psi(1,2)\vert A(1)\otimes I^{(2)} \vert\psi(1,2) \rangle
   \langle \psi(1,2)\vert I^{(1)}\otimes B(2) \vert\psi(1,2) \rangle.
\end{equation}

Note that equation~(\ref{correlazione}) implies that no correlation exists
 between such pairs of observables.
\end{correlation}

\noindent {\it Proof:}

If ${\cal S}_{1}$ is non entangled with ${\cal S}_{2}$ then, according to
 Theorem~\ref{factorizability}, $\vert\psi(1,2) \rangle$ is factorized, from
 which equation~(\ref{correlazione}) follows trivially.

Let us now assume that equation~(\ref{correlazione}) is satisfied by all
 bounded operators $A(1)$ and $B(2)$. Given the state $\vert\psi(1,2)\rangle$
 we consider its biorthonormal decomposition
\begin{equation}
 \label{decomposizione}
  \vert \psi(1,2) \rangle = \sum_{i} p_{i} \vert \varphi_{i}(1) \rangle
  \vert \theta_{i}(2)\rangle.
\end{equation}
Now we choose:
\begin{equation}
 A(1)=\vert \varphi_{r}(1)\rangle \langle \varphi_{r}(1) \vert
 \:\:\:\:\:\:\:\:\:\:\:\:\:\:\:
 B(2)=\vert \theta_{r}(2)\rangle \langle \theta_{r}(2) \vert,
\end{equation}
and we impose equation~(\ref{correlazione}) to hold for the considered state
 and the chosen observables. Since:
\begin{equation}
 \langle \psi(1,2) \vert A(1)B(2) \vert \psi(1,2) \rangle = p_{r}^{2}
 \end{equation}
\[
 \langle \psi(1,2) \vert A(1)\vert \psi(1,2) \rangle
 \langle \psi(1,2) \vert B(2)\vert \psi(1,2) \rangle=p_{r}^{4},
\]
the request~(\ref{correlazione}) implies that $p_{r}=1$ or that $p_{r}=0$
 for any r, which, by taking into account that $\sum_{i} p_{i}^{2}=1$ shows
 that only one term occurs in equation~(\ref{decomposizione}), i.e.:
\begin{equation}
 \vert \psi(1,2)\rangle = \vert \varphi_{k}(1) \rangle \vert \theta_{k}(2)
 \rangle,
\end{equation}
for an appropriate {\it k}.
We have thus proved that if the state is not factorized there is at least a
 pair of observables for which equation~(\ref{correlazione}) is not satisfied.
 $\:\:\:\Box$

The two observables appearing in equation~(\ref{correlazione}) being
 completely arbitrary, the previous theorem holds also for projection
 operators: with such a choice one sees that in the case of entanglement the
 joint probabilities $\langle \psi(1,2)\vert P^{(1)}\otimes P^{(2)}
 \vert\psi(1,2) \rangle$ for outcomes of independent measurement processes
 performed on both subsystems cannot be expressed, in general, as the product
 of the probabilities for the two outcomes.
Now, within quantum mechanics with the completeness assumption, it is easy to
 prove~\cite{refftl1,refftl2,refftl3,refftl4} that the mere fact of performing
 a measurement on one of the two entangled systems cannot alter the
 probability of any given outcome for a measurement on the other.
If one takes into account this fact, the just outlined situation implies
 that at least some probabilities for the outcomes of measurements on
 subsystem ${\cal S}_{1}$ depend on the outcomes of measurements performed
 on ${\cal S}_{2}$. Such a peculiar feature\footnote{For a detailed discussion
 of this point see refs. \cite{re6,re7,re8,re9}} displayed by systems in
 non-separable (i.e. non factorized)  states is usually termed as {\em Outcome
 Dependence}.

As it is well known the simplest example of this curious characteristics of
 entangled quantum systems is represented by the paradigmatic case of the
 singlet state of two distinguishable spin $1/2$-particles when one disregards
 the spatial degrees of freedom:

\begin{equation}
 \vert \psi(1,2) \rangle =\frac{1}{\sqrt{2}}[\, \vert z\uparrow \rangle_{1}
 \vert z\downarrow \rangle_{2} - \vert z\downarrow \rangle_{1}\vert z\uparrow
 \rangle_{2}\,].
\end{equation}

The joint probability, $Pr(\sigma_{1z}=\uparrow,\sigma_{2z}=\downarrow)$, of
 finding the first particle with spin up and the second particle with spin
 down along $z$-direction, when two measurements along the same axis are
 actually performed, equals 1/2. On the other hand the probabilities
 $Pr(\sigma_{1z}=\uparrow)$ and $Pr(\sigma_{2z}=\downarrow)$ of getting the
 indicated outcomes, which are defined as:
\begin{equation}
 Pr(\sigma_{1z}=\uparrow)=
 \sum_{y=\uparrow,\downarrow}Pr(\sigma_{1z}=\uparrow,\sigma_{2z}=y),
\end{equation}
\[
 Pr(\sigma_{2z}=\downarrow)=\sum_{x=\uparrow,\downarrow}
 Pr(\sigma_{1z}=x,\sigma_{2z}=\downarrow),
\]
are both equal to $1/2$ so that:
\begin{equation}
 \label{outcome}
 Pr(\sigma_{1z}=\uparrow)\cdot Pr(\sigma_{2z}=\downarrow)=\frac{1}{2}\cdot
 \frac{1}{2}.
\end{equation}
Thus, in accordance with Theorem~\ref{correlation}, the joint probability
 $Pr(\sigma_{1z}=\uparrow,\sigma_{2z}=\downarrow)=1/2$ does not match the
 value~(\ref{outcome}), proving therefore the {\em Outcome Dependence}
 property displayed by non-separable states.
This naturally leads to the conclusion that the probability distributions
 of the results of measurements on the two separate entangled subsystems in
 the singlet state are dependent from each other.


\section{Entanglement of $N$ distinguishable particles}

In this section we extend the previous analysis to the case in which the
 quantum system under consideration is composed by more than two
 subsystems.
This generalization is not trivial, since it requires to take into account
 all possible correlations between the component particles.
In particular, it may happen that a group of particles, which can be
 entangled or non-entangled among themselves, is not entangled with the
 remaining ones.
After having identified disentangled groups, one must repeat the analysis for
 the members of each group, up to the point in which he has grouped all the
 particles of the system (which here we assume to be all distinguishable from
 each other) into  sets which are disentangled from each other, while no
 further decomposition is possible. Apart from this complication the problem
 can be tackled by following step by step the procedures we have used in the
 previous sections to deal with systems composed of two distinguishable
 constituents.
The extreme case, as we will see, is that of a system in a state corresponding
 to completely non-entangled constituents. Since all arguments and proofs of
 the theorems we will present in this Section can be obtained by repeating
 step by step those of the previous Section for the case of two particles,
 we will limit ourselves to present the relevant definitions and theorems.


\subsection{Entanglement between subsets of the constituents}

We begin by considering the possibility of grouping the particles in two
 subsets which are non-entangled with each other.
Obviously, the particles of each subset may be entangled among themselves.
Let us consider a composite system of $N$ distinguishable quantum particles,
 whose state vector $\vert \psi(1,\dots, N)\rangle$ belongs to the direct
 product of the single particle Hilbert spaces, ${\cal H}=\prod_{i}\otimes
 {\cal H}_{i}$, and let us adopt the following definition:

\begin{mutualentang}
 \label{mutualentang}
 The subsystem ${\cal S}_{(1\dots M)}={\cal S}_{1} + \ldots +{\cal S}_{M}$
 of a composite quantum system ${\cal S}= {\cal S}_{1}+...+{\cal S}_{M}+...
 +{\cal S}_{N}$ in the pure state $\vert\psi(1,...,N)\rangle$, is
 {\bf non-entangled} with the subsystem ${\cal S}_{(M+1\dots N)}={\cal S}_{M+1}
 + \ldots +{\cal S}_{N}$ if there exists a one dimensional projection
 operator $P^{(1\dots M)}$ of ${\cal H}_{1} \otimes \ldots \otimes
 {\cal H}_{M}$, such that:

\begin{equation}
 \label{separabilitaemme}
  Tr^{(1+\ldots+N)}[\, P^{(1\dots M)}\otimes I^{(M+1 \dots N)}
  \vert \psi(1,\dots,N)\rangle \langle \psi(1,\dots,N)\vert \,]=1.
\end{equation}
\end{mutualentang}

This definition implies that we can attribute complete sets of objective
 properties (elements of reality) to at least the two subgroups of the
 particles we have indicated as ${\cal S}_{(1\dots M)}$ and
 ${\cal S}_{(M+1\dots N)}$ without worrying about the other subgroup.
The natural generalization of the corresponding theorem for two particles
 is then easily derived by recalling that the biorthonormal decomposition
 holds in general for the direct product of two Hilbert spaces and by
 repeating step by step its proof:

\begin{mutualfactor}
 \label{mutualfactor}
 If consideration is given to a many-particle quantum system described by
 the pure state $\vert\psi(1,\dots, N)\rangle$ (or by the corresponding
 pure density operator) of the Hilbert space ${\cal H}=\prod_{i}\otimes
 {\cal H}_{i}$, each of the following three conditions is necessary and
 sufficient in order that subsystem ${\cal S}_{(1\dots M)}$ is non-entangled
 with subsystem ${\cal S}_{(M+1\dots N)}$:

\begin{enumerate}
 \item there exists a projection operator $P^{(1...M)}$ onto a
 one-dimensional manifold of ${\cal H}_{1}\otimes ...\otimes {\cal H}_{M}$
 such that $Tr^{(1+...+N)}\left[ P^{(1...M)}\otimes
 I^{(M+1...N)}\rho^{(1...N)}\right]=1;$

 \item the reduced statistical operator $\rho^{(1...M)}=Tr^{(M+1...N)}
 \left[ \rho ^{(1...N)}\right] $ of subsystem $S_{(1...M)}$ is a projection
 operator onto a one-dimensional manifold of ${\cal H}_{1}\otimes ...\otimes
 {\cal H}_{M};$

  \item the state vector $\left| \psi (1,...,N)\right\rangle $ is factorizable,
  i.e., there exist a state $\left| \phi (1,...,M)\right\rangle$ of
  ${\cal H}_{1}\otimes ...\otimes {\cal H}_{M}$ and a state $\left| \xi
  (M+1,...,N)\right\rangle$ of $H_{M+1}\otimes ...\otimes H_{N}$ such that
  $\vert\psi(1,...,N)\rangle=\left| \phi(1,...,M)\rangle \otimes
  \vert \xi(M+1,...,N)\right\rangle$.
\end{enumerate}
\end{mutualfactor}

It is worthwhile to generalize Theorem~\ref{rank} to the present situation of
 $N$ distinguishable particles.
We show once again that there is a strict relation between the existence of a
 projection operator $P^{(1\dots M)}$ satisfying
 condition~(\ref{separabilitaemme}), i.e. between that fact that the particles
 fall into two non-entangled sets, and the range
 ${\cal R}[\, \rho^{(1\dots M)}\,]$ of the reduced statistical operator
 $\rho^{(1\dots M)}$  of the first M particles of the compound system:

\begin{rankN}
\label{rankN}
  A necessary and sufficient condition for the  projection
 operator $P^{(1\dots M)}_{{\cal M}}$ onto the linear manifold of ${\cal
H}_{1}\otimes \dots
 \otimes{\cal H}_{M}$ to satisfy the two following conditions:

\begin{enumerate}
 \item $Tr^{(1\dots M)}[\,P^{(1\dots M)}_{{\cal M}} \rho^{(1\dots M)}\,]=1;$
 \item there is no projection operator ${\tilde P}^{(1\dots M)}_{{\cal N}}$
 of ${\cal H}_{1}\otimes \dots \otimes{\cal H}_{M}$ satisfying the conditions
 ${\tilde P}^{(1\dots M)}_{{\cal N}} < P^{(1\dots M)}_{{\cal M}}$ (i.e. it
 projects onto a proper submanifold ${\cal N}$ of ${\cal M}$) and \\
 $Tr^{(1\dots M)}[\,{\tilde P}^{(1\dots M)}_{{\cal N}} \rho^{(1\dots M)}\,]=1$,
\end{enumerate}

\noindent is that the range ${\cal R}[\,\rho^{(1\dots M)}\,]$ of the reduced
  statistical operator $\rho^{(1\dots M)}$ coincides with ${\cal M}$.
\end{rankN}

\noindent {\it Proof:} it is a straightforward generalization of the proof
 already given for Theorem~\ref{rank}. $\:\:\:\Box$ \\

Leaving aside the case which we have already analyzed in which {\cal M} is
 one-dimensional, once again two possibilities arise:

\begin{enumerate}
 \item ${\cal R}[\, \rho^{(1\dots M)}\,]= {\cal M} \subset {\cal H}_{1}\otimes
 \dots \otimes {\cal H}_{M},$
 \item ${\cal R}[\, \rho^{(1\dots M)}\,]= {\cal H}_{1}\otimes
 \dots \otimes {\cal H}_{M}.$
\end{enumerate}

While in the first case we can say that the two groups of particles are
 {\em partially entangled} since we can attribute to the subsystem
 ${\cal S}_{(1\dots M)}$ some objective properties ($\,$in fact, given an
 operator $\Omega^{(1\dots M)}$ which commutes with $P^{(1\dots M)}_{{\cal M}}$
 we can consider the spectrum ${\cal B}$ of its restriction
 $\Omega_{R}=P^{(1\dots M)}_{{\cal M}}\Omega^{(1\dots M)}
 P^{(1\dots M)}_{{\cal M}}$ to ${\cal M}$, and we can say that subsystem
 ${\cal S}_{(1\dots M)}$ {\em has} the objective and (possibly) unsharp
 properties that $\Omega_{R}$ has a value belonging to ${\cal B}$$\,$), in
 the second case
 subsystem ${\cal S}_{(1\dots M)}$ is {\em totally entangled} since it does
 not possess any objective property whatsoever ($\,$the only projection
 operator $P^{(1\dots M)}$ satisfying condition~(\ref{separabilitaemme})
 being the identity operator$\,$).


\subsection{The case of completely non-entangled constituents}

As already remarked, the constituents of the subsystems ${\cal S}_{(1...M)}$
 and ${\cal S}_{(M+1...N)}$ of the original system may, in turn, be entangled
 or non-entangled among themselves.
One has then to consider the corresponding states $\vert\phi(1,...,M)\rangle$
 and $\vert\xi(M+1,...,N)\rangle$ and to repeat for them an analysis of the
 type we have just described.
We will not go into  details, we will limit ourselves to analyze briefly
 the case in which all $N$ constituents of the original system are
 non-entangled with each other.

\begin{defofentangN}
 \label{defofentangN}
 The pure state $\vert \psi(1,\dots, N)\rangle \in {\cal H}_{1}\otimes
 \dots \otimes {\cal H}_{N}$ is completely non-entangled if there exist $N$
 one-dimensional projection operators $P^{(i)}$ belonging to the Hilbert
 space ${\cal{H}}_{i}$ respectively, such that\footnote{We observe that it
 is sufficient to admit that there exist $N-1$ such projection operators,
 since  from the existence of $N-1$ of them it follows that a further
 operator with the same properties must exist.}:

\begin{equation}
  Tr^{(1+\dots +N)}[\, P^{(i)} \vert \psi(1,\dots, N)\rangle\langle
  \psi(1,\dots, N) \vert\,]=1 \:\:\:\:\:\:\forall i=1\dots N.
\end{equation}

\end{defofentangN}

One can then easily prove the following theorems which are straightforward
 generalizations of those we have  proved for the simpler case of
 two-particle states.

\begin{onedimensionalN}
 \label{onedimensionalN}
 The state $ \vert \psi(1,\dots, N)\rangle$ is completely non-entangled iff
 the $N$ reduced density operators $\rho^{(i)}=Tr^{\forall j\neq i}
 [\, \vert \psi(1,\dots,N) \rangle\langle \psi(1,\dots, N) \vert \,]$, where
 the trace is calculated with the exclusion of subsystem $i$, are one
 dimensional projection operators.
\end{onedimensionalN}

\begin{factorizabilityN}
 \label{factorizabilityN}
 The pure state $\vert \psi(1,\dots, N)\rangle \in {\cal H}_{1}\otimes \dots
 \otimes {\cal H}_{N}$ is completely non-entangled iff it is
 completely factorizable, i.e. there exist $N$ states
 $\vert \varphi(1) \rangle\in {\cal H}_{1}, \dots, \vert \theta(N)\rangle
 \in  {\cal H}_{N}$ such that $\vert \psi(1,\dots, N)\rangle=\vert \varphi(1)
 \rangle\otimes \dots\otimes \vert \theta(N)\rangle$.
\end{factorizabilityN}


\subsection{Correlations between the subsystems}

Obviously, one can repeat also in the present case the considerations of
 Subsection~\ref{quattropuntotre} and one can show that in the case in which
 the first $M$ particles are non-entangled with the remaining $K=N-M$ ones,
 an equation perfectly analogous to~(\ref{correlazione}) holds.
In fact
\begin{corrbetween}
 Subsystem ${\cal S}_{(1...M)}$ is non entangled with subsystem
 ${\cal S}_{(M+1...N)}$ iff, given the pure state $\vert \psi(1...N)\rangle$
 of the composite system, the following equation holds for any pair of
 observables $A(1\dots M)$ of ${\cal{H}}_{1}\otimes \dots \otimes
 {\cal{H}}_{M}$ and $B(M+1 \dots N)$ of ${\cal{H}}_{M+1}\otimes \dots
 \otimes {\cal{H}}_{N}$ such that $\vert \psi(1...N)\rangle$ belongs to their
 domains:
 \begin{equation}
   \langle \psi(1\dots N) \vert \,A \otimes B \, \vert \psi(1\dots N)
   \rangle = \langle \psi(1\dots N) \vert \,A\, \vert \psi(1\dots N)\rangle
   \langle \psi(1\dots N) \vert \,B\, \vert \psi(1\dots N)\rangle.
\end{equation}
\end{corrbetween}
>From the above equation there follows that the joint probabilities for
 outcomes of independent measurements processes on the two non-entangled
 subsystems factorize.


\vspace{1cm}
\begin{center}
 {\LARGE\bf  Part III: Entanglement and properties of quantum systems of
 identical particles in pure states}
\end{center}

This part is devoted to the problem of attributing properties and to the
 analysis of entanglement of composite systems whose constituents are
 identical.
As we will see such a problem is a quite delicate one and requires a detailed
 discussion.
With the exception of few papers~\cite{refexe1,refexe2,refexe3,refexe4}
 this matter has not been adequately discussed in the literature.


\section{Identity, individuality and properties in quantum theory}

The so called ``principle of individuality'' of physical systems has a long
 history in philosophy, the most naive position about it deriving from the
 observation that even two extremely similar objects will always display
 some differences in their properties  allowing to distinguish and to
 individuate the objects.
Leibniz has strongly committed himself to such a position by claiming:
 {\it ``there are never in nature two exactly similar entities in which one
 cannot find an internal difference''}.
However one could try to individuate even absolutely identical objects by
 taking into account that they differ at least for their location in space
 and time.
In the debate about this problem it is generally agreed that the objects we
 are interested in should be regarded as individuals.
But then a quite natural problem arises: can the fundamental ``objects'' of
 current physical theories, such as electrons, protons, etc., be regarded as
 individuals? And what is the status of such a question within the classical
 and quantum schemes?

The first obvious remark is that in both schemes such entities are assumed as
 identical in the sense of possessing precisely all the same intrinsic
 properties, such as rest mass, charge, spin, magnetic moment, and so on.
However, within the framework of classical mechanics each particle follows
 a perfectly defined trajectory and thus it can (at least in principle) be
 distinguished from all the others.

The situation is quite different in quantum mechanics, due to the fact that
 such a theory does not even allow to entertain the idea of particle
 trajectories and implies that wave functions spread, so that, even if one
 could label at a certain time one of the two identical particles as 1 and
 the other as 2, one would not be able, even in principle, to claim, with
 reference to a subsequent act of detection, whether the particle which has
 been detected is the one he has labeled 1 or 2.

The above facts have led some philosophers of science to the conclusion
 that quantum particles cannot be regarded as individuals in any of the
 traditional meanings of such a term.
We will not enter into this relevant question for which we refer the reader
 to some recent interesting
 contributions~\cite{refide1,refide2,refide3,refide4}; we plainly accept, as
 imposed by the formalism, that when one is dealing with assemblies of
 identical quantum systems it is simply meaningless to try to ``name'' them
 in a way or another.

However, the problem of identifying indiscernible objects is not the relevant
 one for this paper.
What concerns us is the possibility of considering some properties as
 objectively possessed by quantum systems.
Accordingly, it goes without saying that, when dealing with the system, e.g.,
 of two electrons, we will never be interested in questions like ``is the
 electron which we have labeled  1 at a certain position or is its spin
 aligned with a given axis?'' but our only concerns will be of the type: on
 the basis of the knowledge of the state vector describing the composite
 system, can one legitimately consider as objective a statement of the kind
 ``there is an electron in a certain region and it has its spin up along a
 considered direction''?
Obviously, in accordance with the position we have taken in
 Section~\ref{sezionetre}, the
 above statement must be read as: does the theory guarantee that if a
 measuring apparatus aimed to reveal an electron and to measure its spin
 along the considered direction would be activated, it will give with
 certainty the considered outcomes?

When one takes such a perspective the problem of considering the constituents
 (we are not interested in which ones) of a system of identical particles as
 possessing objectively definite properties can be tackled in a mathematically
 precise way.
Correspondingly, one can formulate in a rigorous way the idea that identical
 particles are non entangled.

We will analyze first of all the case of two identical particles, which
 represents an ideal arena to point out various subtle aspects of the problem
 under investigation.


\section{Entanglement of two identical particles}
\label{entanglementoftwoidenticalparticles}

As just mentioned, in the case of a system containing identical constituents
  the problem of entanglement has to be reconsidered.
In fact, the naive idea that the two systems being non-entangled requires
 and is guaranteed by the fact that their state vector is the direct product
 of vectors belonging to the corresponding Hilbert spaces, cannot be simply
 transposed to the case of interest.
One can easily realize that this must be the case by taking into account that
 the only allowed states for a system of two identical particles must exhibit
 precise symmetry properties under the exchange of the two particles.
If one would adopt the previous criterion one would be led to conclude
 (mistakenly) that non-entangled states of identical particles cannot
 exist\footnote{We are disregarding here the case of the direct product
 of two identical state vectors for two identical bosons. Such a case will
 be discussed in great details in what follows.}.
The inappropriateness of such a conclusion derives from not taking into
 account various fundamental facts, in particular that identical particles are
 truly indistinguishable, so that one cannot pretend that a particular one
 of them has properties, and that the set of observables for such a system
 has to be restricted to the self-adjoint operators which are symmetric for
 the exchange of the variables referring to the two subsystems.

To prepare the ground for such an analysis, we begin by discussing the case of
 two identical particles which turns out to be simpler.
For the moment we will deal simultaneously with the case of identical fermions
 and bosons. However since some relevant differences occur in the two cases
 we will subsequently consider them separately in
 Subsections~\ref{thefermioncase} and~\ref{thebosoncase}.

The necessary modifications of the definitions and theorems of this section
 when many particles are taken into account will be given at the appropriate
 stage of our analysis, after a critical reconsideration of the problem of
 property attribution will be presented.


\subsection{Entanglement and individual properties of identical constituents}
\label{entanglementandind}

As we have noticed in Section~\ref{generaldefinitions}, in the case of
 distinguishable particles
 the fact that one of the two constituents of a composite system
 possesses a complete set of properties automatically implies that the same
 holds true for the other constituent.
While this happens also for identical fermions, it is not true, in general,
 for systems involving identical bosons.
We will strictly link the idea that two identical particles are non entangled
 to the request that both of them possess a complete set of properties.
Accordingly we give the following definition:

\begin{defofentangidentical}
 \label{defofentangidentical}
  The identical constituents ${\cal S}_{1}$ and ${\cal S}_{2}$ of a composite
  quantum system ${\cal S}={\cal S}_{1}+{\cal S}_{2}$ are {\bf non-entangled}
  when both constituents possess a complete set of properties.
\end{defofentangidentical}

Taking into account this fact, and in order to deal as far as possible
 simultaneously with the fermion and the boson cases, we will first of all
 identify the conditions under which one can legitimately claim that {\em one
 of the constituents possesses a complete set of properties}.
Once this will be done, we will separately deal with the problem of the
 entanglement  by distinguishing the fermion from the boson case.

In accordance with the above analysis, let us begin by identifying the
 necessary and sufficient conditions in order that one of a pair of identical
 particles possesses a complete set of properties.
\begin{defofentangidentical2}
 \label{defofentangidentical2}
 Given a composite quantum system ${\cal S}={\cal S}_{1}+{\cal S}_{2}$ of
 two identical particles described by the pure density operator $\rho$, we
 will say that one of the constituents has a complete set of properties iff
 there exists a one dimensional projection operator $P$, defined on the Hilbert
 space ${\cal H}^{(1)}$ of each of the subsystems, such that:

\begin{equation}
\label{wittgenstein}
  Tr^{(1+2)}[\,E(1,2)\rho \,]=1
\end{equation}
\noindent where
\begin{equation}
  \label{operatorediproiezione}
   E(1,2)=P^{(1)}\otimes I^{(2)} + I^{(1)}\otimes P^{(2)}  - P^{(1)}\otimes
   P^{(2)}.
\end{equation}
\end{defofentangidentical2}
We stress that the operator $E(1,2)$ is symmetric under the exchange of the
 labels of the two particles and that it is a projection operator:
 $\,[E(1,2)\,]^{2}=E(1,2)$.
Furthermore $Tr^{(1+2)}[\,E(1,2)\rho \,]$ gives the probability of finding
 {\em at least} one of the two identical particles in the state onto which the
 one-dimensional operator $P$ projects as is immediately checked by noticing
 that $E(1,2)$ can also be written as\footnote{We remark that one could drop
 the last  term in the expression~(\ref{proiettore}) getting an operator whose
 expectation value would give the probability of precisely one particle having
 the properties associated to $P$.
In the case of identical fermions this would make no difference but for
 bosons it would not cover the case of both particles having precisely the
 same properties.}

\begin{equation}
  \label{proiettore}
  E(1,2)=P^{(1)}\otimes [\,I^{(2)}-P^{(2)}\,] + [\,I^{(1)}-P^{(1)}\,]\otimes
  P^{(2)} + P^{(1)}\otimes P^{(2)}.
\end{equation}

\noindent When $E(1,2)$ is multiplied by $\rho$, the trace of the first
 term in eq.~(\ref{proiettore}) gives the probability that particle 1 has the
 property associated to $P$ while the second one does not have such a property,
 the trace of the second term gives the same probability with particle 1 and
 2 interchanged and the trace of the third term gives the probability that
 both particles have the considered property.
Since the above occurrences are mutually exclusive, condition
 $Tr^{(1+2)}[\,E(1,2)\rho\,]=1$ implies that at least one particle has the
 property under consideration.

It is interesting to relate the fact that one constituent possesses a complete
 set of properties to the explicit form of the state vector. This is specified
 by the following theorem:

\begin{factorizabilityidentical}
 \label{factorizabilityidentical}
 One of the identical constituents of a composite quantum system
 ${\cal S}={\cal S}_{1} + {\cal S}_{2}$, described by the pure normalized
 state $\vert \psi(1,2) \rangle$ has a complete set of properties
 iff $\vert \psi(1,2) \rangle$ is obtained by symmetrizing or antisymmetrizing
 a factorized state.
\end{factorizabilityidentical}

\noindent {\it Proof:}
 If $\left| \psi (1,2)\right\rangle $ is obtained by symmetrizing or
 antisymmetrizing a factorized state of two identical particles:
\begin{equation}
  \label{statononortogonale}
  |\psi (1,2)\rangle=N[|\varphi^{(1)}\rangle\otimes|\chi^{(2)}\rangle\pm
  |\chi^{(1)}\rangle\otimes|\varphi^{(2)}\rangle],
  \label{Na}
\end{equation}
expressing the state $|\chi^{(i)}\rangle$ as follows
\begin{equation}
 |\chi^{(i)}\rangle=\alpha |\varphi^{(i)}\rangle +\beta
 |\varphi_{\bot }^{(i)}\rangle,\qquad \langle \varphi^{(i)}
 |\varphi_{\bot }^{(i)}\rangle=0
 \label{Nb}
\end{equation}
and choosing $P=|\varphi \rangle \langle\varphi |$ one gets immediately
  \begin{equation}
   Tr^{(1+2)}[E(1,2)\rho ]\equiv \langle\psi (1,2)|E(1,2)|\psi (1,2)\rangle =
   \frac{2(1\pm |\alpha |^{2})}{2(1\pm |\alpha |^{2})}=1
   \label{Nc}
  \end{equation}
Alternatively, since $E(1,2)$ is a projection operator:
  \begin{equation}
  \label{condizione6.6}
   \Big[ \langle \psi (1,2)|E(1,2)|\psi (1,2)\rangle =1 \Big] \Rightarrow
   \Big[ \left\| E(1,2)|\psi (1,2)\rangle \right\| =1 \Big] \Rightarrow
   \Big[ E(1,2)|\psi (1,2)\rangle = |\psi (1,2)\rangle \Big].
  \label{Nd}
  \end{equation}
\noindent If one chooses a complete orthonormal set of single particle states
 whose first element $|\Phi_{0}\rangle$ spans the one-dimensional linear
 manifold onto which the one dimensional projection operator $P$ projects,
 writing
\begin{equation}
  \label{statonorma}
   \vert \psi (1,2) \rangle= \sum_{ij} c_{ij} \vert \Phi^{(1)}_{i}\rangle
   \otimes \vert \Phi^{(2)}_{j}\rangle\:\:\:\:\:\:\:\:\:\:\:\sum_{ij} \vert
   c_{ij} \vert^{2}=1,
\end{equation}
and, using the explicit expression for $E(1,2)$ in terms of such a $P$, one
 gets:
  \begin{equation}
  \label{pippo}
   E(1,2)|\psi (1,2)\rangle =
   |\Phi_{0}^{(1)}\rangle \otimes \left[ \sum_{j\neq 0}c_{0j}
   |\Phi_{j}^{(2)}\rangle \right] +\left[ \sum_{j\neq 0}c_{j0}
   |\Phi_{j}^{(1)}\rangle \right] \otimes
   |\Phi_{0}^{(2)}\rangle + c_{00}|\Phi _{0}^{(1)}\rangle \otimes
   |\Phi_{0}^{(2)}\rangle.
  \end{equation}
Imposing condition~(\ref{condizione6.6}), i.e. that the r.h.s. of
 equation~(\ref{pippo}) coincides with $\vert \psi(1,2) \rangle$ as given by
 equation~(\ref{statonorma}), we obtain $c_{ij}=0$ when both $i$ and $j$ are
 different from zero.
Taking into account that for identical particles $c_{0j}=\pm c_{j0}$, the
 normalization condition of the state $\vert \psi(1,2) \rangle$ becomes
  \begin{equation}
  \label{normalization}
  \vert c_{00} \vert^{2} + 2\sum_{j\neq 0} \vert c_{oj} \vert^{2} =1.
  \end{equation}
We have then shown that:
  \begin{equation}
  \label{generale}
   \vert \psi(1,2) \rangle =
   |\Phi_{0}^{(1)}\rangle \otimes \left[ \sum_{j\neq 0}c_{0j}
   |\Phi_{j}^{(2)}\rangle \right] +\left[ \sum_{j\neq 0}c_{j0}
   |\Phi_{j}^{(1)}\rangle \right] \otimes
   |\Phi_{0}^{(2)}\rangle + c_{00}|\Phi_{0}^{(1)}\rangle \otimes
   |\Phi_{0}^{(2)}\rangle.
  \end{equation}
In the case of fermions $c_{00}=0$.
Then, introducing a normalized vector $\vert \Xi^{(i)}\rangle =
 \sqrt{2}\sum_{j\neq 0} c_{0j}|\Phi_{j}^{(i)}\rangle$ one has
  \begin{equation}
  \label{fermioni}
   |\psi (1,2)\rangle = \frac{1}{\sqrt{2}}\, \left[\,
   |\Phi_{0}^{(1)}\rangle|\Xi ^{(2)}\rangle -
   |\Xi ^{(1)}\rangle|\Phi_{0}^{(2)}\rangle\,\right],
  \end{equation}
with $|\Phi_{0}^{(k)} \rangle$ and $|\Xi^{(k)} \rangle$ orthogonal.

For bosons, defining the following normalized vector
  \begin{equation}
   \vert \Theta^{(i)} \rangle = \sqrt{\frac{4}{2- |c_{00}|^{2}}} \, \left[\,
   \sum_{j\neq 0} c_{oj} |\Phi_{j}^{(i)} \rangle  + \frac{c_{00}}{2}
   |\Phi_{0}^{(i)} \rangle \,\right],
  \end{equation}
the two-particle state vector~(\ref{generale}) becomes
  \begin{equation}
  \label{duebosoni}
   \vert \psi(1,2) \rangle = \sqrt{\frac{2- |c_{00}|^{2}}{4}}\, \left[\,
   \vert \Phi_{0}^{(1)}\rangle \vert \Theta^{(2)} \rangle +
   \vert \Theta^{(1)} \rangle\vert \Phi_{0}^{(2)}\rangle\, \right].
  \end{equation}
Note that in this case the states $\vert \Phi_{0}^{(k)} \rangle$ and
 $\vert \Theta^{(k)} \rangle$ are orthogonal iff the coefficient $c_{00}$
 is zero. $\:\:\:\:\:\:\:\:\:\:\:\Box$

There follows that the process of symmetrization or antisymmetrization of a
 factorized quantum state describing a system composed of identical
 particles does not forbid to attribute a complete set of physical properties
 to the subsystems: the only claim that one cannot make is to attribute the
 possessed property to one rather than to the other constituent.

At this point it is appropriate to deal separately with the case of
 identical fermions and identical bosons.
We will denote the operator $E(1,2)$ of equation~(\ref{operatorediproiezione})
  as $E_{f}(1,2)$ and $E_{b}(1,2)$ in the two cases, respectively.


 \subsubsection{The fermion case}
\label{thefermioncase}

We analyze first of all the case of two identical fermions, in which, since
 $P^{(1)}\otimes P^{(2)}=0$ on the space of totally antisymmetric functions,
 one can drop such a term in all previous formulae.
Accordingly $E_{f}(1,2)=P^{(1)}\otimes I^{(2)}+I^{(1)}\otimes P^{(2)}$.
Some remarks are appropriate.

As one sees from equation~(\ref{fermioni}) and in accordance with
 Definition~\ref{defofentangidentical2}, due to the orthogonality of
 $\vert \Phi_{0} \rangle$ and $\vert \Xi \rangle$, for such a state one can
 claim not only that there is one fermion possessing the complete set of
 properties identified by the state $\vert \Phi_{0} \rangle$, but also one
 fermion possessing the complete set of properties identified by the state
 $\vert \Xi \rangle$.

According to Definition~\ref{defofentangidentical}, we have thus proved
 the following theorem for a two fermion system:

\begin{factorizabilityidentical2}
 \label{factorizabilityidentical2}
The identical fermions ${\cal S}_{1}$ and ${\cal S}_{2}$ of a composite
 quantum system ${\cal S}={\cal S}_{1}+{\cal S}_{2}$ described by the pure
 normalized state $\vert \psi(1,2) \rangle$  are non-entangled iff
 $\vert \psi(1,2) \rangle$ is obtained by antisymmetrizing a factorized state.

\end{factorizabilityidentical2}

If, with reference to expression~(\ref{fermioni}),
 we call $P=\vert \Phi_{0} \rangle \langle \Phi_{0}\vert$ and
 $Q=\vert \Xi \rangle \langle \Xi \vert$ and we
 define in terms of them the projection operators $E_{f}(1,2)=P^{(1)}\otimes
 I^{(2)}+I^{(1)}\otimes P^{(2)}$ and $F_{f}(1,2)=Q^{(1)}\otimes
 I^{(2)}+I^{(1)}\otimes Q^{(2)}$, we see that
  \begin{equation}
  \label{correlafermioni}
   \left\{
   \begin{array}{ll}
   Tr^{(1+2)} \left[\, E_{f}(1,2)\, \vert \psi (1,2) \rangle \langle
   \psi (1,2) \vert \,\right] & =  1,  \\
   & \\
   Tr^{(1+2)} \left[\, F_{f}(1,2)\, \vert \psi (1,2) \rangle \langle
   \psi (1,2) \vert \,\right] & =  1,
   \end{array}
   \right.
  \end{equation}
Moreover, $E_{f}(1,2)\cdot F_{f}(1,2)$ is a projection operator
 onto a one-dimensional manifold of the Hilbert space of the two fermions, it
 coincides with the operator $P^{(1)}\otimes Q^{(2)}+Q^{(1)}\otimes P^{(2)}$
 and, as a consequence of the relations~(\ref{correlafermioni}), it satisfies:
  \begin{equation}
  \label{C}
   \langle \psi (1,2)\vert E_{f}(1,2) \cdot F_{f}(1,2) \vert \psi(1,2)\rangle
   \equiv \langle \psi (1,2) \vert P^{(1)} \otimes Q^{(2)}+Q^{(1)} \otimes
   P^{(2)} \vert \psi (1,2) \rangle =1.
  \end{equation}
Before concluding this Subsection we judge it extremely relevant to call
 attention to the existence of a certain arbitrariness concerning the
 properties one can consider as objectively possessed by the constituents of
 a non entangled state of two fermions.
In fact, suppose the state $\vert \psi (1,2) \rangle$ has the
 expression~(\ref{fermioni}). Then, if consideration is given to the two
 dimensional manifold spanned by the single particle states $\vert \Phi_{0}
 \rangle$ and $\vert \Xi \rangle$, one immediately sees that if one chooses
 any two other orthogonal single particle states $\vert \Lambda \rangle$ and
 $\vert \Gamma \rangle$ spanning the same manifold, then $\vert \psi (1,2)
 \rangle$ can also be written (up to an overall phase phactor) as:
  \begin{equation}
   \vert \psi (1,2)\rangle= \frac{1}{\sqrt{2}} \,[\, \vert \Lambda^{(1)}
   \rangle
   \vert \Gamma^{(2)} \rangle - \vert \Gamma^{(1)} \rangle
   \vert \Lambda^{(2)} \rangle \,].
  \end{equation}
Obviously, such an expression makes legitimate the assertion that the two
 fermions have the complete sets of properties associated to $\vert \Lambda
 \rangle$ and $\vert \Gamma \rangle$.

Such a fact might appear as rather puzzling.
However as we will see in Section~\ref{deepeningthe}, it does not give rise
 to conceptual
 problems, but it requires to analyze more deeply the situation, to make
 perfectly clear the subtle interplay between the identity of the
 constituents and the problem of attributing objective properties to them.


\subsubsection{The boson case}
\label{thebosoncase}

Let us consider now the boson case. As one sees from
 equations~(\ref{statononortogonale}) and~(\ref{duebosoni}), once more, the
 requirement that one of the two identical bosons possesses a complete set of
 properties implies that the state is obtained by symmetrizing a factorized
 state.
However, there are some remarkable differences with respect to the fermion
  case.
With reference to the expression~(\ref{duebosoni}) we see that now three cases
 are possible:
\begin{itemize}
\item  $\vert \Theta^{(i)} \rangle \propto \vert \Phi^{(i)}_{0} \rangle$.
  In such a case the state is $\vert \psi(1,2) \rangle = \vert \Phi_{0}^{(1)}
   \rangle \otimes \vert \Phi_{0}^{(2)} \rangle$ and one can claim that
  ``there are two bosons with the complete set of properties associated to
   $P=\vert \Phi_{0} \rangle \langle \Phi_{0}\vert$''.

\item  $\langle \Theta^{(i)} \vert \Phi^{(i)}_{0} \rangle =0$, i.e.
  $c_{00}=0$. One can then consider the operators $E_{b}(1,2)$ and
  $F_{b}(1,2)$ and their product  $E_{b}(1,2)F_{b}(1,2) = P^{(1)}\otimes
  Q^{(2)}+Q^{(1)} \otimes P^{(2)}$ where $P= \vert \Phi_{0} \rangle \langle
  \Phi_{0} \vert$ and $Q= \vert \Theta \rangle \langle \Theta\vert$.
Exactly the same argument of the case of two identical fermions makes then
 clear that one can legitimately claim that ``state $\vert \psi(1,2)\rangle$
 represents a system where one of the two bosons has the properties
 associated to the projection operator $P$ {\em and} one those associated to
 the projection operator $Q$''.

\item  Finally it can happen that $\langle \Theta^{(i)} \vert
 \Phi^{(i)}_{0}\rangle \neq 0$ but $\vert \Theta^{(i)} \rangle $ is not
 proportional to $\vert \Phi^{(i)}_{0} \rangle$. Then, even though we can
 state that ``there is a boson with the properties associated to the
 projection operator $P$'' as well as ``there is a boson with the properties
 associated to the projection operator $Q$'', we cannot state that
 ``the state~(\ref{duebosoni}) describes a system in which one of the bosons
 has the properties associated to $P$ {\em and} the other those associated
 to $Q$''.
Actually in the considered case there is a non vanishing probability of
 finding both particles in the same state.
\end{itemize}
According to our Definition~\ref{defofentangidentical} which, as already
 stated, we adopt completely in general for systems of identical particles,
 we see that in the last of the just considered cases we cannot assert that
 the two bosons are non entangled, while we can do so for the first two cases.
The following theorem has thus been proved:

\begin{defofentangidentical3}
 \label{defofentangidentical3}
  The identical bosons of a composite quantum system ${\cal S}={\cal S}_{1}
  + {\cal S}_{2}$ described by the pure normalized state $\vert \psi(1,2)
  \rangle$ are {\bf non-entangled} iff either the state is obtained by
  symmetrizing a factorized product of two orthogonal states or if it is the
  product of the same state for the two particles.
\end{defofentangidentical3}
 Before concluding, we point out that in the boson case and when $\vert \psi
 (1,2) \rangle$ is obtained by symmetrizing two orthogonal vectors, contrary
 to what happens for two fermions, the two states are perfectly defined (up
to a phase factor),
 i.e., there are no other orthogonal states $\vert \alpha
 \rangle$ and $\vert \beta \rangle $ differing from $\vert \Phi_{0} \rangle$
 and $\vert \Theta \rangle$, such that one can write $\vert \psi (1,2)\rangle$
 in the form:
\begin{equation}
 \label{nonsipuo}
 \vert \psi(1,2)\rangle =\frac{1}{\sqrt{2}}\left[ \vert \alpha^{(1)}
 \rangle\vert \beta^{(2)} \rangle + \vert \beta^{(1)} \rangle\vert \alpha^{(2)}
 \rangle \right].
\end{equation}
The proof is easily derived along the following lines.
Writing\footnote{Note that given $\vert\alpha^{(i)}\rangle$ we have chosen a
 precise phase factor in defining $\vert\beta^{(i)}\rangle$}
\begin{equation}
   \left\{
   \begin{array}{ll}
   \vert \alpha^{(i)} \rangle = a\vert \Phi_{0}^{(i)} \rangle + b \vert
   \Theta^{(i)} \rangle & \\
   \vert \beta^{(i)} \rangle = -b^{\star}\vert \Phi_{0}^{(i)} \rangle +
   a^{\star} \vert \Theta^{(i)} \rangle &
   \end{array}
   \right.
\end{equation}
with $\vert a \vert^{2} +\vert b \vert^{2}=1$, we get from
 equation~(\ref{nonsipuo}):
  \begin{equation}
   \vert \psi(1,2) \rangle = \frac{1}{\sqrt{2}} [ \,- 2ab^{\star}
   \vert \Phi_{0}^{(1)} \rangle \vert \Phi_{0}^{(2)}\rangle +
   (\vert a \vert^{2} -
   \vert b \vert^{2} )( \vert \Phi_{0}^{(1)} \rangle \vert \Theta^{(2)} \rangle
   + \vert \Theta^{(1)} \rangle \vert \Phi_{0}^{(2)}\rangle)
   + 2ba^{\star} \vert \Theta^{(1)} \rangle \vert \Theta^{(2)} \rangle \,]
  \end{equation}
which can coincide with
  \begin{equation}
   \vert \psi(1,2) \rangle = \frac{1}{\sqrt{2}} [\, \vert \Phi_{0}^{(1)}\rangle
   \vert \Theta^{(2)} \rangle + \vert \Theta^{(1)} \rangle \vert
   \Phi_{0}^{(2)}\rangle]
  \end{equation}
iff $a=0$ or $b=0$ implying that $\vert \alpha \rangle$ is $\vert \Phi_{0}
 \rangle$ and $\vert \beta \rangle$ is $\vert \Theta \rangle$ or viceversa.
   $\:\:\:\:\Box$


 \subsubsection{Concluding remarks}

The conclusion of this section is that the concept of entanglement can be
 easily generalized to the case of two identical quantum subsystems provided
 one relates it to the possibility of attributing complete sets of objective
 properties to both constituents.
The theorems of Section~\ref{entanglementandind} make absolutely precise the
 mathematical aspects
 characterizing the states which can be considered as describing non-entangled
 systems, i.e. the fact that: i) the states for the whole system must be
 obtained by appropriately (anti)symmetrizing a factorized state of the two
 particles; ii) the factors of such states must be orthogonal in the fermion
 case and they can be either orthogonal or equal in the boson case.

Obviously, the above conclusion implies that entangled states of two identical
 particles can very well occur.
Just to give an example we can consider the following state of two spin-$1/2$
 particles:
  \begin{equation}
   \label{proiezionesututto}
   \vert \psi(1,2) \rangle = \frac{1}{\sqrt{2}}\,[\, \vert \vec{n}\uparrow
   \rangle_{1}\vert \vec{n}\downarrow \rangle_{2} - \vert \vec{n}\downarrow
   \rangle_{1} \vert \vec{n}\uparrow \rangle_{2}
   \,]\otimes \vert \omega(1,2) \rangle
  \end{equation}
$\vert \omega(1,2) \rangle$ being a symmetric state of ${\cal L}(R^{3})
 \otimes {\cal L}(R^{3})$. State~(\ref{proiezionesututto}) cannot be written
 as a symmetrized product of two orthogonal states, and, consequently no
 constituent possesses any conceivable complete set of (internal and spatial)
 properties.


 \subsection{Sharp and unsharp properties}

In our discussion concerning the properties of one of a pair of identical
 constituents we have focussed our attention on complete set of properties.
The formalization of this idea consists in assuming that there exists a single
 particle projection operator $P$ onto a one-dimensional manifold such that
 the projection operator of equation~(\ref{operatorediproiezione}) satisfies
 condition~(\ref{wittgenstein}).
Obviously we could have played an analogous game by considering a projection
 operator $P_{\cal M}$ of the single particle Hilbert space ${\cal H}^{(1)}$
 projecting onto a multidimensional submanifold ${\cal M}$ of such a space.
Suppose that, with this choice, the corresponding operator:
  \begin{equation}
   E_{\cal M}(1,2) = P_{\cal M}^{(1)}\otimes (I^{(2)} - P_{\cal M}^{(2)})
   + (I^{(1)} - P_{\cal M}^{(1)}) \otimes P_{\cal M}^{(2)} +
   P_{\cal M}^{(1)} \otimes P_{\cal M}^{(2)}
  \end{equation}
satisfies:
  \begin{equation}
  \label{sharp}
   Tr[\,E_{\cal M}(1,2) \vert \psi(1,2) \rangle \langle \psi(1,2)
   \vert\,]=1.
  \end{equation}
As it is immediate to see one can  consider an arbitrary single particle
 self-adjoint operator $\Omega$ commuting with $P_{\cal M}$, and consider the
 restriction $\omega({\cal M})$ of its spectrum to the manifold ${\cal M}$.
Then the validity of~(\ref{sharp}) implies that we can state that one of
 the particles has the property that the value of $\Omega$ belongs to
 $\omega({\cal M})$.

Thus, in the considered case, even though we cannot attribute to any of the
 particles a complete set of properties, we can, in general, attribute to it
 unsharp properties.
If the only manifold ${\cal M}$ for which the above situation holds is the
 whole Hilbert space, we can state that the particles possess no sharp
 or unsharp properties at all.

In complete analogy to what has been said in
 Subsection~\ref{entanglementandpropertiesoftwo} concerning the
 various degrees of entanglement occurring in systems of distinguishable
 particles, it is worthwhile to summarize the situation in the following
 way, where $\vert \psi(1,2) \rangle$ represents an arbitrary state of
 two identical particles:
\begin{itemize}
\item there exists a one dimensional projection operator $P$ of
 ${\cal H}^{(1)}$ such that\\
 $Tr[\, E(1,2) \vert \psi(1,2) \rangle \langle \psi(1,2) \vert\,]=1$
 $\:\:\Rightarrow \:\:$ one subsystem possesses a complete set of properties;

\item there exists a projection operator $P_{\cal M}$ projecting onto a
 proper submanifold ${\cal M} \subset {\cal H}$ of dimension greater than
 $1$, such that $Tr[\, E_{\cal M}(1,2) \vert \psi(1,2) \rangle
 \langle \psi(1,2) \vert\,]=1$ $ \:\:\Rightarrow \:\:$ one subsystem
 possesses some properties, but not a complete set of them;

\item  there exists no projection operator $P_{\cal M}$ projecting onto a
 proper submanifold of ${\cal H}^{(1)}$ such that
 $Tr[\, E_{\cal M}(1,2) \vert \psi(1,2) \rangle \langle \psi(1,2) \vert\,]
 =1$ $\:\:\Rightarrow \:\:$ one subsystem (actually both of them) does not
 possess any property at all.
\end{itemize}
%


\subsection{Correlations in the case of two identical particles}
\label{correlationsinthecase}

In this subsection we reconsider briefly the problem of the correlations of
 two particles in entangled or non-entangled states in the case in which they
 are identical.
Before coming to a detailed analysis let us stress that the problem under
 consideration has a particular relevance in the specific case in which the
 two particles are in different spatial regions, since this is the case in
 which the problem of the nonlocal aspects of the formalism emerges as a
 central one.
Let us then consider two identical particles with space and internal degrees
  of freedom and let us denote as ${\cal H}_{sp}(1,2)$ and
  ${\cal H}_{int}(1,2)$ the corresponding Hilbert spaces.
The Hilbert space for the whole system is, obviously, the appropriate
 symmetric or antisymmetric submanifold ${\cal H}_{S,A}(1,2)$ of the space
 ${\cal H}_{sp}(1,2)\otimes $ ${\cal H}_{int}(1,2)$.
Let us also assume that the pure state associated to the composite system is
 obtained by (anti)symmetrizing a factorized state of the two particles
 corresponding to their having different spatial locations.
To be explicit, we start from a state:
  \begin{equation}
  \label{correlation1}
   |\psi_{fact}(1,2)\rangle=|\varsigma \rangle_{1} |R\rangle_{1}
   |\chi \rangle_{2} |L\rangle_{2},
  \label{G1}
  \end{equation}
where $|\varsigma \rangle$ and $|\chi \rangle$ are two arbitrary states of
 the internal space of a particle and $|R\rangle$ and $|L\rangle$ are two
 orthogonal states whose spatial supports are compact, disjoint and
 far away from each other.
This situation is the one of interest for all experiments about the non-local
 features of quantum states.
>From the state~(\ref{G1}) we pass now to the properly (anti)symmetrized one:
  \begin{equation}
   |\psi (1,2) \rangle=\frac{1}{\sqrt{2}}\left[ |\varsigma \rangle_{1}
   |R\rangle_{1} |\chi\rangle_{2} |L\rangle_{2} \pm |\chi \rangle_{1}
   |L\rangle_{1} |\varsigma \rangle_{2} |R\rangle_{2}\right] .
  \label{G2}
  \end{equation}

Note that we already know that if consideration is given to the operators
  \begin{eqnarray}
   E(1,2) &=&P^{(1)}\otimes I^{(2)}+I^{(1)}\otimes P^{(2)}
   - P^{(1)}\otimes P^{(2)},\:\:\:\:\: P=|\varsigma \rangle |R\rangle
 \langle R| \langle \varsigma | \nonumber \\
   F(1,2) & =& Q^{(1)}\otimes I^{(2)}+I^{(1)}\otimes Q^{(2)}
    - Q^{(1)}\otimes Q^{(2)}, \:\:\:\:\:
   Q=|\chi \rangle |L\rangle \langle L\vert \langle\chi |
   \label{G3}
  \end{eqnarray}
the following equations hold:
  \begin{equation}
   Tr^{(1+2)}\left[ E(1,2) |\psi (1,2)\rangle \langle \psi (1,2)|\right] =1,
  \qquad
   Tr^{(1+2)}\left[ F(1,2)|\psi (1,2)\rangle \langle \psi (1,2)|\right] =1,
  \label{G4}
  \end{equation}
which guarantee that the properties related to the projection operators $P$
 and $Q$ can be considered as objectively possessed. However, here we are
 interested in what the theory tells us concerning the correlations between
 the outcomes of measurement processes on the constituents.
To this purpose, we consider  two arbitrary observables
 $\Omega ^{(1)}$ and $\Sigma ^{(2)}$ of the internal space of the particles
 and we evaluate the expectation value:
\begin{eqnarray}
   &&\!\!\!\!\!\!\!\!\!\!\!\!\!
   \langle \psi |\left[ \Omega ^{(1)}|R\rangle_{1 1}\langle R|\otimes
   I^{(2)} + I^{(1)}\otimes \Omega^{(2)} |R\rangle_{2 2}\langle R|\right]
   \left[ \Sigma^{(1)} |L\rangle_{1 1}\langle L| \otimes I^{(2)}
   + I^{(1)} \otimes \Sigma^{(2)} |L\rangle_{2 2} \langle L|\right] |\psi
 \rangle \nonumber \\
   &=&\langle \varsigma |\Omega |\varsigma \rangle \langle \chi |\Sigma
   |\chi \rangle.
  \label{G5}
  \end{eqnarray}
Equation~(\ref{G5}) shows that the ``properties referring to the internal
 degrees of freedom" factorize, just as in the case of two distinguishable
 particles.
Obviously, the same conclusion does not hold when the state is not of the
 considered type, e.g., when it is a genuinely entangled state such as:
\begin{equation}
 \vert \psi (1,2) \rangle = \frac{1}{2} \left[ |\varsigma \rangle_{1}
 |\chi \rangle_{2} - |\chi \rangle_{1} |\varsigma \rangle_{2}\right]
 \otimes \left[ |R \rangle_{1} |L \rangle_{2} \pm |L \rangle_{1}
 |R \rangle_{2}\right] .
 \label{G6}
\end{equation}
The conclusion should be obvious: also from the point of view of the
 correlations, and consequently of the implications concerning nonlocality,
 the non-entangled states of two identical particles have the same nice
 features of those of two distinguishable particles.


  \subsection{Deepening the investigation}
\label{deepeningthe}

The analysis of the previous subsections has clarified the situation
 concerning systems of two identical particles, making precise which is the
 appropriate way to pose the problem of attributing objective properties and
 of the entanglement within such a context.
The relations between the physical and the formal aspects of such a problem
 have also been discussed.
However, some delicate questions which deserve a further analysis have been
 naturally raised. This subsection is devoted to deal which such matters.
We will begin by trying to make clear, resorting to elementary physical
 examples, some subtle points which could give rise to misunderstandings.
Subsequently we will reconsider the problems we have already mentioned,
 arising from the arbitrariness about the properties which can be considered
 as possessed in the case of identical fermions.


\subsubsection{Clarifying the role of the spatial and internal degrees of
 freedom}

Let us consider a system of two identical spin 1/2 particles.
We stress that if one would confine his attention to the spin degrees of
 freedom alone, then, following our definitions and theorems, one would be led
 to conclude that the singlet state,which can be obtained by antisymmetrizing,
 e.g., the state $\vert z \uparrow \rangle_{1}\vert z\downarrow\rangle_{2},$
 would be a non entangled state.
How does this fit with our previous remarks and the general (and correct)
 position that such a state is, in a sense, the paradigmatic case of an
 entangled two body system? We have already called attention to the
 necessity of taking also into account, e.g., the position of the
 constituents, to legitimately raise the relevant questions about their
 properties.
But the matter must be analyzed on more general grounds.
In analogy with state~(\ref{correlation1}) of the previous Subsection we
 consider a factorized state of the type
  \begin{equation}
   \label{quisotto}
   \vert \psi(1,2)\rangle = \vert z \uparrow \rangle_{1}\vert R_{i} \rangle_{1}
   \vert z \downarrow \rangle_{2} \vert R_{j} \rangle_{2},
  \end{equation}
with $\langle R_{i} \vert R_{j} \rangle=\delta_{ij}$.
Now we can make our point: even though it is meaningless (within a quantum
 context) to speak of particle 1 as distinguishable from particle 2, we can
 ``individuate'' the identical objects by resorting to the different spatial
 quantum numbers $i$ and $j$.
Concerning the state obtained from~(\ref{quisotto}) by the antisymmetrization
 procedure we are sure that one particle (we do not know which one) has the
 spatial property associated to the quantum number $i$ and one has the
 property associated to the quantum number $j$.
It is then meaningful to raise the question of the relations existing between
 the internal properties and the spatial properties. In the considered case
 we can use each of the differing quantum numbers to ``individuate'' the
 constituents and raise e.g. the question: has the particle identified by the
 quantum number $i$, definite spin properties?
The answer is obviously affirmative; in our case it definitely has spin up
 along the {\it z}-axis.
Note that we could also have used the spin quantum numbers to ``individuate''
 the particles and we could have raised the question: does the particle with
 spin up along {\it z} have precise spatial properties?
And the answer would have been yes: it has the spatial properties associated
 to the state $\vert R_{i}\rangle$.

On the contrary, for a state like
\begin{equation}
\label{veroepr}
 \vert \psi(1,2)\rangle = \frac{1}{\sqrt{2}}\, [\, \vert z \uparrow \rangle_{1}
 \vert z \downarrow \rangle_{2} - \vert z \downarrow \rangle_{1}
 \vert z \uparrow \rangle_{2}\,] \otimes [\, \vert R_{i} \rangle_{1}
 \vert R_{j} \rangle_{2}+ \vert R_{j} \rangle_{1}\vert R_{i} \rangle_{2}\,],
\end{equation}
which is not obtainable by antisymmetrizing a factorized state, it is not
 possible, for example, to attribute any definite spin property to the particle
 identified by the index $i$ and equivalently no definite spatial property
 can be attributed to the particle with spin up.
In the case where $\vert R_{i}\rangle $ and $\vert R_{j}\rangle $ correspond
 to two distant spatial location, the vector~(\ref{veroepr}) represents the
 paradigmatic state considered in the usual EPR argument and in the
 experiments devised to reveal the non-local features of quantum mechanics.

The picture should now be clear: no state of two fermions in the
 singlet spin state can be obtained by antisymmetrizing a factorized wave
 function, when also the remaining degrees of freedom are taken into account.
In this sense, and paying the due attention to the subtle problems we have
 discussed, one can understand how there is no contradiction between the usual
 statement that the singlet state is entangled and the fact that, if one
 disregards the spatial degrees of freedom, it can be obtained by
 antsymmetrizing a factorized spin state.


\subsubsection{More about the case of two identical particles}

In Section~\ref{entanglementandind} we have shown that, in the case of
 identical particles, property attribution is legitimate iff the state is
 obtained by symmetrizing or antisymmetrizing an appropriate factorized state.
However, in the fermion case the request that the state can
 be written in the form
  \begin{equation}
  \label{moreabout1}
   \vert \psi(1,2) \rangle = \frac{1}{\sqrt{2}} [\, \vert \Lambda^{(1)}
   \rangle \vert \Gamma^{(2)} \rangle - \vert \Gamma^{(1)} \rangle
   \vert \Lambda^{(2)} \rangle \,],
  \end{equation}
where $\vert \Lambda \rangle$ and $\vert \Gamma \rangle$ are two arbitrary
 orthogonal vectors of the single particle Hilbert space ${\cal H}^{(1)}$
 leaves some indefiniteness concerning the possessed properties and compels
 us to face the problem arising from this arbitrariness.

In fact, on  the one hand, according to the position we have taken in this
 paper, which is perfectly in line with the one of Einstein, i.e., that ``when
 one can predict the outcome of a prospective measurement with certainty,
 there is an element of physical reality associated to it'' all claims of the
 type ``one fermion possesses the complete set of properties associated
 to $\vert \Gamma \rangle$ and the other those associated to $\vert \Lambda
 \rangle$" are perfectly legitimate for the state~(\ref{moreabout1}).
On the other hand this might appear, at first sight, quite embarrassing when
 one takes into account that the properties we are considering, when we change
 the states in terms of which we express the unique state $|\psi (1,2)\rangle$
 of the composite system, may be very well incompatible among themselves, in
 the quantum mechanical sense.

However, there are at least two reasons for which one can ignore this, at
 first sight, puzzling situation, one of formal and physical nature, the
 second having more to do with the laboratory practice.
The general reason derives from the fact that within quantum mechanics it
 may very well happen that incompatible observables have common eigenstates.
For instance, with precise reference to the case under discussion, if
 consideration is given to the infinitely many noncommuting number operators
 $N_{\lambda}=a_{\lambda }^{\dagger }a_{\lambda }$ counting the number of
 fermions in an arbitrary single particle state $|\lambda \rangle$ of the
 two dimensional manifold spanned by $\vert\Lambda\rangle$ and
 $\vert\Gamma\rangle$, the state $\vert \psi (1,2)\rangle$ is a simultaneous
 eigenstate of all the $N_{\lambda }$'s belonging to the eigenvalue 1.
This implies, according to the quantum mechanical rules that any apparatus
 devised to measure whether there are fermions in such a state, will give
 with certainty the outcome 1, i.e., it will allow to conclude that ``there
 is one fermion in such a state''.
No matter how peculiar this situation might appear, it is a clear cut
 consequence of the formalism and of the criterion for attributing properties
 to physical systems\footnote{We consider it appropriate to call attention
 to a fact that makes the situation even less embarrassing than it might
 appear.
To this purpose, let us consider two arbitrary non-orthogonal vectors
 $\vert \lambda \rangle$ and $\vert \gamma \rangle$ of our two dimensional
 manifold and the associated projection operators $P_{\lambda}$ and
 $P_{\gamma}$.
In terms of them we build the projection operators $E_{\lambda}(1,2)$ and
 $E_{\gamma}(1,2)$, with obvious meaning of the symbols.
Suppose now we perform a measurement aimed to ascertain whether there is
 one fermion in state $\vert \lambda \rangle$, i.e.,  we measure
 $E_{\lambda}(1,2)$, or, equivalently, we measure the observable $N_{\lambda}$.
We then get for sure the eigenvalue 1  and, and this is the crucial point,
 the measurement does not alter in any way the state vector.
This means that the probability of finding, in a subsequent measurement, one
 fermion in the state $\vert \gamma \rangle$ is still equal to one, and has
 not been influenced by the first measurement.}.

Coming now to the relevant practical aspects of the problem we stress once more
 that, for what concerns entanglement and property attribution, the
 physically most significant and interesting aspect is the one of the non-local
 correlations between distant and noninteracting particles occurring in
 connection with entangled systems.
To illustrate this point we can make reference to the following state of two
 identical fermions:
  \begin{equation}
   \label{esempiononepr}
   |\psi (1,2)\rangle=\frac{1}{\sqrt{2}}\left[ |z\uparrow \rangle_{1}
   |R\rangle_{1} |z\downarrow\rangle_{2} |L\rangle_{2} - |z\downarrow
  \rangle_{1}
   |L\rangle_{1} |z\uparrow \rangle_{2} |R\rangle_{2}\right] .
  \label{L4}
  \end{equation}
For such a state, as we have already discussed in
 Subsection~\ref{correlationsinthecase}, it is
 perfectly legitimate, if we attach a prominent role to  positions,
 to claim that ``there is a particle at R and it has spin up along the
 $z-$axis'' as well as that ``there is a particle at L and it has spin down
 along the $z-$axis''.
The possibility of making such a claim is the characteristic feature which
 makes such a state basically different from
 the state~(\ref{proiezionesututto}) or from the singlet state of the EPR
 set-up.
One cannot however avoid recognizing that, if some meaning would be attached
 to single particle states like:
  \begin{equation}
   \label{statistrani}
   |\Gamma \rangle=\frac{1}{\sqrt{2}}\left[ |z\uparrow \rangle |R\rangle +
   |z\downarrow \rangle |L\rangle\right],\qquad|\Lambda
   \rangle=\frac{1}{\sqrt{2}}\left[ |z\uparrow \rangle |R\rangle -|z\downarrow
   \rangle |L\rangle\right]
  \end{equation}
which do not correspond either to definite locations or to definite spin
 properties of a particle, then one could claim that in
 state~(\ref{esempiononepr}) ``there is a particle with the properties
 associated to $\vert \,\Gamma\, \rangle$ {\em and} one with the properties
 associated to $|\Lambda\rangle$''.
It goes without saying that measurements involving states like those of
 equation~(\ref{statistrani}) are extremely difficult to perform and of no
 practical interest.


\subsection{Concerning some  misconceptions about entanglement for
 systems of identical particles}

In the literature one can find some inappropriate statements about
 entanglement in the case of systems whose constituents are identical.
Such misconceptions derive from not having appropriately taken into account
 the real physical meaning and implications of entanglement.
In its essence, the characteristic trait of entanglement derives from the fact
 that a system which is composed of two subsystems is associated to a
 state vector such that the subsystems have only ``potentialities" concerning
 most or even all conceivable observables, potentialities which are
 immediately actualized when one performs a measurement on one of the two
 subsystems (the most striking situation being connected to instantaneous
 actualization at-a-distance).
For distinguishable particles, as we have seen, such an occurrence is strictly
 related  to the fact that the state vector be nonfactorized.

It is therefore not surprising that some authors  have been inappropriately
 led to identify entanglement with factorizability.
However, suppose that in the case of two distinguishable particles,
 one starts from a state like~(\ref{quisotto}) which is manifestly factorized
 and, consequently, makes legitimate to state that particle 1 has spin up along
 the $z$-axis and is in the eigenstate $\vert\,R_{i}\,\rangle$ of an
 appropriate observable $\Omega^{(1)}$ pertaining to the eigenvalue
 $\omega_{i}$, while particle 2 has spin down along the $z$-axis and is in the
 eigenstate $\vert\,R_{j}\,\rangle$ of the observable $\Omega^{(2)}$
 pertaining to the
 eigenvalue $\omega_{j}$. Suppose now that the two particles are identical
 fermions, so that  one properly antisymmetrizes the above state.
Then, the resulting state vector is formally no longer factorized, but it is
 non-entangled since, as we stress once more taking the risk of being
 pedantic, it makes perfectly legitimate to make the joint statement
 that ``there is one fermion with spin up and the property $\Omega=\omega_{i}$
 and one fermion with spin down and the property $\Omega=\omega_{j}$".
Moreover, the act of measuring one of the two ``properties" does not change
 in any way the fact that the other property can be considered as objectively
 possessed both before and after the measurement process.
This is the reason for which the state {\em must be claimed} to be
 non-entangled.

We call attention to the fact  that the state being non-entangled is an
 intrinsic property characterizing it, as  follows from the analysis of the
 previous sections, and does not depend in any way on the basis or the
 formal apparatus we choose to describe it.

It is just due to a failure of fully appreciating the above facts that one
 can be led to make misleading statements.
As an example, in~\cite{GHZ1} it is stated that {\it one may not draw
 conclusions about entanglement in configuration space by looking at the
 states in Fock space}.
This statement is based on the fact that, according to the authors, in the
 case of two bosons, the state:
  \begin{equation}
   \vert \phi \rangle = \vert 1_{\vec{k}}\, 1_{\vec{l}} \rangle,
   \end{equation}
which describes two identical particles  with momenta $\vec{k}$ and $\vec{l}$,
 is a factorizable state in Fock space, being instead an entangled one if one
 looks at its form in terms of the momentum basis for single particles.
This argument is clearly in contradiction with what we have just pointed out.
In fact, the state $\vert \phi \rangle$, if we indicate with
 $a^{\dagger}(\vec{k})$ the creation operator of a boson with momentum
 $\vec{k}$, has the following Fock representation and, equivalently, the
 following expression in terms of a single particle momentum basis:
  \begin{equation}
   \vert \phi \rangle = \vert 1_{\vec{k}}\, 1_{\vec{l}} \rangle
   \equiv \frac{1}{\sqrt{2}}\,a^{\dagger}(\vec{k})a^{\dagger}(\vec{l})
   \vert 0 \rangle\:\:\:\:\:\Longleftrightarrow\:\:\:\:\:
   \frac{1}{\sqrt{2}}[\, \vert \vec{k} \rangle_{1}\vert \vec{l} \rangle_{2}
   + \vert \vec{l} \rangle_{1}\vert \vec{k} \rangle_{2}\, ]
  \end{equation}
As the formula shows, the state is obtained by a process of symmetrization
 of a factorized state of two ``orthogonal'' states, and as such it is
 non-entangled.
In particular it is perfectly legitimate to claim that there exists with
 certainty a boson with momentum $\vec{k}$ and one with momentum $\vec{l}$.

This means that states which are ``factorized" in Fock space
 have precisely the same physical properties as those following from their
 explicit form in  configuration or momentum  space. Being non-entangled, in
 the case of two identical particles is, just as in the case of two
 distinguishable particles, a property which has nothing to do with the way
 one chooses to express the state vector.


\section{Entanglement of $N$ indistinguishable particles}

We analyze here the case of $N$ indistinguishable particles.
In particular we will be concerned with the analog of the question we have
 discussed in the case of $N$ distinguishable particles: can one single out a
 ``subgroup'' of the constituents (obviously we cannot identify them) to
 which one can attach a complete set of properties as objectively possessed?
The problem is conceptually a rather delicate one and requires  remarkable
 care.
Moreover, it has to be stressed that it has a great conceptual and practical
 relevance.
For instance we can be naturally led to face a situation like the following:
 there is a Helium atom here and a Lithium atom there (in a distant region).
We then must pretend that a claim of the kind ``this one is a Helium atom"
 (or, as we will see, in general, one coinciding with it to an extreme - and
 controllable - degree of accuracy) can be made consistently, in spite of the
 fact that the correct wave function is totally antisymmetric under the
 exchange of the electrons of the Helium and Lithium atoms.

Besides these physical aspects we are mainly interested in defining in a
 conceptually correct way the idea that the set of $N$ identical particles we
 are dealing with can be partitioned into two ``subsets" of cardinality $M$
 and $K$, which are non-entangled with each other\footnote{The authors of
 ref.~\cite{refide3} have remarked that, in spite of their
indistinguishability,
 ``{\it the electrons of an atom, taken as a whole, possess some properties
 which are characteristic of a set.}
{\it For instance, they have a cardinality, even if we cannot count them,
 hence we cannot make an ordinal number to correspond to each electron}".
For this reason they have appropriately introduced the terminology quaset
 (abbreviation for quasi-set) for a collection of quantum elements which are
 indistinguishable from each other.
The authors have also called attention to the possibility of
 considering subquasets, by identifying their elements on the basis of their
 sharing a specific single-particle property. As a typical example they
 consider the electrons in the shell $2p$ of an atom as a subquaset.
In the analysis we are performing, we will deal with a strictly analogous
 situation, and so, to be rigorous, we should speak of quasets and subquasets.
Moreover, for a satisfactory description of the situation we are going to
 tackle, we should enlarge the idea of subquasets by making reference to a
 ``subgroup of particles" which are related only as a whole to a precise
 property.
In a sense, we will not make reference to the individual elements which have
 precise individual properties, but to the subquaset which has a global
 property.
Our generalization is, at any rate, strictly related to the one of the authors
 of ref. \cite{refide3}, since, as we will see, when we will split a quaset,
 e.g., of $N$ fermions into two subquasets of cardinality $M$ and $K$,
 respectively, we will be lead to assume that there exist an appropriate
 single particle basis such that the two quasets involve two disjoint subsets
 of the elements of this single particle basis.
In the paper, for simplicity, we will not use systematically the appropriate
 terminology of quasets, and we will speak, quite loosely, of ``groups of
 particles".
We believe the reader will have clear what we have in mind, allowing us to
avoid
 resorting to the use of a terminology which is yet not usual in the analysis
 of entanglement.}.
By following strictly the procedure we have introduced for the case of two
 particles we will do this by considering the possibility of attributing a
 complete set of properties to each subset and we will give the following
 definition:
  \begin{definizionedient}
  \label{definizionedient}
   Given a quantum system of $N$ identical particles  described by a pure
   state $\vert\psi^{(N)}\rangle$ we will say that it contains two {\bf
   non-entangled}
   ``subgroups" of particles of cardinality $M$ and $K$ $(M+K=N)$, when both
   subgroups possess a complete set of properties.
  \end{definizionedient}
The conditions under which it will be possible to attribute a complete set of
 properties to a quantum subsystem will be made mathematically precise in the
 following subsections.

As we will see, to make statements of the sort we are interested in, i.e.,
 that objective properties can be attached to the subsets associated to a
 partition of the particles or, equivalently, that such subsets are
 non-entangled among themselves, we have to impose quite strict constraints
 on the state vector of the whole system.
After having identified them in a very precise manner,  we will be able to
 evaluate how well they are satisfied in practice and, correspondingly, we
 will be in the position of judging the degree of legitimacy of our claims
 concerning precise physical situations.

Here we will deal, from the very beginning, separately with the fermion and
 boson cases.


\section{Identical fermions}

To analyze this problem it is appropriate to begin by fixing our notation
 and by deriving some simple results which we shall need in what follows.
We will deal with a system of $N$ identical fermions and with subsystems of
 such a system.


\subsection{Some mathematical preliminaries}

We  denote as ${\cal{H}}_{A}^{(R)}$ the Hilbert space which is appropriate
 for a system of $R$ identical fermions, i.e., the space of the totally
 skew-symmetric states $\vert \psi (1,...,R) \rangle$ of the variables
 (e.g spatial and internal) of the constituents.
Obviously ${\cal{H}}^{(1)}$ is the space of single particle states. Let us
 denote as $\left\{ \vert \varphi_{i} \rangle \right\}$ a complete orthonormal
 set in such a space.
A basis for ${\cal{H}}_{A}^{(N)}$ is then obtained by antisymmetrizing and
 normalizing the product states
 $\left\{\, \vert \varphi_{i_{1}}(1) \rangle\otimes ...\otimes \vert
 \varphi_{i_{N}}(N) \rangle \,\right\}$ which, when the subscripts take all the
 allowed values, are a basis of  ${\cal{H}}^{(1)}\otimes ...\otimes
 {\cal{H}}^{(1)}$.
For simplicity let us introduce, as usual, the linear antisymmetrization
 operator $A$ which acts in the following way on the states
 $\left\{ \,\vert \varphi_{i_{1}}(1) \rangle\otimes...\otimes\vert
 \varphi_{i_{N}}(N) \rangle \,\right\}$:
  \begin{equation}
   \label{antisimmetrizzatore}
   A\left\{ \, \vert \varphi_{i_{1}}(1) \rangle \dots \vert \varphi_{i_{N}}(N)
   \rangle \, \right\} \equiv \sum_{P}(-)^{p}P \left\{ \, \vert
  \varphi_{i_{1}}(1)
   \rangle \dots \vert \varphi_{i_{N}}(N) \rangle \,\right\},
  \end{equation}
where the sum is extended to all permutations $P$ of the variables $(1,...,N)$
 - or equivalently of the subscripts $(i_{1},...,i_{N})$ - and $p$ is the
 parity of the permutation $P$.
As it is well known $A\left\{ \vert \varphi_{i_{1}}(1)\rangle \dots \vert
 \varphi_{i_{N}}(N) \rangle \right\}$ can be simply expressed as the
 determinant of an appropriate matrix.
The states~(\ref{antisimmetrizzatore}) are not normalized, their norm being
 equal to $\sqrt{N!}$, so that  the basis generated in the above way is
 given by the states $\frac{1}{\sqrt{N!}}A\left\{ \vert \varphi_{i_{1}}(1)
 \rangle...\vert \varphi_{i_{N}}(N)\rangle \right\}$.

We will not use directly such states to express the most general state of
 ${\cal{H}}_{A}^{(N)}$, but we will write it as
\begin{equation}
 \vert \Psi (1,...,N)\rangle = \sum_{i_{1},...,i_{N}}a_{i_{1}...i_{N}} \vert
 \varphi_{i_{1}}(1)\rangle \dots \vert \varphi_{i_{N}}(N) \rangle,
 \label{GC2}
\end{equation}
where the coefficients $a_{i_{1}...i_{N}}$ are totally skew-symmetric and
 are chosen in such a way that $\vert \Psi (1,...,N)\rangle$ turns out to
 be normalized, i.e. they satisfy:
  \begin{equation}
   a_{P(i_{1}...i_{N})} = (-)^{p} a_{i_{1}...i_{N}};\qquad \:\:\:\:\:\:
   \sum_{i_{1},...,i_{N}} \vert  a_{i_{1}...i_{N}} \vert^{2}=1.
   \label{GC3}
  \end{equation}
In the first of the above relations $P$ represents an arbitrary permutation
 of the subscripts of $a_{i_{1}...i_{N}}$, and $p$ the parity of the considered
 permutation.

>From now on we will deal with ${\cal{H}}_{A}^{(N)}$ and we will be
interested
 in ``splitting'' the $N$ identical constituents into two ``subsets'' (with
 reference to their cardinality) of $M$ and $K=N-M$ particles.
We begin by recalling a trivial fact, i.e, that the Hilbert space
 ${\cal{H}}_{A}^{(N)}$ is a closed linear submanifold of the direct product
 ${\cal{H}}_{A}^{(M)}\otimes{\cal{H}}_{A}^{(K)}$.
This follows trivially from  Laplace's formula for determinants which can be
 written as:
\begin{equation}
 A\left\{\, \vert \varphi_{i_{1}}(1) \rangle...\vert \varphi_{i_{M}}(M)
 \rangle \vert \varphi_{r_{1}}(M+1)\rangle... \vert \varphi_{r_{K}}(N)
 \rangle\, \right\} =
\end{equation}
\[
  {\cal{G}} \biggl[ \, A \left\{ \vert \varphi_{i_{1}}(1)\rangle ...\vert
  \varphi_{i_{M}}(M)\rangle \right\}
  A\left\{ \vert \varphi_{r_{1}}(M+1) \rangle ...\vert \varphi_{r_{K}}(N)
  \rangle \right\} \biggr]
 \label{GC4}
\]
where the symbol ${\cal{G}}$ at the r.h.s. indicates that one has to
 sum over all the permutations between the first M particles and the
 remaining ones, attaching to the various terms the appropriate sign.
The above formula shows that the elements of a basis of ${\cal{H}}_{A}^{(N)}$
 can be expressed in terms of the direct products of the elements of two
 orthonormal complete sets of ${\cal{H}}_{A}^{(M)}$ and ${\cal{H}}_{A}^{(K)}$.
Since, when consideration is given to two states which have common
 single-particle indices the antisymmetrization procedure yields the zero
 vector $\vert \omega \rangle$ of ${\cal{H}}_{A}^{(N)}$, the claim that
 ${\cal{H}}_{A}^{(N)}\subset {\cal{H}}_{A}^{(M)}\otimes {\cal{H}}_{A}^{(K)}$
 follows.


\subsubsection{Defining an appropriate single particle basis with reference
 to a given state of $M$ fermions}
\label{defininganappropriate}

Given our system of $N$ identical fermions, we pick up $M$ of them, and we
 consider a state $\vert \Pi^{(M)}(1...M) \rangle \in{\cal{H}}_{A}^{(M)}$
 whose Fourier decomposition on the product basis of the single particle
 states $\left\{ \vert \varphi_{i} \rangle \right\}$ is:
  \begin{equation}
   \label{espansione}
   \vert \Pi^{(M)}(1...M)\rangle  =  \sum_{i_{1}...i_{M}} a_{i_{1}...i_{M}}
   \vert \varphi_{i_{1}}(1)\rangle ...\vert \varphi_{i_{M}}(M)\rangle.
  \end{equation}
We then choose an arbitrary normalized single particle state $\vert
 \Phi^{(1)}\rangle$, we represent it on the chosen single particle basis,
  \begin{equation}
   \label{defining1}
   \vert \Phi^{(1)} \rangle = \sum_{t} b_{t} \vert \varphi_{t} \rangle,
  \end{equation}
and, with reference to the state~(\ref{espansione}), we define the following
 subset $V_{\perp }^{\Pi 1}$ of ${\cal{H}}^{(1)}$:
  \begin{equation}
   \label{defining2}
   V_{\perp }^{\Pi 1}\equiv \left\{\, \vert \Phi^{(1)}\rangle\, \vert \:
   \sum_{t}\, b_{t}^{*}a_{ti_{2}...i_{M}}=0, \:\:\: \forall\:\:\:
   i_{2},...,i_{M}\right\}
  \end{equation}
We note that $V_{\perp }^{\Pi 1}$ is independent from the single particle
 basis we have used to identify it and from the index which is saturated in
 equation~(\ref{defining2}).

The reader will have no difficulty in realizing that $V_{\perp }^{\Pi 1}$ is
 a closed linear submanifold\footnote{It must be noted that it may very well
 happen that such a manifold turns out to contain only the zero vector of
 ${\cal{H}}^{(1)}$. To see this we consider for simplicity the case $M=2$ and
 we write $\vert\Pi^{(2)}(1,2)\rangle$ as in equation~(\ref{espansione}),
 $\vert\Pi^{(2)}(1,2)\rangle = \sum_{ij}a_{ij}\vert \varphi_{i}(1) \rangle
 \vert\varphi_{j}(2)\rangle, a_{ij}=-a_{ji}$. A vector $\vert\phi\rangle =
 \sum_{t}b_{t}\vert\varphi_{t}\rangle$ belongs to $V_{\perp }^{\Pi 1}$ iff
 it satisfies $\sum_{j}a_{ij}b_{j}^{*}=0$. If one considers a linear operator
 $A$ whose representation is given, in the considered basis, by the matrix
 $a_{ij}$, then $\vert\phi\rangle\in{V_{\perp }^{\Pi 1}}$ iff $A$ admits the
 zero eigenvalue, the vector $(b_{1}^{*},...,b_{k}^{*},...)$ being the
 associated eigenvector.
The reader will have no difficulty in realizing that an operator $A$ whose
 matrix elements satisfy $a_{ij}=-a_{ji}$ and does not admit the zero
 eigenvalue is easily constructed. Actually the Pauli matrix $\sigma_{y}$ is
 such an operator.
The conclusion is that the request - which we will do in what follows - that
 $V_{\perp }^{\Pi 1}$ does not reduce to the
 zero vector, implies by itself some constraints for the state
 $\vert\Pi^{(M)}\rangle$. If such constraints are not satisfied, then the
 procedure we are going to present cannot be developed and the state
 $\vert\Pi^{(M)}\rangle$ cannot be combined with another state
 $\vert\Phi^{(K)}\rangle$ ($K=N-M$) to generate a state of
 ${\cal{H}}^{(N)}_{A}$ such that there are ``subsets" of $M$ and $K$ particles
 possessing definite properties.} of ${\cal{H}}^{(1)}$.
It is useful to mention that another way to characterize $V_{\perp }^{\Pi 1}$
 is the following.
Suppose we use the shorthand notation $\int dX$ to denote an integral over
 the space and a summation over the internal variables of the  fermion $X$.
Then eq.~(\ref{defining2}) can be written as
  \begin{equation}
   \label{defining3}
   V_{\perp }^{\Pi 1}\equiv \left\{\, \vert \Phi^{(1)}\rangle \,| \: \int
   d1 \Phi^{(1)*}(1)\Pi^{(M)}(1,2,...,M) = 0\right\} ,
  \end{equation}
0 being the function of the variables $(2,...,M)$ which vanishes almost
 everywhere.
It goes without saying that, due to the skew-symmetry of $\Pi^{(M)}(1,...,M)$
 in its arguments, the same condition~(\ref{defining3}) can be written by
 saturating an arbitrary variable, i.e., by imposing that $\vert \Phi^{(1)}
 \rangle$ satisfies $\int dX \Phi^{(1)*}(X) \Pi^{(M)}(1,...,X-1,X,X+1,...,M)
 =0.$

Since $V_{\perp }^{\Pi 1}$ is a closed linear submanifold of ${\cal{H}}^{(1)}$
 we can now consider its orthogonal complement $V^{\Pi 1}$:
\begin{equation}
  {\cal{H}}^{(1)}=V^{\Pi 1}\oplus V_{\perp }^{\Pi 1},  \label{M4}
\end{equation}
and we can choose a complete orthonormal set $\left\{\, \vert \phi_{i}
 \rangle\,\right\}$ of single particle states such that, splitting the whole
 set of nonnegative integers into two disjoint subsets $\Delta $ and
 $\Delta_{\perp }$, one has:
\begin{equation}
 \label{kant}
   \vert \phi_{i} \rangle \in V^{\Pi 1} \Leftrightarrow \quad i\in
   \Delta; \qquad \:\:\:\:\:\:\:\:
   \vert\phi_{i} \rangle \in V_{\perp }^{\Pi 1}
   \Leftrightarrow \quad i\in \Delta_{\perp }.
   \label{L5}
\end{equation}
The following theorem will be useful in what follows:
\begin{sviluppodipi}
The vector $\vert \Pi^{(M)}(1,...,M) \rangle$ can be written as:
  \begin{equation}
   \vert \Pi^{(M)}(1,...,M) \rangle = \sum_{i_{1}...i_{M}\in \Delta }
   c_{i_{1}...i_{M}} \vert \phi_{i_{1}}(1) \rangle ...
   \vert \phi_{i_{M}}(M) \rangle,
  \end{equation}
where all the indices $i_{1},...,i_{M}$ belong to $\Delta $ and moreover all
 single particle states whose indices belong to $\Delta $ actually appear
 (in the sense that some nonvanishing coefficients characterized by them
 occur) at the r.h.s. of the above equation.

In other words, the Fourier expansion of $\vert \Pi^{(M)}\rangle $ in terms
 of the states of the basis $\left\{ \vert \phi_{i} \rangle \right\}$
 involves all single particle states spanning $V^{\Pi1}$ and no single
 particle state spanning $V_{\perp}^{\Pi1}$.
\end{sviluppodipi}
\textit{Proof}: Suppose there exists an index $k$ belonging to
  $\Delta_{\perp}$ such that $c_{ki_{2}...i_{M}}\neq 0$ for at least one choice
  of the indices $i_{2}\dots i_{M}$.
On the other hand, since $k\in\Delta_{\perp}$, the single particle basis vector
  $\vert \phi_{k} \rangle$ belongs to $V^{\Pi1}_{\perp}$ and, as such, it
  satisfies
\begin{equation}
\label{stranezza}
  \langle \phi_{k} \vert \Pi^{(M)} \rangle = \sum_{i_{2}\dots i_{M}}
  c_{ki_{2}...i_{M}} \vert \phi_{i_{2}}(2) \rangle \dots \vert \phi_{i_{M}}
 (M) \rangle = 0.
\end{equation}
Since the vectors $\vert \phi_{i_{2}}(2) \rangle \dots \vert \phi_{i_{M}}
 (M) \rangle$ are linearly independent for any given choice of the indices
 $i_{2}\dots i_{M}$, eq.~(\ref{stranezza}) implies $c_{ki_{2}\dots i_{M}} = 0,
 \:\:\:\forall \:i_{2}\dots i_{M}$, which is contrary to the hypothesis.

On the other hand, let us suppose that there exists an index $j$ belonging to
 $\Delta$ such that $c_{ji_{2}...i_{M}}= 0\:\:\:\forall i_{2}... i_{M}$.
This means that
  \begin{equation}
   \langle \phi_{j} \vert \Pi^{(M)} \rangle = \sum_{i_{2}\dots i_{M}}
   c_{ji_{2}...i_{M}} \vert \phi_{i_{2}}(2) \rangle \dots \vert \phi_{i_{M}}
   (M) \rangle = 0
  \end{equation}
implying that the vector $\vert \phi_{j} \rangle$ belongs to
 $V^{\Pi1}_{\perp}$ which is absurd. $\:\:\:\:\Box$

Summarizing, choosing any vector $\vert \Pi^{(M)}(1,...,M)\rangle $ of
 ${\cal{H}}_{\cal{A}}^{(M)}$ such that $V^{\Pi 1}_{\perp}$ differs from the
 zero vector of ${\cal{H}}^{(1)}$, uniquely identifies two closed linear
 submanifolds of ${\cal{H}}^{(1)}$ whose direct sum coincides with
 ${\cal{H}}^{(1)}$ itself,
 and, correspondingly, a complete orthonormal set of single particle states
 which is the union of two subsets $\left\{ \vert \phi_{i}\rangle\right\},i\in
 \Delta$ and $\left\{ \vert \phi_{j}\rangle\right\},j\in \Delta_{\perp}$ such
 that all and only the states $\left\{ \vert\phi_{i}\rangle\right\}$ for which
 $i\in \Delta$ enter into the Fourier expansion of $ \vert \Pi^{(M)}(1,...,M)
 \rangle $ in terms of the basis generated by the antisymmetrized and
 normalized products of the set $\left\{\vert \phi_{i}\rangle\right\}$.

We pass now from the Hilbert space ${\cal{H}}^{(1)}$ to the spaces
 ${\cal{H}}_{A}^{(M)}$ and ${\cal{H}}_{A}^{(K)}$, which, as already stated,
 are those we are interested in.
Having partitioned the complete set of single particle states $\left\{\vert
 \phi_{j} \rangle\right\}$ into two subsets according to their indices
 belonging to $\Delta$ or $\Delta_{\perp}$, we consider now two important
 (for our purposes) proper submanifolds $V^{\Pi M}$ of ${\cal H}_{A}^{(M)}$
 and  $V^{\Pi K}_{\perp}$ of ${\cal H}_{A}^{(K)}$, respectively.
They are simply the manifolds spanned by the states:
  \begin{equation}
  \label{M5}
   V^{\Pi M}:\:\:\:\:\:\frac{1}{\sqrt{M!}}\: A \,\left\{\, \vert
   \phi_{i_{1}} \rangle,...,\vert \phi_{i_{M}} \rangle \right\}, \qquad
   i_{1},...,i_{M}\in \Delta
  \end{equation}
  \[
   V^{\Pi K}_{\perp}:\:\:\:\:\:\frac{1}{\sqrt{K!}}\: A \,\left\{\, \vert
   \phi_{j_{1}} \rangle ,...,\vert
   \phi_{j_{K}} \rangle \right\}, \qquad j_{1},...,j_{K}\in \Delta_{\perp}.
    \nonumber
  \]
In brief, $V^{\Pi M}$ is the set of all the states of  ${\cal H}_{A}^{(M)}$
 such that their Fourier expansion in terms of the single particle states
 $\left\{ \vert \phi_{i}\rangle\right\}$ contains only states whose indices
 belong to $\Delta$, and $V_{\perp }^{\Pi K}$ is the set of all states of
 ${\cal{H}}_{A}^{(K)}$ such that their Fourier expansions contains only
 states whose indices belong to $\Delta_{\perp}$.

In virtue of our definition of the two manifolds $V^{\Pi M}$ and
 $V_{\perp}^{\Pi K}$, there follows trivially that the saturation of any
 variable of a state $\vert \Phi^{(K)} \rangle \in V_{\perp }^{\Pi K}$ with
 any variable of a state $\vert \Sigma^{(M)} \rangle \in V^{\Pi M} $ gives
 the null function of the unsaturated variables:
  \begin{equation}
   \label{int}
   \int dX\, \Sigma ^{(M)}(1,...,X,...M)\, \Phi ^{(K)*}(M+1,...,X,...N)=0.
  \end{equation}
We will call a pair of states for which~(\ref{int}) holds ``one-particle
 orthogonal".
Analogously, when we have two closed linear manifolds such that
 condition~(\ref{int}) is satisfied for any pair of vectors taken from one
 and the other of them, we will say that the manifolds themselves are
 ``one-particle orthogonal".

With reference to eq.~(\ref{int}) we would like to call attention to the fact
 that, taking into account the preceding arguments, one can easily prove the
 following theorem:
\begin{giancarlo}
 Given any pair of states which are ``one-particle orthogonal", one can find
 an appropriate complete orthonormal single particle basis such that the
 Fourier expansions of the two states involve disjoint subsets of the states of
 this single particle basis.
\end{giancarlo}
The proof is easily obtained by noticing, first of all, that eq.~(\ref{int}),
 if one fixes the value of all the variables appearing in $\Phi^{(K)}$
 different from $X$, shows that the manifold $V^{\Sigma 1}_{\perp}$ (with
 obvious meaning of the symbol) does not
 reduce to the zero vector.
One then can follow the previous procedure to build the appropriate single
 particle basis satisfying the above theorem.


\subsubsection{Antisymmetrized products of appropriate states of
 ${\cal{H}}_{A}^{(K)}$ and a given state of ${\cal{H}}_{A}^{(M)}$}

In this subsection we consider a fixed state $\vert \Pi^{(M)}(1,...,M)
 \rangle$ of ${\cal{H}}_{A}^{(M)}$ and an arbitrary
 state $\vert \Phi^{(K)}(M+1,...,N) \rangle$ of $V_{\perp}^{\Pi K}$.
We take the direct product of the two and we totally antisymmetrize
 it, i.e., we consider the non-normalized state
\begin{eqnarray}
 \label{paperino}
 \vert \tilde{\psi}^{(N)} (1,...,N) \rangle & = &
 P_{A} [\:\vert \Pi^{(M)}(1,...,M) \rangle \otimes  \vert
 \Phi^{(K)}(M+1,...,N) \rangle\: ] \\ & = & \frac{1}{N!} \, A\,[\:
 \vert \Pi^{(M)}(1,...,M) \rangle \otimes  \vert \Phi^{(K)}(M+1,...,N)
 \rangle ]. \nonumber
 \end{eqnarray}
 where the linear operator
 $P_{A}=\frac{1}{N!}A$ is the projection operator on the submanifold
 ${\cal{H}}_{A}^{(N)}$ of ${\cal{H}}^{(1)}\otimes \dots \otimes
 {\cal{H}}^{(1)}$.

To evaluate its norm as well as to prove a theorem which will be
 useful in what follows, it is convenient to resort to a simple
 trick by dividing the permutations of the $N$ particles implied by
 the symbol $A$ in the above equation, into two families ${\cal{F}}$
 and ${\cal{G}}$, where ${\cal{F}}$ contains all the permutations
 which exchange the first $M$ and/or the second $K$ variables among
 themselves, while ${\cal{G}}$ contains only permutations which
 exchange {\it at least} one variable $(1,...,M)$ with the remaining
 ones.
It holds :
\begin{equation}
\label{paperone}
 A \equiv [\: \sum_{{\cal F}} (-1)^{f}  F + \sum_{{\cal G}} (-1)^{g} G\:],
\end{equation}
 $f$ and $g$ being the parity of the corresponding permutations.

Note that, since
\begin{equation}
\label{paperina}
 F[\,\vert \Pi^{(M)}(1,...,M) \rangle \otimes \vert \Phi^{(K)}(M+1,...,N)
  \rangle ] = (-1)^{f}[\vert \Pi^{(M)}(1,...,M) \rangle \otimes \vert
  \Phi^{(K)}(M+1,...,N) \rangle ],
\end{equation}
we have
\begin{equation}
\label{mafalda}
\sum_{{\cal F}} (-1)^{f} F[\vert
 \Pi^{(M)} (1,...,M) \rangle \otimes \vert \Phi^{(K)} (M+1,...,N)
  \rangle ] = \\
\end{equation}
\[ \sum_{{\cal F}}[\vert \Pi^{(M)} (1,...,M) \rangle \otimes \vert \Phi^{(K)}
 (M+1,...,N) \rangle ] =  M!K! \:[\vert \Pi^{(M)} (1,...,M) \rangle \otimes
 \vert \Phi^{(K)} (M+1,...,N) \rangle ].
\]
Before going on we remark that if $\vert \chi^{(K)} \rangle$ belongs to
 $V_{\perp}^{\Pi K}$ one has
\begin{equation}
\label{cavolo}
\langle \chi^{(K)} (M+1,...,N) \vert  \:\sum_{{\cal G}} (-1)^{g} G\: \vert\,
 \Pi^{(M)} (1,...,M) \rangle
 \vert \Phi^{(K)} (M+1,...,N) \rangle =0,
\end{equation}
where zero denotes the function of the variables $(1,...,M)$ which vanishes
  almost everywhere.
In fact any individual term of the sum over
 ${\cal G}$ has at least one of the variables from $M+1$ to $N$ which
 belongs to the state $\vert\Pi^{(M)}\rangle$ and, as such, it
 involves single particle state indices confined to the set $\Delta$.
Since the same variable belongs to the state
 $\vert\chi^{(K)}\rangle$, and therefore it is associated to single
 particle states whose indices belong to $\Delta_{\perp}$, the
 integration over such a variable gives the result zero.

Coming back to our unnormalized state~(\ref{paperino}), taking into
 account that $P_{A}^{2}= P_{A}$, we have
 \[
 \label{gastone}
 \langle \tilde{\psi}^{(N)} \vert \tilde{\psi}^{(N)} \rangle =
 \]
 \[
  = \frac{1}{N!} \langle \Pi^{(M)}(1,...,M) \Phi^{(K)}(M+1,...,N)
   \vert \: [\: \sum_{{\cal F}} (-1)^{f} F + \sum_{{\cal G}} (-1)^{g}
   G \:] \:\vert \Pi^{(M)}(1,...,M) \Phi^{(K)}(M+1,...,N) \rangle
 \]
\begin{equation}
 = \frac{1}{N!} \sum_{{\cal F}} \langle \Pi^{(M)}(1,...,M)
 \Phi^{(K)}(M+1,...,N) \vert \Pi^{(M)}(1,...,M)
 \Phi^{(K)}(M+1,...,N) \rangle = \frac{K!M!}{N!}.
\end{equation}
In deriving the above equation we have taken into account the fact
 that, since $\vert\Phi^{(K)}(M+1,...,N)\rangle \in V_{\perp}^{\Pi K}$,
 the sum over ${\cal G}$ does not contribute, in accordance
 with~(\ref{cavolo}).
The correctly normalized state we are interested in is then
\begin{equation}
\label{qui}
 \vert \psi^{(N)} (1,...,N) \rangle = \sqrt{ { N \choose K}} P_{A}\: [\:
   \vert \Pi^{(M)}(1,...,M)\rangle \otimes \vert \Phi^{(K)}(M+1,...,N)
   \rangle \:]
\end{equation}
Having identified the closed linear manifold $V_{\perp }^{\Pi K}$ of
 ${\cal{H}}_{A}^{(K)}$ we consider now a complete orthonormal set
 $\left\{ \vert \Theta_{\perp i}^{(K)}(M+1,...,N)\rangle \right\} $ which
 spans such a manifold.
In terms of these states and of the state $\vert
 \Pi^{(M)}(1,...,M)\rangle $ we build up the orthonormal set
 $\left\{ \vert \omega_{\perp i}^{(N)} (1,...,N) \rangle \right\} $
 of states of ${\cal{H}}_{A}^{(N)}$ according to:
\begin{equation}
\label{bassotti}
 \vert \omega_{\perp i}^{(N)}(1,...,N) \rangle
   \equiv \sqrt{ { N \choose K}} P_{A} \left[\: \vert
   \Pi^{(M)}(1,...,M) \rangle \otimes \vert \Theta_{\perp i}^{(K)}(M+1,...,N)
 \rangle\: \right].
\end{equation}
We already know that such states are normalized, while their orthogonality is
  easily proved by taking into account that $P_{A}^{2}=P_{A},$ and
  eqs.~(\ref{paperone}) and~(\ref{cavolo}).

There follows that the operators $\vert \omega_{\perp i}^{(N)}
 \rangle \langle \omega_{\perp i}^{(N)} \vert$ are a set of
 orthogonal projection operators and, consequently, the operator
\begin{equation}
\label{lucilla}
 E_{A\perp }^{\Pi (N)} = \sum_{i}
   \vert  \omega_{\perp i}^{(N)} \rangle \langle \omega_{\perp i}^{(N)} \vert
\end{equation}
is also a projection operator of ${\cal{H}}_{A}^{(N)}$.


\subsubsection{Some useful technical details about the formal procedure of
 the previous subsections}
\label{someuseful}

The identification of the ``one-particle orthogonal" linear manifolds
 $V^{\Pi M}$ and $V^{\Pi K}_{\perp}$ has been made starting from the
 consideration of a precise state $\vert\Pi^{(M)}\rangle$ of
 ${\cal{H}}_{A}^{(M)}$. However, since we will be interested in states
 like~(\ref{qui}) which are obtained by antisymmetrizing a product of a state
 $\vert\Pi^{(M)}\rangle$ and a state $\vert\Phi^{(K)}\rangle$ which are
 ``one-particle orthogonal", we could have followed the opposite line of
 approach, by assigning the prominent role to the state
 $\vert\Phi^{(K)}\rangle$.
In doing this we would have been led to identify two ``one-particle
 orthogonal" linear manifolds $V^{\Phi K}$ and $V^{\Phi M}_{\perp}$, which
 differ, in general, from those mentioned above.

Note, however, that just as  $\vert\Pi^{(M)}\rangle \in V^{\Pi M}$ and
 $\vert\Phi^{(K)}\rangle \in V^{\Pi K}_{\perp}$, it also happens that
 $\vert\Pi^{(M)}\rangle\in V^{\Phi M}_{\perp}$ and
 $\vert\Phi^{(K)}\rangle \in V^{\Phi K}$.

The simplest example of the above situation is represented, in the case of
 two identical fermions, by the state:
\begin{equation}
 \vert \psi (1,2)\rangle = \frac{1}{\sqrt{2}}\,[\, \vert\Lambda(1)\rangle
  \vert\Gamma(2)\rangle - \vert\Gamma(1)\rangle\vert\Lambda(2)\rangle\,]
\end{equation}
\noindent with $\langle\Lambda(i)\vert\Gamma(i)\rangle=0$. In such a case,
 if we start with the state $\vert\Lambda \rangle$, the manifold
 $V^{\Lambda 1}_{\perp}$ is the one spanned by $\vert\Gamma \rangle$ and by
 an orthonormal set of states $\vert\Theta_{\perp i}\rangle$ which spans
 the manifold orthogonal to both $\vert\Gamma\rangle$ and
 $\vert\Lambda \rangle$.
If we identify the manifolds by the corresponding projection operators, we
 have:
  \begin{equation}
   \begin{array}{rl}
   V^{\Lambda 1}&\Longleftrightarrow P=\vert \Lambda \rangle \langle \Lambda
   \vert \\
   V^{\Lambda 1}_{\perp}&\Longleftrightarrow
   P=\vert\Gamma\rangle\langle\Gamma\vert+\displaystyle\sum_{i} \vert
   \Theta_{\perp i}\rangle\langle\Theta_{\perp i}\vert.
   \end{array}
  \end{equation}
On the contrary, if we choose the state $\vert\Gamma\rangle$ to set up our
 procedure, we would end up with the two manifolds:
\begin{equation}
 \begin{array}{rl}
 V^{\Gamma 1}&\Longleftrightarrow P=\vert\Gamma\rangle\langle\Gamma\vert \\
 V^{\Gamma 1}_{\perp}&\Longleftrightarrow
   P=\vert\Lambda\rangle\langle\Lambda\vert+\displaystyle\sum_{i}
   \vert\Theta_{\perp i}\rangle\langle\Theta_{\perp i}\vert.
  \end{array}
\end{equation}
It goes without saying that in such a case one could consider two other one
 particle orthogonal manifolds $W$ and $W_{\perp}$ as follows:
\begin{equation}
 \begin{array}{rl}
 W&\Longleftrightarrow
   P=\vert\Lambda\rangle\langle\Lambda\vert+\displaystyle\sum_{i\in\delta}
   \vert\Theta_{\perp i}\rangle\langle\Theta_{\perp i}\vert \\
   W_{\perp}&\Longleftrightarrow
   P=\vert\Gamma\rangle\langle\Gamma\vert+\displaystyle\sum_{i\in\delta^{'}}
   \vert\Theta_{\perp i}\rangle\langle\Theta_{\perp i}\vert,
   \end{array}
  \end{equation}
\noindent $\delta$ and $\delta^{'}$ representing a partition of the positive
 integers.

It is very easy to understand the formal reasons of the just considered
 situation.
When dealing with a system of $N$ fermions, we started with the state
 $\vert\Pi^{(M)}\rangle$, and we have identified the single particle
 submanifold $V^{\Pi 1}_{\perp}$ characterized by the orthonormal set
 $\vert\phi_{j}\rangle$, $j\in\Delta_{\perp}$.
We have then defined $\Delta$ as the complement of $\Delta_{\perp}$ and we
 have shown that the Fourier expansion of $\vert\Pi^{(M)}\rangle$ involves
 all single particle states $\vert\phi_{i}\rangle$ for which $i\in\Delta$.
We have also taken into account a state $\vert\Phi^{(K)}\rangle\in
 V^{\Pi K}_{\perp}$.
We remark now that there is no reason why the Fourier expansion of
 $\vert\Phi^{(K)}\rangle$ should involve all states for which
 $j\in\Delta_{\perp}$.
Suppose it actually involves only a subset $\Delta_{\Phi}$ of $\Delta_{\perp}$.
If this is the case, we are naturally led to consider the following
 eigenmanifolds:
\begin{itemize}
 \item the manifold $V_{\Delta}^{(M)}$ which is spanned by the basis vectors
  \begin{equation}
   \frac{1}{\sqrt{M!}}\: A \,\left\{\, \vert \phi_{i_{1}} \rangle ,...,\vert
   \phi_{i_{M}} \rangle \right\}, \qquad i_{1},...,i_{M}\in \Delta;
  \end{equation}
 \item the manifold $V_{\Delta_{\Phi}}^{(K)}$ which is spanned by the basis
  vectors
 \begin{equation}
 \frac{1}{\sqrt{K!}}\: A \,\left\{\, \vert \phi_{j_{1}} \rangle ,...,\vert
 \phi_{j_{K}} \rangle \right\}, \qquad j_{1},...,j_{K}\in \Delta_{\Phi}\subset
 \Delta_{\perp};
  \end{equation}
\item the manifolds $V_{\Delta^{'}}^{(M)}$ and $V_{\Delta^{'}}^{(K)}$ which
 are spanned by the basis vectors
  \begin{equation}
   \frac{1}{\sqrt{M!}}\: A \,\left\{\, \vert \phi_{r_{1}} \rangle ,...,\vert
   \phi_{r_{M}} \rangle \right\}, \qquad r_{1},...,r_{M}\in \Delta^{'},
  \end{equation}
  and
\begin{equation}
 \frac{1}{\sqrt{K!}}\: A \,\left\{\, \vert \phi_{s_{1}} \rangle ,...,\vert
 \phi_{s_{K}} \rangle \right\}, \qquad s_{1},...,s_{K}\in \Delta^{'},
 \end{equation}
\end{itemize}

\noindent respectively, where $\Delta^{'}$ contains all single particle
 indices which do not belong to $\Delta$ or $\Delta_{\Phi}$.

Despite  the fact that there seems to be a certain degree of freedom in
 choosing a couple of ``one-particle orthogonal'' manifolds, the
 appropriateness of the above remarks will appear clearly when, in
 Subsection~\ref{nonentangledsubsets}, we will discuss the physical meaning
 of our requirements
 concerning complete sets of properties and/or the non-entangled
 character of appropriate subsets of a system of identical constituents.
We will in fact show that the one particle orthogonality is a necessary
 condition in order that one can do the physics within each such
 manifold by disregarding the other.


\subsection{Entanglement and Properties for systems of $N$ identical
 fermions}

Bearing in mind the mathematical formalism we have introduced in the previous
 sections, we can now formalize the idea of non-entangled states of a system
 of identical fermions.


\subsubsection{States of many identical fermions and their properties}

We begin by characterizing in a mathematically precise way the fact that a
 subgroup, or, better a ``subquaset" of $M$ of the $N$ identical fermions
 possesses a complete set of properties:
\begin{defentanggroupfermion}
  Given a system $S^{(N)}$ of $N$ identical fermions in a pure state $\vert
  \psi^{(N)}\rangle$ of ${\cal{H}}_{A}^{(N)}$ we will claim that two subsets
  of cardinality $M$ and $K$ ($N=M+K$), respectively, both possess a complete
  set of properties iff there exists a
  state $\vert \Pi^{(M)} \rangle$ of ${\cal{H}}_{A}^{(M)}$ such that
 \begin{equation}
   \label{condizionedinonsep}
   \langle \psi^{(N)} \vert E_{A\perp }^{\Pi (N)}(1,...,N) \vert \psi^{(N)}
   \rangle =1,
  \end{equation}
where $E_{A\perp }^{\Pi (N)}(1,...,N) = \sum_{s} \vert \omega_{\perp s}^1{(N)}
 \rangle \langle \omega_{\perp s}^{(N)} \vert$ is the projection operator of
 ${\cal{H}}_{A}^{(N)}$ given by eq.~(\ref{lucilla}), which is uniquely
 identified, according to the previous procedure, by the assignment of the
 state $\vert \Pi^{(M)}\rangle$.
\end{defentanggroupfermion}

Condition~(\ref{condizionedinonsep}) assures that, given the state
 $\vert \psi^{(N)}\rangle$ of $N$ identical fermions and the specific single
 particle basis $\left\{ \vert \phi_{k}\rangle \right\}$ of
 equation~(\ref{kant}), the probability of finding ``a group" of $M$ particles
 described by the state $\vert \Pi^{(M)} \rangle$ {\em and} the remaining ones
 in single particle states which do not appear in the Fourier decomposition of
 $\vert \Pi^{(M)}\rangle$, equals one.
In the next subsections we will discuss the precise physical reasons
 for which, when the above conditions are satisfied, it is correct to claim
 that ``there is a set of $M$ fermions which is non entangled with
 the remaining set of $K$ particles''.

Here we note the significant fact that $E_{A\perp}^{\Pi(N)}(1,...,N)$ turns
 out to be the restriction to the totally antisymmetric manifold of $N$
 indistinguishable fermions of the projection operator
 $E(1,...,N)=\tilde{\cal{S}} \:[\: \vert \Pi^{(M)} \rangle \langle\Pi^{(M)}
 \vert \otimes \sum_{i} \vert \Theta_{\perp i}^{(K)} \rangle \langle
 \Theta_{\perp i}^{(K)}\vert\:]$, where we have indicated with
 $\tilde{\cal{S}}$ the sum of the ${ N \choose K}={N \choose M}$ terms in
 which one or more of the first $M$ indices are exchanged with the remaining
 ones.

In fact, by resorting to the projection operator $P_{A}=\frac{1}{N!} A$
 onto the manifold ${\cal{H}}_{A}^{(N)}$, we have:
\[
 \label{restrizproiettore}
 P_{A} E(1,...,N) P_{A} =
\]
\[
   \frac{1}{(N!)^{2}} A\,[\: \tilde{\cal{S}} \:[\:
   \vert \Pi(1\dots M) \rangle \langle\Pi(1\dots M) \vert \otimes
   \sum_{i} \vert  \Theta_{\perp i}(M+1\dots N) \rangle \langle
   \Theta_{\perp i}(M+1 \dots N) \vert\:]\,\:]A =
\]
\[
   \frac{1}{(N!)^{2}} {N \choose K} \sum_{i} A\,[\: \vert \Pi(1\dots M)
   \rangle \vert \Theta_{\perp i}(M+1 \dots N) \rangle \:]
   [\: \langle \Pi(1\dots M) \vert \langle \Theta_{\perp i}(M+1 \dots N)
   \vert\:]\,A =
\]
\begin{equation}
  =\sum_{i} \vert \omega_{\perp i}^{(N)} \rangle \langle \omega_{\perp i}^{(N)}
 \vert \equiv E_{A \perp}^{\Pi (N)},
\end{equation}

\noindent where the coefficient ${ N \choose K}$ corresponds to the number of
 permutations produced by the operator $\tilde{\cal{S}}$, while the last
 ine follows directly from the definition of the states~(\ref{bassotti}).

In the special case of two identical fermions,
 we notice that the operator $E_{A\perp}^{\Pi(2)}=\sum_{i}
 \vert \omega_{\perp i}^{(2)}\rangle \langle \omega_{\perp i}^{(2)} \vert$
 reduces to the one of equation~(\ref{proiettore}) with $P^{(1)}\otimes
 P^{(2)}=0$, which we have used to define the properties of two fermion
 states.

In fact, if we denote with $P$ the projection operator of ${\cal{H}}^{(1)}$
 which projects onto the one-dimensional manifold spanned by the state
 $\vert \Pi^{(1)} \rangle$, it is easy to see that $E_{A\perp}^{\Pi(2)}$
 turns out to be:
\begin{equation}
 E_{A\perp}^{\Pi(2)} = \sum_{i}\vert \omega_{\perp i}^{(2)}\rangle
  \langle\omega_{\perp i}^{(2)} \vert =
\end{equation}
\[
  \frac{1}{2} P^{(1)} \otimes
  (I^{(2)}-P^{(2)})+\frac{1}{2}(I^{(1)}-P^{(1)})\otimes P^{(2)}
  - \frac{1}{2}\sum_{i} [\, \vert \Theta_{\perp i}(1) \rangle \vert \Pi(2)
 \rangle - \vert \Pi(1) \rangle \vert \Theta_{\perp i}(2) \rangle\, ].
\]
Since it is easy to prove that the last term, when acting on
 ${\cal H}_{A}^{(2)}$, coincides with the sum of the first two terms, we have
 proved that:
\begin{equation}
 \label{riduzioneproiettore}
 E_{A\perp}^{\Pi(2)} = (I^{(1)}-P^{(1)})\otimes P^{(2)} + P^{(1)} \otimes
 (I^{(2)}-P^{(2)}).
\end{equation}
Equation~(\ref{riduzioneproiettore}), shows that the general formalism we
 have developed in order to deal with the properties of $N$ identical fermions
 reduces to the one we have already used in
 Section~\ref{entanglementoftwoidenticalparticles} when dealing with two
 identical quantum constituents of a composite system.


\subsubsection{A relevant theorem}

It is now extremely easy to prove the following theorem which identifies the
 mathematical properties to be satisfied by a state vector in order that it
 describes subsets of identical constituents possessing a complete set of
 properties.
\begin{relevant}
\label{relevant}
 A necessary and sufficient condition in order that a state $\vert
 \psi^{(N)}  \rangle$ of the Hilbert space of $N$ identical fermions allows
 the identification of two subsets of cardinality $M$ and $K$ ($N=M+K$) of
 particles which possess a complete set of properties is that
 $\vert \psi^{(N)}\rangle$ be obtained by antisymmetrizing and
 normalizing the direct product of two states $\vert \Pi^{(M)}(1,...,M)
 \rangle$ and $\vert \Phi^{(K)}(M+1,...,N)\rangle$ of ${\cal{H}}_{A}^{(M)}$
 and ${\cal{H}}_{A}^{(K)}$ respectively, where
 $\vert \Phi^{(K)}(M+1,...,N)\rangle$ belongs to $V^{\Pi K}_{\perp}$.
\end{relevant}
{\it Proof}: Suppose that:
\begin{equation}
 \label{relevant1}
 \vert \psi^{(N)} \rangle = \sqrt{ { N \choose K}} P_{A}\:[\: \vert \Pi^{(M)}
   (1,...,M) \rangle \otimes \vert \Phi^{(K)}(M+1,...,N) \rangle\: ],
\end{equation}
with $\vert \Phi^{(K)}(M+1,...,N) \rangle \in V_{\perp }^{\Pi K}$.
Then one can write:
  \begin{equation}
   \label{relevant2}
   \vert \Phi^{(K)}(M+1,...,N) \rangle = \sum_{s}a_{s} \vert
   \Theta_{\perp s}^{(K)}(M+1,...,N) \rangle.
  \end{equation}
Replacing~(\ref{relevant2}) in (\ref{relevant1}) one has:
  \begin{equation}
   \vert \psi^{(N)} \rangle = \sqrt{ { N \choose K}} \sum_{s}a_{s} P_{A}
   \left[\, \vert \Pi ^{(M)}(1,...,M) \rangle \otimes \vert
   \Theta_{\perp s}^{(K)}(M+1,...,N)\rangle \,\right]
    = \sum_{s}a_{s} \vert \omega_{\perp s}^{(N)} \rangle,
  \label{W5}
  \end{equation}
There follows:
   \begin{equation}
   E_{A\perp }^{\Pi (N)} \vert  \psi^{(N)} \rangle = \sum_{rs} a_{s} \vert
   \omega_{\perp r}^{(N)} \rangle \langle \omega _{\perp r}^{(N)} \vert
   \omega_{\perp s}^{(N)} \rangle = \sum_{s}a_{s} \vert  \omega_{\perp s}^{(N)}
   \rangle = \vert  \psi^{(N)} \rangle,
  \label{W6}
  \end{equation}
which implies
  \begin{equation}
  \langle \psi^{(N)} \vert E_{A\perp }^{\Pi (N)} \vert \psi ^{(N)} \rangle =1.
  \label{W7}
  \end{equation}
Conversely, since $E_{A\perp }^{\Pi (N)}$ is a projection operator, the
 condition $\langle \psi^{(N)} \vert E_{A\perp }^{\Pi (N)} \vert \psi^{(N)}
 \rangle =1$ implies:
  \begin{equation}
   E_{A\perp }^{\Pi (N)} \vert \psi ^{(N)} \rangle = \vert \psi^{(N)}\rangle,
   \label{W8}
  \end{equation}
i.e., putting $b_{s}=\langle \omega _{\perp s}^{(N)} \vert \psi^{(N)}\rangle$:
  \begin{eqnarray}
   \vert  \psi^{(N)} \rangle  & = & \sum_{s} \vert \omega _{\perp s}^{(N)}
   \rangle \langle \omega _{\perp s}^{(N)} \vert \psi^{(N)} \rangle =
   \sqrt{ { N \choose K}} P_{A} \sum_{s}b_{s} \left[ \vert \Pi ^{(M)}
  \rangle
   \otimes \vert \Theta _{\perp s}^{(K)} \rangle \right]
    \label{W9} \\
  &=&\sqrt{ { N \choose K}} P_{A} \left[ \vert \Pi ^{(M)} \rangle \otimes
   \sum_{s}b_{s} \vert \Theta_{\perp s}^{(K)} \rangle \right] =
   \sqrt{ { N \choose K}} P_{A} \left[ \vert \Pi^{(M)} \rangle \otimes \vert
   \Phi^{(K)} \rangle \right],  \nonumber
  \end{eqnarray}
where $\vert \Phi ^{(K)} \rangle = \sum_{s}b_{s}\vert \Theta_{\perp s}^{(K)}
 \rangle \in V_{\perp }^{\Pi K}. \:\:\:\:\:\Box $


  \subsubsection{Non-Entangled subsets of $N$ identical fermions}
\label{nonentangledsubsets}

As we have just shown, the requirement that a ``subset" of $M$ fermions of a
 system of $N$ identical fermions in the pure state $\vert\psi^{(N)}\rangle$
 possesses a complete set of properties requires that the state vector of the
 system has the form~(\ref{relevant1}) and this happens iff
 equation~(\ref{condizionedinonsep}) is satisfied.
We now note that, in the case under consideration, the two ``factors"
 $\vert\Pi^{(M)}\rangle$ and $\vert\Phi^{(K)}\rangle$, play a perfectly
 symmetrical role.
It follows that it is possible to define a projection operator
 $E^{\Phi N}_{A\perp}$, which plays precisely the same role as
 $E^{\Pi N}_{A\perp}$ and, correspondingly it allows us to claim that ``a
subset
 of $K$ fermions has the complete set of properties associated to the pure
 state $\vert\Phi^{(K)}(M+1,...,N)\rangle$''.

Since, according to Definition~\ref{definizionedient}, we have related the
 possibility of
 claiming that the system of $N$ fermions contains two non-entangled
 ``subsets" of cardinality $M$ and $K$, respectively, to the possibility of
 attributing a complete set of properties to the two subsets under
 examination, the following theorem has been proved:
  \begin{teodianti}
  \label{teodianti}
   Given a system $S^{(N)}$ of $N$ identical fermions in a pure state $\vert
   \psi^{(N)}\rangle$ of ${\cal{H}}_{A}^{(N)}$ it contains two non-entangled
   subsets of cardinality $M$ and $K$ iff  $\vert\psi^{(N)}\rangle$ can be
   written as:
  \begin{equation}
   \vert \psi^{(N)} \rangle = \sqrt{ { N \choose K}} P_{A}\:[\: \vert
   \Pi^{(M)} (1,...,M) \rangle \otimes \vert \Phi^{(K)}(M+1,...,N) \rangle\: ]
  \end{equation}
  where the states $\vert\Pi^{(M)}(1,...,M) \rangle$ and $\vert \Phi^{(K)}
   (M+1,...,N) \rangle$ are ``one-particle orthogonal" among themselves.

  \end{teodianti}


  \subsubsection{The physical motivations for the conditions that a subset be
  non-entangled}
\label{thephysical}

In this subsection we present precise physical arguments which should clarify
 the formal constraints we have imposed to claim that a subset
  of a set of identical particles is non-entangled
 with the remaining ones, in spite of the fact that the whole state satisfies
 the requirement of being totally antisymmetric. To this purpose, let us
 take into account a complete orthonormal set of single particle states and
 let us consider, in strict analogy with what we have done in
 Subsection~\ref{someuseful}, two disjoint
 subsets\footnote{Note that we do not assume that the union of the states
 with indices belonging to such subsets coincides with the complete
 orthonormal single particle basis. As a clarifying example one can
consider two denumerable set of states
spanning the manifold of the square integrable functions with disjoint
supports $A$ and $B$, respectively,
such that $A\cup B$ is strictly contained in ${\bf R}^{3}$.}$\Delta$ and
$\Delta^{*}$. We consider, as usual, the two closed submanifolds
$V^{(M)}_{\Delta}$ and
 $V^{(K)}_{\Delta^{*}}$ of ${\cal H}^{(M)}_{A}$ and ${\cal H}^{(K)}_{A}$,
 respectively, with obvious meaning of the symbols: they are the manifolds
 such that the Fourier development of their states involve only single
 particle states whose indices belong to $\Delta$, or to $\Delta^{*}$,
 respectively.
In place of a precise state $\vert\Pi^{(M)}\rangle$ (as we did up to now),
 here we consider arbitrary states of $V^{(M)}_{\Delta}$ and
 $V^{(K)}_{\Delta^{*}}$, and two orthonormal bases
 $\{\vert\Upsilon^{(M)}_{j}\rangle\}$ and $\{\vert\Xi^{(K)}_{l}\rangle\}$
 spanning such manifolds.

By repeating the calculations of the previous sections, one immediately
 proves that, if consideration is given to an arbitrary pair of states
 $\vert\chi^{(M)}\rangle$ and $\vert\tau^{(M)}\rangle$ of $V^{(M)}_{\Delta}$
 or to another pair of states $\vert\mu^{(K)}\rangle$ and
 $\vert\nu^{(K)}\rangle$ of $V^{(K)}_{\Delta^{*}}$, the following relations
 hold:
  \begin{equation}
  \label{mio}
   \sum_{l}\left|\left[\sqrt{ { N \choose K}}\langle \chi^{(M)}\vert\langle
  \Xi^{(K)}_{l}
   \vert P_{A}\right]\cdotp \left[\sqrt{ { N \choose K}} P_{A} \vert
  \tau^{(M)}\rangle
  \vert \nu^{(K)}\rangle \right] \right|^{2}=\vert \langle \chi^{(M)} \vert
  \tau^{(M)}
  \rangle \vert^{2}
   \end{equation}

  \noindent and
  \begin{equation}
  \label{tuo}
   \sum_{j}\left \vert \left[ \sqrt{ {N \choose K}} \langle
  \Upsilon^{(M)}_{j}\vert\langle
   \mu^{(K)}\vert P_{A}\right]\cdotp \left[\sqrt{{ N \choose K}} P_{A} \vert
   \tau^{(M)}\rangle\vert\nu^{(K)}\rangle\right]\right|^{2}=\vert \langle
  \mu^{(K)} \vert
  \nu^{(K)} \rangle \vert^{2}
  \end{equation}
\noindent These equations show that, provided two one-particle orthogonal
 manifolds $V^{(M)}_{\Delta}$ and $V^{(K)}_{\Delta^{*}}$ can be identified,
 and provided the interactions between the particles determining the
 subsequent evolution do not alter the specific features of the state vector,
 then one can {\it do the physics within each manifold by disregarding the
 other one}, even though the appropriate antisymmetrization requests for the
  whole set of fermions are respected\footnote{With reference to
 eq.(\ref{mio}), we stress that the one-particle orthogonality of the states
 $\vert\tau^{(M)}\rangle$ and $\vert\nu^{(K)}\rangle$, and
 $\vert\chi^{(M)}\rangle$ and $\vert\Xi^{(K)}\rangle$ as well as the
 corresponding ones for the states appearing in eq.~(\ref{tuo}), is absolutely
 fundamental - as the reader can check - in order that the (physically
 important) equality sign between the expressions at the left and right hand
 sides of the equations hold.}.
These considerations should have made clear the appropriateness of adopting
 our criteria (as given in Theorem~\ref{teodianti}) for the attribution of
 complete sets
 of properties associated to $\vert\Pi ^{(M)}\rangle$ and
 $\vert\Phi ^{(K)}\rangle$ and for the identification of non-entangled subsets
 of a system of $N$ identical fermions.

A concluding remark. Obviously, (see also the discussion by A. Messiah in
 his book \cite{re10}), the most significant instance of the above
 situation is the one in which the set of states whose indices belong to
 $\Delta $ are single particle states whose wave functions have compact
 support within a region $A$, while those whose indices belong to
 $\Delta^{*}$ have support confined to a region B disjoint from A.
In such a case we can claim that ``there are M fermions in region $A$ with
 precise properties",  in spite of the fact that the presence of the remaining
 identical fermions (confined within a different region) has been rigorously
 taken into account.

The particular relevance of the case in which the ``subsets" of our analysis
 are related to single particle states with disjoint supports, emerges clearly
 from the previous considerations.
The special situation according to which we  can ``look at a part of the
 universe" disregarding the rest of it (which however has precise implications
 for the total state vector) occurs easily (at least to an extremely high
 degree of accuracy - see the next section) when the manifolds
 $V^{(M)}_{\Delta}$ and $V^{(K)}_{\Delta^{*}}$ correspond precisely to single
 particle states with compact disjoint supports.

With reference to this fact, we point out that in this paper we have dealt with
 the general case and we have avoided to put, from the very beginning,
 limitations to the specific structure of our one-particle orthogonal
 manifolds, to stick as far as possible to a rigorous and general mathematical
 treatment.
However, we can now call attention to the fact that if the identification
 of the two considered manifolds would be related, e.g., to  the indices
 $\Delta$ and $\Delta^{*}$ corresponding to single particle states having
 different parity under space reflections, then, even though for a
 $N$-particle state like $\sqrt{ { N \choose K}}P_{A}\vert\Pi^{(M)}\rangle
 \vert\Phi^{(K)}\rangle$ (which is built in terms of a pair of one-particle
 orthogonal states $\vert\Pi^{(M)}\rangle$ and $\vert\Phi^{(K)}\rangle$) we
 can attribute complete sets of properties
 to the appropriate subsets of cardinality $M$ and $K$, almost every
 interaction between the particles will destroy the ``factorizability" as
 well as the one-particle orthogonality of the factors of the complete state.

Once more, as we have repeatedly outlined in this paper, particle positions
 play an absolutely prominent role in making physically interesting and
 meaningful our analysis.

The above considerations should have clarified our line of thought in
  approaching the problem of identifying non-entangled states in the case of
  systems with identical constituents.

\subsubsection{An important specification about non-entangled states}

Our definition of non-entangled subsets of systems of identical fermions might
be considered not fully appropriate by some readers, due to the fact that it
does not imply the {\em local} factorizability of position
probabilities~\footnote{We are grateful to the refereee for having called our
attention on the fact that this point deserves a detailed analysis.}.
To discuss this question it seems useful to limit, for the moment, our
considerations to the system of two identical fermions and to derive two simple
theorems.
We consider a single particle complete othonormal set $\left\{ \vert
 \varphi_{i} \rangle \right\} $ and two disjoint subsets $\Delta$ and
 $\Delta^{\star}$ of the ensemble ${\bf N}$ of the natural numbers.
 We do not require that the union of $\Delta$ and $\Delta^{\star}$ exhausts the
 whole ${\bf N}$.
We denote as $V_{\Delta}$ and $V_{\Delta^{\star}}$ the two orthogonal
 submanifolds of the single particle Hilbert space ${\cal H}$ spanned by the
 vectors $\left\{ \vert \varphi_{i} \rangle \right\}$, with $i\in \Delta$, and
 $\left\{ \vert \varphi_{j} \rangle \right\}$, with $j\in \Delta^{\star}$,
 respectively.
We consider two orthogonal projection operators $P_{\delta}$ and
$Q_{\delta^{\star}}$ onto the submanifolds $V_{\delta} \subseteq V_{\Delta}$
 and $V_{\delta^{\star}} \subseteq V_{\Delta^{\star}}$ of ${\cal H}$, and, in
 terms of them we define the following projection operator:
 \begin{equation}
 \label{animportant1}
 E(1,2) = P_{\delta}(1)\otimes Q_{\delta^{\star}}(2) + Q_{\delta^{\star}}(1)
 \otimes P_{\delta}(2)
 \end{equation}
 of the Hilbert space of the system of two identical fermions.
 Finally we take into account a non-entangled (according to our definition)
 state vector of our two identical fermions:
 \begin{equation}
 \label{animportant2}
 \vert \psi(1,2) \rangle = \frac{1}{\sqrt{2}}\, [ \vert \Pi(1) \rangle \vert
 \Phi(2) \rangle - \vert \Phi(1) \rangle \vert \Pi(2) ]
 \end{equation}
 where
 $\vert \Pi \rangle$ belongs to $ V_{\Delta}$ and $\vert \Phi \rangle$
 belongs to $ V_{\Delta^{\star}}$.
With these premises we can now formulate the following theorem:
\begin{localfact1}
Given the state (\ref{animportant2}), the joint probability distribution of
finding one particle in a state belonging to $V_{\delta}$ and one particle in a
state belonging to $V_{\delta^{\star}}$ factorizes into the product of the
probabilities of the single events.
\end{localfact1}
\noindent {\em Proof.} : the proof is straightforward. According to our
analysis
of Section 7.1 it amounts simply to verify that:
 \begin{equation}
 \label{animportant3}
 \langle \psi (1,2) \vert E(1,2) \vert \psi(1,2) \rangle =
 \langle \Pi \vert P_{\delta} \vert \Pi \rangle \cdot
 \langle \Phi \vert Q_{\delta^{\star}} \vert \Phi \rangle
 \end{equation}
where
 \begin{equation}
 \label{animportant4}
 \langle \Pi \vert P_{\delta} \vert \Pi \rangle =  \langle \psi (1,2) \vert
 P_{\delta}(1)\otimes ( I(2) -P_{\delta}(2) ) +( I(1) -P_{\delta}(1) )
  \otimes P_{\delta}(2)\vert \psi(1,2) \rangle
 \end{equation}
 \begin{equation}
 \label{animportant5}
 \langle \Phi \vert P_{\delta} \vert \Phi \rangle =  \langle \psi (1,2) \vert
 Q_{\delta^{\star}}(1)\otimes ( I(2) -Q_{\delta^{\star}}(2) ) +
 ( I(1) - Q_{\delta^{\star}}(1) ) \otimes Q_{\delta^{\star}}(2)\vert \psi(1,2)
 \rangle
 \end{equation}
and this completes the proof. $\Box$

We need also another elementary theorem.
Suppose we consider observables $A$ and $B$ which have non vanishing matrix
elements between states belonging to $ V_{\Delta}$ and $ V_{\Delta^{\star}}$.
Then we have the following theorem:
\begin{localfact2}
For pairs of observables connecting the two manifolds $V_{\Delta}$ and
$V_{\Delta^{\star}}$,
 in general, the joint probability of getting a pair of eigenvalues
when the composite system is in state~(\ref{animportant2}) does not
factorize into
the product of the  probabilities of the single events.
\end{localfact2}
\noindent {\em Proof.} : this theorem is easily proved by considering the two
elementary operators $A= \vert \varphi_{r} \rangle \langle
 \varphi_{s} \vert + \vert \varphi_{s} \rangle \langle \varphi_{r} \vert$
 and $B = i\,[\,\vert \varphi_{r} \rangle \langle
 \varphi_{s} \vert - \vert \varphi_{s} \rangle \langle \varphi_{r} \vert\,]$
with $r\in \Delta$ and $s\in \Delta^{\star}$.
The eigenvectors of these operators are
 \begin{equation}
 \label{animportant6}
 A= \pm 1 \:\:\Rightarrow \:\:\frac{1}{\sqrt{2}} \,[\,\vert \varphi_{r}\rangle
  \pm \vert\varphi_{s} \rangle \, ] \:\:\:\:\:\:\:\:\:\:\:\:
 B= \:\:\pm 1 \:\:\Rightarrow \frac{1}{\sqrt{2}} \,[\,\vert \varphi_{r}\rangle
  \pm i \vert\varphi_{s} \rangle \, ]
 \end{equation}
Then it is trivial to see that if $\vert \Pi \rangle$ contains the state
 $\vert \varphi_{r} \rangle$ and $\vert \Phi \rangle$ contains the state
 $\vert \varphi_{s} \rangle$:
 \begin{equation}
\label{animportant7}
 Pr(A=1 \& B=-1) \neq Pr(A=1)\cdot Pr(B=-1)
 \end{equation}
with obvious meaning of the symbols. $\Box$

With these premises we can now be more precise about our line of thought.
As we have discussed in great detail before, our definition of a non-entangled
 state of two identical particles uniquely identifies two orthogonal
 submanifolds $V_{\Delta}$ and $V_{\Delta^{\star}}$.
Given a pair of observables $A$ and $B$ commuting with $P_{\Delta}$ and with
 $Q_{\Delta^{\star}}$ respectively, let us consider their restrictions
 $\tilde{A}$ and $\tilde{B}$ to $V_{\Delta}$ and $V_{\Delta^{\star}}$.
The first of the theorems we have proved implies then the factorizability
of the
 joint probabilities of getting a pair of eigenvalues of $\tilde{A}$ and
 $\tilde{B}$.
The second one shows that for operators connecting states of $V_{\Delta}$ and
 $V_{\Delta^{\star}}$ this does not hold, in general.

Now we can tackle the problem of the local factorizability of probabilities.
To fully understand it, it is sufficient to identify $V_{\Delta}$ and
 $V_{\Delta^{\star}}$ with the Hilbert spaces spanned by vectors which, in the
 configuration representation, have compact support in the two disjoint
subsets
 $\Delta$ and $\Delta^{\star}$ of the real axis.
In this case, if  we identify $V_{\delta}$ and $V_{\delta^{\star}}$
 with the Hilbert spaces of the square integrable functions of compact
support in
the indicated
 intervals, our theorem tells us that the joint probability of finding one
 particle in the space interval $\delta$ and one in $\delta^{\star}$ factorizes
provided the wave function is non-entangled, with factors having the
appropriate
supports, one $\Delta$ and one in $\Delta^{\star}$.

This makes precise that the non-entangled state~(\ref{animportant1})
exhibits, in
 the considered case, local factorizability of position probabilities.
It has however to be remarked that, according to our second theorem, if
 consideration is given to two observables $A$ and $B$ such that their matrix
 elements $\langle x \vert A \vert x^{\star} \rangle$ and $\langle x \vert B
 \vert x^{\star} \rangle$ do not vanish for $x\in \Delta$ and $ x^{\star}\in
 \Delta^{\star}$, then, in general, in  the considered state:
 \begin{equation}
\label{animportant8}
 Pr(A=a_{i} \& B=b_{j}) \neq Pr(A=a_{i})\cdot Pr(B=b_{j} ),
 \end{equation}
i.e. the joint probability of getting the outcomes $a_{i}$ and $b_{j}$ for
such observables
does not factorize. The situation should now be clear. The  non-entangled
character
  of the state $\vert \psi (1,2)\rangle$ identifies precise
 manifolds and, correspondingly, observables commuting with the projection
 operators onto them.
The restrictions of such observables to the considered manifolds identify the
 relevant  properties related to the state.
It is just the joint probabilities referring to such properties which
factorize.
We perfectly agree that the physically really interesting case is the one in
 which the manifolds $V_{\Delta} $ and $V_{\Delta^{\star}} $, as well as the
 submanifolds $V_{\delta} $ and $V_{\delta^{\star}} $, are associated to
disjoint sets of position variables, in which case local factorizability
holds in
the situation described above. Actually as we have already done and as we
are going
to discuss in the next
 section we are inclined to attach a particularly relevant role to this case
 since we are convinced that position variables must be endowed with a
 priviliged status.
 However, from the point of view we have taken in this paper, which makes
systematic
 reference to the problem of the legitimacy of attributing properties to the
 constituents, it seems perfectly appropriate and correct to adopt the
 definition we have chosen to identify non-entangled states of identical
 particles. In the general case, the probabilities which factorize are
related to
observables different from positions, and as such they are not local.

 A final remark. All previous considerations can be easily generalized to
 non-entangled states of $N$ identical fermions. In such a case, if one
deals with
  the typical non-entangled state:
 \begin{equation}
 \label{animportant9}
 \vert \psi(1,\dots,N) \rangle = \sqrt{ {N \choose K}} P_{A} \left[ \vert
 \Pi(1,\dots,M) \, \rangle \vert \Phi(M+1,\ldots,N) \rangle \right]
 \end{equation}
 and  if one considers two arbitrary projection operators $P(1,\ldots,M)$
 {\em and}
 $Q(M+1,\ldots,N)$, which project onto the two {\em one-particle orthogonal}
 submanifolds $V_{\Delta}^{(M)}$ and $V_{\Delta^{\star}}^{(K)}$ containing
 $\vert \Pi \rangle$ and $\vert \Phi \rangle$, respectively, one can easily
prove that
 \begin{equation}
 \label{animportant10}
 \langle \psi \vert P_{A} \tilde{S} [ P(1,\dots,M)\otimes Q(M+1,\dots,N)
]P_{A}
 \vert \psi \rangle =
 \end{equation}
 \[
 \langle \psi \vert P_{A} \tilde{S} [ P(1,\dots,M)\otimes I(M+1,\dots,N)
]P_{A}
 \vert \psi \rangle \cdot
 \langle \psi \vert P_{A} \tilde{S} [ I(1,\dots,M)\otimes Q(M+1,\dots,N)
]P_{A}
 \vert \psi \rangle
 \]
 which is the obvious generalization of our previous result
(\ref{animportant3}) for two
identical fermions.
 Moreover, a similar argument can be developed  for the boson case.

  \subsection{Some useful remarks concerning almost perfect non-entanglement}

Having made precise the idea of a ``group of particles'' of a system of
 identical particles being non-entangled with the remaining ones (or, more
 precisely, the idea of complete set of properties objectively associated
 to the ``subquasets'' of a ``quaset'' of $N$ identical particles) we can
 reconsider the situation we have envisaged at the beginning of this Section,
 i.e., the case in which a Helium atom is here (at the origin $O$ of our
 reference frame) and a Lithium atom is there, let us say at a distance $d$
 from $O$.
Our worries concerned the legitimacy of claiming ``there is a Helium atom at
 the origin'' or ``there is a Lithium atom at a distance $d$ from the origin"
 when one takes into account, e.g., the identity of the electrons of the two
 systems which requires  the state vector to be totally skew-symmetric under
 their exchange.
To discuss the conceptually relevant aspects of this problem we will confine,
 for simplicity, our considerations only to the electrons which are involved,
 disregarding the nuclei of the atoms - and the necessary antisymmetrization
 concerning the protons and the neutrons.

The puzzling question we have to face is the following: since the electrons
 are indistinguishable,  in which sense can  we state that two of them are
 around the origin (to make the Helium atom which is there) and three at a
 distance $d$? And then, in which sense  can one claim that ``there is a
 Helium atom at the origin''?
The answer, as we have stressed in this section, can be only given by
 paying the due attention to the total state vector of the complete system
 ``Helium+Lithium''.
Such a state vector has the form:
  \begin{equation}
   \label{eliolitio}
   \vert \psi^{(5)} \rangle \propto {\cal{G}} \left[ \vert
  Helium^{(2)}here\rangle
   \otimes \vert Lithium^{(3)}there \rangle \right] .
  \end{equation}
And now we are in trouble. The factors $\vert Helium^{(2)}here \rangle$
 and $\vert Lithium^{(3)}there\rangle $ of the state inside the square
 brackets at the r.h.s. do not satisfy exactly our fundamental request of
 being one-particle orthogonal.

However, we can explicitly evaluate integrals like the one of
 equation~(\ref{int}), which, when they vanish, make legitimate precise claims
 concerning the objective properties of the considered subquasets.
We notice that the modulus of the relevant integral is of the order of the
 overlap integrals of the electronic wave functions.
Since they decrease exponentially outside a region of $10^{-8}cm$ from the
 corresponding nuclei, we immediately see that, for a distance between the
 two nuclei of the order of $1cm,$ the relevant integral
 turns out to have a value of the order of $10^{-10^{16}}$.
It is easy to convince oneself that this figure represents also the
 probability that, if one has an apparatus devised to check whether there is
 an Helium atom around the origin, it will not detect such an atom.

So, our claim ``there is a Helium atom around the origin'' is, strictly
 speaking, not perfectly correct but has only an approximate validity.
However, as appropriately pointed out by the authors of ref.~\cite{refide3},
 {\em ``all assertions of physics have that kind of approximation.
When we state that the heat passes spontaneously only from a hotter to a
 cooler body in contact, we really mean that in a real case it is extremely
 probable that it should do so''}.

Concluding, in the considered case the formal conditions which are
 necessary for attributing consistently objective properties to ``a group of
 particles'' are satisfied to an extremely high degree of accuracy so that,
 precisely in the same way as we consider valid all the (unavoidably
 approximate) assertions about physical systems, we can confidently say that
 ``there is a Helium atom here {\em and} a Lithium atom there''.

A concluding remark is appropriate. The analysis we have performed has
 played an important role in making clear what are the formal features which
 make legitimate, in a rigorous or in an extremely well approximate way, to
 consider, in the case of a system of identical constituents, two subgroups
 of them as disentangled from each other and as possessing objectively
 precise properties.

To fully appreciate the real relevance of our considerations we invite the
 reader to consider the case in which, in place of the state~(\ref{eliolitio})
 one is dealing with a state like:
  \begin{equation}
   \label{eliolitioent}
   \vert \psi^{(5)} \rangle \propto {\cal{G}} \:\left[
   \vert Helium^{(2)}here \rangle
   \otimes \vert Lithium^{(3)}there \rangle +\vert Lithium^{(3)}here \rangle
   \otimes \vert Helium^{(2)}there \rangle \right]
  \end{equation}
which is perfectly possible and relatively easy to prepare, and which would
 not make legitimate, in any way whatsoever, to make claims about what is here
 being a Helium rather than a Lithium atom. We stress that the
 embarrassment with a state like the one we have just considered does not
 arise from the fact that the strict conditions which would make our claims
 absolutely rigorous are not exactly satisfied, but from the fact that the
 state of the whole system is genuinely entangled due to the fact that it is
 obtained by antisymmetrizing a nonfactorized state.
Even if the  states of the Helium and Lithium atoms  would satisfy
 our strict requirements of one-particle orthogonality, no objective property
 referring to the region around the origin (and the one at a distance $d$
 from it) could be identified and claimed to be possessed.

Finally, we mention that, in the case of the state~(\ref{eliolitio}), the
 approximation of disregarding the overlap of the electronic (or nucleonic)
 wave functions associated to the two atoms, is practically equivalent to
ignoring  the request of
 totally antisymmetrizing the state vector under the exchange of the electrons
 of the Helium and Lithium atoms\footnote{What we have in mind should
 become clear if one makes reference, e.g., to the evaluation of exchange
 effects using perturbation theory.
In such a case, for example, the matrix elements of the Hamiltonian between
 states which are products of states with disjoint supports vanish, so that
 one can avoid performing the antisymmetrization procedure}.
Once more the legitimacy of doing so can be explicitly evaluated by taking
 into account to which extent one can disregard ({\it de facto} they
 are absolutely negligible) the exchange effects.
However, giving up the antisymmetrization request, amounts to considering the
 electrons of the Helium as distinguishable from those of the Lithium atom.
If one makes this step, then one sees that the conclusions we have drawn
 concerning systems of identical constituents reduce to those we have derived
 for the case in which they are distinguishable.


\subsection{Completely non-entangled indistinguishable fermions}

As in the case of $N$ distinguishable particles, once one has identified two
 ``groups" of particles which are non-entangled with each other, one can
 raise the question of whether also the``members" of each subset can be
 subdivided into non-entangled subsubsets.
We will limit ourselves to give the definition and the
 associated theorem (which is easily proved) which characterizes the states
 corresponding to completely non-entangled identical fermions.
\begin{defofentangNidentical}
 \label{defofentangNidentical}
 The pure state $\vert \psi(1,\dots, N)\rangle \in {\cal H}^{(N)}_{A}$
 describing a system of $N$ indistinguishable
 fermion particles, is completely non-entangled if there exist
  $N$ mutually orthogonal one-dimensional projection operators $P_{i}$,
  $i=1\dots N$, such that:
  \begin{equation}
   \label{nparticelleidentiche}
   Tr^{(1+\dots +N)}[\, E_{i}\, \vert \psi(1,\dots, N)\rangle\langle
   \psi(1,\dots, N) \vert\,]=1 \:\:\:\:\:\:\:\:\:\forall i=1\dots N
  \end{equation}
\begin{flushleft}
 where $E_{i}=I^{(1)}\otimes\dots \otimes I^{(N)} - (I^{(1)}-P_{i}^{(1)})
 \otimes \dots \otimes (I^{(N)}-P^{(N)}_{i})$.
\end{flushleft}
\end{defofentangNidentical}
The quantity $Tr^{(1+\ldots+N)}[\,E_{i}\vert\psi^{(N)}\rangle \langle
 \psi^{(N)} \vert \,]$, where the projection operators $E_{i}$ are totally
 symmetric under the exchange of two arbitrary particles, gives the
 probability of finding one fermion in a well definite one-dimensional
 manifold, the one onto which $P_{i}$ projects.
Therefore, in a completely non-entangled physical system composed of identical
 constituents, equations~(\ref{nparticelleidentiche}) guarantee that all the
 $N$ particles possess a complete set of objective properties.

The content of the following theorem, which is a generalization of the one we
 have already proved in the simpler case of two particles, shows the
 relevance of Definition~\ref{defofentangNidentical}:
\begin{factorizabilityNidentical}
 \label{factorizabilityNidentical}
 A system ${\cal{S}}= {\cal{S}}_{1}+...+{\cal{S}}_{N}$ of $N$ identical
 half-integer spin particles described by the pure state
 $\vert \psi(1,\dots,N)\rangle$ is completely non-entangled iff it
 can be obtained by antisymmetrizing a completely factorized
 state. Note that the factors can be assumed to be orthogonal among themselves
 without any loss of generality.
\end{factorizabilityNidentical}


  \section{Identical bosons}

We have now to face the problems of property attribution and entanglement in
 the case of a system of $N$ indistinguishable bosons.
The main difference with respect to the case of identical fermion systems
 derives from the fact that, when one splits the set of the $N$ particles
 into two or more subsets which have a complete set of properties, it may
 happen that two such subsets containing the same number $L$ of particles are
 associated to the same state $\vert\Gamma^{(L)}\rangle$.
Alternatively, as we will see, the various subsets must be associated to
 states (which may very well contain a different number of particles) which
 are one-particle orthogonal among themselves in the precise sense defined in
 Subsection~\ref{defininganappropriate}.
These are the only two cases which can give rise to disentangled subsets of
 the whole set of particles.

For simplicity, we will confine our considerations to the possible occurrence
 of only two disentangled subsets (we shall suggest subsequently how one has
 to proceed in the general case) and we will distinguish the cases of
 one-particle orthogonal and identical factors of the factorized state we have
 to symmetrize.

Taking into account the analysis of Subsection~\ref{thebosoncase} concerning
 the two-boson
 case, it is easy to see that the two above considered instances (i.e., the
 appearance of identical or one-particle orthogonal state vectors) cannot occur
 together if one requires the system to contain two subgroups possessing
 simultaneously a complete set of properties, i.e. which are non-entangled
 with each other.


\subsection{Boson subsets corresponding to different properties}

We recall that, in accordance with our Definition~\ref{definizionedient},
 in order to be allowed
 to speak of two non-entangled subsets of particles it must be possible to
 attach a complete set of properties to both subsets.
As the reader can easily understand, the case we are  interested  in here
 - i.e. the one in which we exclude that the subsets have precisely the same
 properties - can be dealt with by repeating step by step the analysis we have
 performed for the fermion case.
Accordingly, we will limit ourselves to recall the appropriate
 definitions and theorems, with the due changes, without going through
 detailed arguments and proofs.

In analogy with the fermion case, we will denote as ${\cal H}^{(R)}_{S}$ the
 Hilbert space of the state vectors which are totally symmetric for the
 exchange of all variables of $R$ identical bosons.
Moreover we define the projection operator $P_{S}$ on the totally symmetric
 submanifold  ${\cal H}^{(N)}_{S}$ of ${\cal H}^{(N)}$ as
 $P_{S}=\frac{1}{N!}S$, where $S$ is the linear operator which acts in the
 following way on a $N$-single particle basis:
\begin{equation}
 S \left\{\vert \varphi_{i_{1}}(1) \rangle \dots \vert \varphi_{i_{N}}(N)
 \rangle \right\} =
   \sum_{P}P \left\{ \vert \varphi_{i_{1}}(1) \rangle \dots \vert
   \varphi_{i_{N}}(N) \rangle\right\}.
\end{equation}
\noindent where the sum is extended to all permutations P of the variables
 $(1,\dots,N)$.

With reference to a state of $N=L+J$ such bosons, we consider a state
 $\vert\Gamma^{(L)}(1,...,L)\rangle$ describing $L$ such particles and, by
 the procedure of Subsection~\ref{defininganappropriate}, we define the single
 particle
 manifolds $V^{\Gamma 1}$ and $V^{\Gamma 1}_{\perp}$ as well as the manifolds
 $V^{\Gamma L}$ and $V^{\Gamma J}_{\perp}$, with obvious meaning of the
 symbols.
The very procedure to identify such manifolds guarantees that
 $\vert\Gamma^{(L)}(1,...,L)\rangle\in V^{\Gamma L}$ and that the closed
 linear manifolds $V^{\Gamma L}$ and $V^{\Gamma J}_{\perp}$ are one-particle
 orthogonal.

It is easy to see that the properly normalized state vector
 $\vert \psi^{(N)}(1,...,N)\rangle$ obtained by symmetrizing the direct
 product of $\vert\Gamma^{(L)}(1,...,L)\rangle$ and an arbitrary vector
 $\vert\Lambda^{(J)}(1,...,J)\rangle\in V^{\Gamma J}_{\perp}$ is:
  \begin{equation}
  \label{bosonsubset1}
   \vert \psi^{(N)}(1,...,N)\rangle = \sqrt{ {N \choose L}}P_{S} \left[ \vert
   \Gamma^{(L)}(1,...,L)\, \rangle \otimes \vert \Lambda^{(J)}(L+1,...,N)
   \rangle \right].
  \end{equation}
Once again, having identified the closed linear manifold
 $V^{\Gamma J}_{\perp}$ of ${\cal H}^{(J)}_{S}$, we can consider a complete
 orthonormal set spanning such a manifold, namely the set $\left\{ \vert
 \Omega^{(J)}_{\perp j}(L+1,\dots, N) \rangle \right\}$, and build the
 orthonormal set $\left\{ \vert \epsilon^{(N)}_{\perp j}(1,\dots, N)
 \rangle \right\}$ of states of ${\cal H}_{S}^{(N)}$:
\begin{equation}
\label{bosonsubset2}
 \vert \epsilon^{(N)}_{\perp j}(1,\dots, N) \rangle =
 \sqrt{ N \choose L} P_{S} \left[ \vert \Gamma^{(L)}(1,\dots L) \rangle
 \otimes \vert \Omega^{(J)}_{\perp j}(L+1,\dots, N) \rangle \right].
\end{equation}
These states, which are the equivalent for the boson case of the ones we
 have introduced in equation~(\ref{bassotti}), are properly normalized and
 mutually orthogonal and therefore the operator
\begin{equation}
 \label{bosonsubset3}
 E^{\Gamma (N)}_{S\perp} = \sum_{j} \vert \epsilon^{(N)}_{\perp j}(1,\dots,N)
 \rangle \langle \epsilon^{(N)}_{\perp j}(1,\dots, N) \vert
\end{equation}
is a projector of ${\cal H}_{S}^{(N)}$.

It is now possible to characterize in a mathematically definite way, as we did
 when dealing with identical fermions, the fact that each of two
 subgroups of bosons possesses a complete set of different properties.
\begin{setcompletobosoni}
 Given a system $S^{(N)}$ of $N$ identical bosons in a pure state $\vert
 \psi^{(N)}\rangle$ of ${\cal{H}}_{S}^{(N)}$ we will claim that two subsets
 of cardinality $L$ and $J$ $(N=L+J)$, respectively, both possess a
 complete set of different properties iff there exists a
 state $\vert \Gamma^{(L)} \rangle$ of ${\cal{H}}_{S}^{(L)}$ such that
\begin{equation}
 \label{identicalboson1}
 \langle \psi^{(N)} \vert E_{S\perp }^{\Gamma (N)}(1,...,N) \vert \psi^{(N)}
 \rangle =1
\end{equation}
where $E_{S\perp }^{\Pi (N)}(1,...,N) = \sum_{r} \vert \epsilon_{\perp r}^{(N)}
 \rangle \langle \epsilon_{\perp r}^{(N)} \vert$ is the projection operator of
 ${\cal{H}}_{S}^{(N)}$ given by eq.~(\ref{bosonsubset3}), which is uniquely
 identified, according to the previous procedure, by the assignment of the
 state $\vert \Gamma^{(L)}\rangle$.
\end{setcompletobosoni}
From this definition one can easily derive the following remarkable
 results, whose proofs can be obtained by the same arguments leading to
 Theorems~\ref{relevant} and~\ref{teodianti}:
\begin{relevantboson}
 A necessary and sufficient condition in order that a state $\vert \psi^{(N)}
 \rangle$ of the Hilbert space of $N$ identical bosons allows the
 identification of two subsets of cardinality $L$ and $J$ $(N=L+J)$ of
 particles which possess a complete set of different properties
 is that $\vert \psi^{(N)}\rangle$ be obtained by symmetrizing and
 normalizing the direct product of two states $\vert \Gamma^{(L)}(1,...,L)
 \rangle$ and $\vert \Delta^{(J)}(L+1,...,N)\rangle$ of ${\cal{H}}_{S}^{(L)}$
 and ${\cal{H}}_{S}^{(J)}$ respectively, where
 $\vert \Delta^{(J)}(L+1,...,N)\rangle$ belongs to $V^{\Gamma L}_{\perp}$.
\end{relevantboson}
\begin{relevantboson2}
 Given a system $S^{(N)}$ of $N$ identical bosons in a pure state $\vert
 \psi^{(N)}\rangle$ of ${\cal{H}}_{S}^{(N)}$ a sufficient condition in
 order that it contains two {\bf non-entangled} subsets of cardinality
 $L$ and $J$ is that  $\vert\psi^{(N)}\rangle$ can be written as:
\begin{equation}
 \vert \psi^{(N)} \rangle = \sqrt{ { N \choose L}} P_{S}\:[\: \vert
 \Gamma^{(L)} (1,...,L) \rangle \otimes \vert \Delta^{(J)}(L+1,...,N)
 \rangle\: ]
\end{equation}
where the states $\vert\Gamma^{(L)}(1,...,L) \rangle$ and $\vert \Delta^{(J)}
 (L+1,...,N) \rangle$ are ``one-particle orthogonal'' among themselves.
\end{relevantboson2}
The previous definition and theorems yield clear and mathematically
precise
 conditions under which it is possible to consider two subquasets of bosons
 as disentangled from each other.
The condition of ``one-particle orthogonality'' is necessary in order to be
 allowed i) to attribute a set of complete and different properties to each
 subgroup of the whole system and ii) to extend the arguments of
 Subsection~\ref{thephysical}, concerning the possibility of {\em doing
 physics within each manifold by disregarding the other one}, also to the case
 of bosons.
Thus, we have identified a first class of non-entangled boson states as the
 ones which are obtained by symmetrizing direct products of two
 ``one-particle orthogonal'' vectors.


\subsection{Boson subsets corresponding to identical properties}

Let us pass now to characterize the second class of $N$-boson states which,
 according to our general Definition~\ref{definizionedient}, can be considered
 as non-entangled.
In this case there is no need at all to resort to any complicated procedure to
 identify ``one-particle orthogonal'' manifolds;  we can limit ourselves to
 claim that if the state $\vert \psi(1,\dots,N)\rangle$ - $N$-even - is
 obtained by symmetrizing and normalizing a product state of two identical
 factors, i.e. if $\vert \psi(1,\dots,N)\rangle \propto S [\, \vert \Gamma
 (1,\dots, N/2)\rangle \vert \Gamma (N/2+1,\dots, N)\rangle\,]$, then it can be
 considered for sure as non-entangled.
It is in fact apparent that we can attribute to both  subgroups of $N/2$
 particles the complete set of properties associated to the state
 $\vert \Gamma\rangle$.

Though mathematically clear, this situation may appear a bit problematic if
 one tries to develop considerations analogous to those of
 Subsection~\ref{thephysical}.
In fact it is straightforward to show that now it is no longer possible to
 perform a physical measurement on a subgroup of $N/2$ particles without
 affecting, to some extent, the remaining ones.
In this peculiar situation, in spite of the possibility of attributing a
 complete set of properties to both the component subgroups, it is
 practically impossible to devise any measurement process on one subgroup
 whose results will not depend on the presence of the other.
We could say that there are correlations of a certain type which are
 intrinsically due to the fact that the identical particles  are described
 by precisely the same state.

However, this fact does not give rise to any serious problem for two reasons.
First of all, the ``unavoidable correlations'' we have just mentioned are
 related more to the fact that the subgroups are ``truly identical - i.e.,
 in precisely the same state" than to the Hilbert space description of the
 system.
In a sense these effects are analogous to those one meets, even within a
 classical picture, when one compares the implications of Maxwell-Boltzmann
 statistics and those of Bose-Einstein statistics.
Moreover, and much more important, it has to be stressed that since, as
 repeatedly remarked, the most interesting features of the entanglement
 and/or non-entanglement of identical constituents emerge in the case in
 which one has two subgroups confined within different spatial regions, the
 case of the product of two identical states has not a specific physical
 relevance.

In order to illustrate better this situation, let us resort to a simple
 physical example and let us consider a couple of spin zero particles
 described by the following state vector:
\begin{equation}
 \label{esempietto1}
 \vert \psi(1,2) \rangle = \vert \varphi_{\Delta}(1)\rangle
 \vert \varphi_{\Delta}(2)\rangle
\end{equation}
where the normalized wave function associated to the ket
 $\vert \varphi_{\Delta}\rangle$ is defined on the (one-dimensional) real
 axis and has the following form:
\begin{equation}
\label{esempietto2}
 \varphi_{\Delta}(x) = \langle x \vert \varphi_{\Delta} \rangle=
 \left\{
 \begin{array}{ll}
 \frac{1}{\sqrt{\Delta}}  & x \in \Delta \\
  0  & x \notin \Delta
 \end{array}
 \right.
\end{equation}
It is our purpose to show that every conceivable position measurement
 we perform on a particle of the system, will unavoidably alter the
 whole wave function and therefore will modify the probabilistic predictions
 of subsequent measurements.
Let us ask, for example, which is the probability of finding one of the two
 particles inside the closed interval $\Delta_{1} \subset \Delta$ once
 precisely
 one particle has been found in a previous measurement to lie in the disjoint
 interval
 $\Delta_{2}\subset \Delta$, and let us compare this result with the
 probability of finding precisely one particle within $\Delta_{1}$ when no
 previous measurement has been performed.
If we suppose to have found precisely one particle within $\Delta_{2}$, the
 collapsed wave function is obtained by applying to the
 state~(\ref{esempietto1}) the usual operator $P_{\Delta_{2}}\otimes
 (I-P_{\Delta_{2}}) + (I-P_{\Delta_{2}})\otimes P_{\Delta_{2}}$, where
 $P_{\Delta_{2}}$ projects onto the interval $\Delta_{2}$, and then
 normalizing it:
\begin{equation}
\label{esempietto3}
 \vert \tilde{\psi}(1,2) \rangle =
 \frac{1}{\sqrt{ \frac{2\Delta_{2}}{\Delta} (1-\frac{\Delta_{2}}{\Delta})}}
 \biggr[
 \vert {\tilde{\varphi}}_{\Delta_{2}} \rangle \vert \varphi_{\Delta}\rangle
 + \vert \varphi_{\Delta}\rangle \vert {\tilde{\varphi}}_{\Delta_{2}} \rangle
-2 \vert {\tilde{\varphi}}_{\Delta_{2}} \rangle
\vert {\tilde{\varphi}}_{\Delta_{2}} \rangle
 \biggl].
\end{equation}
In the previous equation the wave function
 ${\tilde{\varphi}}_{\Delta_{2}}(x)$ associated to the non-normalized
 state vector $\vert {\tilde{\varphi}}_{\Delta_{2}} \rangle$ has the
 following form:
\begin{equation}
\label{esempietto3.5}
 {\tilde{\varphi}}_{\Delta_{2}}(x) = \langle x \vert
 {\tilde{\varphi}}_{\Delta_{2}} \rangle=
 \left\{
 \begin{array}{ll}
 \frac{1}{\sqrt{\Delta}}  & x \in \Delta_{2} \\
  0  & x \notin \Delta_{2}
 \end{array}
 \right.
\end{equation}
In such a case the desired probability of finding the other particle
 within the interval $\Delta_{1}\subset \Delta$ is easily obtained as:
\begin{equation}
\label{esempietto4}
 \langle  \tilde{\psi}(1,2) \vert P_{\Delta_{1}}\otimes (I-P_{\Delta_{1}})
 + (I-P_{\Delta_{1}})\otimes P_{\Delta_{1}} \vert \tilde{\psi}(1,2) \rangle =
 \frac{\Delta_{1}}{(\Delta - \Delta_{2})}
\end{equation}
On the contrary the probability of finding precisely one particle inside the
 interval $\Delta_{1}$ when no previous measurement has been performed, is:
\begin{equation}
\label{esempietto5}
 \langle  \psi(1,2) \vert P_{\Delta_{1}}\otimes (I-P_{\Delta_{1}})
 + (I-P_{\Delta_{1}})\otimes P_{\Delta_{1}} \vert \psi(1,2) \rangle =
 \frac{2\Delta_{1}}{\Delta} \biggl( 1- \frac{\Delta_{1}}{\Delta} \biggr)
\end{equation}
Since the two probabilities are clearly not equal for arbitrary choices
 of the disjoint intervals $\Delta_{1}$ and $\Delta_{2}$, one could be
 tempted to consider this fact as a manifestation of the outcome dependence
 which is typical of all entangled states.
However, this argument is not correct since the strict and
 unavoidable correlations between position measurements, are simply due to
 the fact that the quantum particles are
 truly identical, and there is no need to invoke a special role played by their
 quantum nature.
It is in fact possible to build up a very simple classical model, consisting
 of two particles which cannot be experimentally distinguished, which
 displays exactly the same correlated properties of
 our quantum pair.

In fact, let us consider a one-dimensional interval of length $\Delta$
 and a couple of indistinguishable classical particles; assuming that each
 particle can, in principle, be found with the same probability in any
 finite subinterval of $\Delta$ of a given amplitude, we evaluate the
 probability distributions corresponding to the above quantum example,
 once we have randomly put the particles inside the interval.
We find (not surprisingly) that the mere fact of having found precisely
 one particle
 inside, for example, the interval $\Delta_{2}$, modifies the probability
 of finding the other particle within $\Delta_{1}$, and this is due to the
 fact that the first information restricts the set of all possible ways of
 distributing the two particles within the interval.
Moreover, it is easy to show that all the probability distributions of
 arbitrary position measurements coincide with those holding for the quantum
 case of two bosons in the same state.

Therefore, it is possible to interpret the peculiar correlations arising
 when dealing with bosons in the same state, as classical correlations
 which are simply due to the truly indistinguishable nature of the
 particles involved, the same situation occurring for a set of classical
 identical particles.


  \subsection{Some remarks about property attribution}

Contrary to what happens in the case of fermions, in which the possibility of
 attributing a complete set of properties to one subgroup automatically
 implies that also the remaining subgroup has a complete set of properties,
 for systems of identical bosons this is no longer true.
This parallels strictly the situation we have already discussed in the case
 of two identical bosons when we have considered the state~(\ref{duebosoni})
 obtained
 by symmetrizing the product of two non-orthogonal factors, and this
 is due to the fact that ``one particle orthogonality'', and not the standard
 orthogonality of the factors, is necessary to claim that both subgroups have
 different properties.
To clarify once more this point, let us consider a state obtained by
 symmetrizing and normalizing two states which are not ``one-particle
 orthogonal" and which contain an equal number of bosons:
\begin{equation}
 \label{gio}
 \vert\Psi^{(N)}(1,...,N)\rangle={\cal N}S[\vert\Gamma^{(N/2)}(1,...,N/2)
 \rangle\vert\Lambda^{(N/2)}(N/2+1,...,N)\rangle].
\end{equation}
If one evaluates the scalar products of such a state with, e.g., the state
 ${\cal N}_{\Gamma}S[\vert\Gamma^{(N/2)}(1,...,N/2)\rangle\\
 \vert\Gamma^{(N/2)}(N/2+1,...,N)\rangle]$, or with
 ${\cal  N}_{\Lambda}S[\vert\Lambda^{(N/2)}(1,...,N/2)\rangle \vert
 \Lambda^{(N/2)}(N/2+1,...,N) \rangle]$, one sees that such scalar products
 are, in general, not equal to zero even if the states $\vert\Gamma\rangle$
 and $\vert\Lambda\rangle$ are orthogonal in the usual sense.
This is sufficient to conclude that in a state like~(\ref{gio}), in which
 $\vert \Lambda \rangle$ and $\vert \Gamma \rangle$ are not one-particle
 orthogonal, there is a nonzero probability of finding ``two subgroups" in
 the same state.
This in turn obviously implies that one cannot state that there are two
 subgroups possessing complete and different properties, while, in turn, the
 very fact that $\vert\Gamma^{N/2}(N/2+1,...,N)\rangle\neq \vert
 \Lambda^{N/2}(N/2+1,...,N)\rangle$ does not permit the  claim that the two
 subgroups possess the same properties.


  \section{A comment on entangled entanglement}

In this section we want to spend few words on a question  addressed in
 ref.~\cite{ref4}, i.e., whether the entanglement itself should be considered
 as an objective property of a given physical system.
Since we have made sharply precise the distinction between entangled and
 non-entangled states, our answer is, obviously, affirmative.
On the contrary, in the above paper the authors suggest that the
 non-separability property displayed by some quantum states {\em ``is not
 independent of the measurement context''}, concluding that the entanglement
 is a contextual property.
We would like, first of all, to stress that the real problem addressed  by
 the authors is completely different  from the one they seem to be interested
 in discussing, and that the conclusions they reach are absolutely obvious and
 rather trivial.
What they actually discuss is whether subjecting a constituent of a many
 ($\geq {3}$) - particle system to a measurement process one can leave the
 remaining particles in an entangled or non-entangled state, depending on the
 measurement one chooses to perform.

Their argument goes as follows. They consider a quantum state describing a
 three spin-1/2 particles, like the one considered by Greenberger, Horne and
 Zeilinger~\cite{ref5}:
  \begin{equation}
   \vert \psi \rangle = \frac{1}{\sqrt{2}}
   [\, \vert z\uparrow \rangle_{1}
   \vert z\uparrow \rangle_{2}
   \vert z\uparrow \rangle_{3} +
   \vert z\downarrow \rangle_{1}
   \vert z\downarrow \rangle_{2}
   \vert z\downarrow \rangle_{3} \,],
  \end{equation}
which is undoubtedly entangled. Then they remark that it is possible to
 leave the two non-measured particles in factorized or entangled states,
 depending on the different measurements which one chooses to perform on the
 third particle ($\,$the choice of measurement corresponding to what they
 call the {\em measurement context}$\,$).
We remark that by measuring the spin of the third particle along the
 $z$-direction one leaves the remaining particles in a factorized state,
 while every other conceivable spin measurement leads to an entangled state.
However, this conclusion has nothing to do with the objective entanglement
 of the initial state.
It is absolutely obvious that the first two particles are sometimes left
 in an entangled state and sometimes not: this is a trivial consequence of
 the reduction postulate of standard quantum mechanics.

Actually, one can easily devise even more general (but always trivial)
 situations in which the fact that two particles of a three-particle system
 are entangled or not  after the third has been subjected to a measurement may
 even depend, not only on the measurement one performs, but on the outcome
 one obtains.
To this purpose let us consider the following entangled state of three
 distinguishable spin one-half particles:
  \begin{equation}
   \vert \psi \rangle = \frac{1}{\sqrt{3}} \,
   \left[ \, \vert z \uparrow \rangle_{1}
   \vert z \uparrow \rangle_{2} \vert \omega_{a} \rangle_{3} +
   (\:\vert z \uparrow \rangle_{1} \vert z \downarrow \rangle_{2} +
   \vert z \downarrow \rangle_{1} \vert z \uparrow \rangle_{2}\:)
   \vert \omega_{b} \rangle_{3} \: \right],
  \end{equation}
where $\vert \omega_{a} \rangle$ and $\vert \omega_{b} \rangle$ are two
 eigenvectors belonging to different eigenvalues of an operator associated to
 an observable $\Omega^{(3)}$ of the third particle (which here, for
 simplicity we consider as distinguishable from the other two).
It is evident that a measurement process of $\Omega^{(3)}$ performed on
 the third particle, produces a final state for the first two particles whose
 entanglement depends strictly on the measurement outcome: if the result
 $\Omega= \omega_{a}$ is obtained, particles $1$ and $2$ are described by a
 factorized state, while in case of $\Omega= \omega_{b}$ the two particles
 are left in an entangled state.
Once more, this is due to the external intervention on the system and to
 wave function collapse, and therefore there is no need at all to attribute a
 special role to the measurement context to characterize the separability
 properties of a quantum system.
The state after the measurement is, as always, completely different from the
 one before it.


\vspace{1cm}
\begin{center}
  {\Large\bf  Part IV: Non pure states}\\
\end{center}

This section is devoted to discuss briefly some problems which are
relevant in connection with the locality issue and the validity of
Bell's inequality. We will limit our
 considerations to the case of distinguishable particles.


  \section{Correlations and Bell's inequality}

Let us consider, for simplicity,  a system of two distinguishable particles
 in a non-pure state which is a statistical mixture, with weights $p_{j}$, of
 factorized states $|\varphi_{j}(1)\rangle |\theta_{j}(2)\rangle$.
As implied by Theorem 4.3, for each of the states appearing in the
 mixture the expectation value of the direct product of two observables
 $A(1)$ of $H_{1}$ and $B(2)$ of $H_{2}$ also factorizes:
  \begin{equation}
   \label{correlationone}
   \langle \varphi_{j}(1)|\langle \theta_{j}(2) |A(1)\otimes B(2)|
   \varphi_{j}(1)\rangle |\theta_{j}(2)\rangle =
   \langle \varphi_{j}(1)|A(1) |\varphi_{j}(1)\rangle \cdot \langle
   \theta_{j}(2)|B(2)| \theta_{j}(2)\rangle.
  \end{equation}
It follows that the expectation value of $A(1)\otimes B(2)$ in the
 non-pure state can be written as:
\begin{equation}
 \label{correlation2}
 \langle A(1)\otimes B(2)\rangle = \sum_{j}p_{j}A_{j}B_{j},
  \:\:\:\:\:\:\:\:\:
  p_{i} >0, \:\:\:\:\:\:\: \sum_{i} p_{i} =1,
\end{equation}
where we have put $A_{j}=\langle \varphi_{j}(1)|A(1)|\varphi_{j}(1)\rangle $
 and $B_{j}=\langle \theta_{j}(2)|B(2)|\theta_{j}(2)\rangle.$

We can now compare the expression~(\ref{correlation2}) with the one giving
 the expectation value of the direct product of two observables in a hidden
 variable theory:
  \begin{equation}
   \label{correlation3}
   \langle A(1)\otimes B(2)\rangle = \int d\lambda \rho
   (\lambda )A(\lambda )B(\lambda),\qquad\:\:\:\:
   \rho(\lambda) >0, \:\:\:\:\:\: \int d\lambda \rho (\lambda )=1.
  \end{equation}
It is obvious that the two expressions~(\ref{correlation2})
 and~(\ref{correlation3}) have exactly the same formal structure.
Just as one can prove, starting from eq.~(\ref{correlation3}) and assuming
 that $\vert A(\lambda)\vert$ and $\vert B(\lambda)\vert$ are less than or
 equal to one, that Bell's inequality is satisfied, one can do the same
 starting from eq.~(\ref{correlation2}).
The conclusion should be obvious, and it has been stressed for the first time
 in ref.~\cite{ult2}: a non-pure state which is a statistical mixture of
 factorized states cannot lead to a violation of Bell's
 inequality\footnote{We note that A.Shimony at al.~\cite{ult1} have claimed
 that no one had proved explicitly this fact before 1989 when it has been
 proved by Werner~\cite{wer}. This is incorrect as one can check by reading
 ref.~\cite{ult2}}.
The converse is obviously not true, once more for the simple reason that the
 correspondence between statistical ensembles and statistical operators is
 infinitely many to one. It could therefore easily happen that a non-pure state
 describing a statistical mixture of non-factorized (i.e. entangled) states is
 associated to the same statistical operator of a statistical mixture of
 factorized states.
Since all expectation values depend only on the statistical operator, also in
 this last case one is lead to the same conclusion, i.e., that no violation of
 Bell's inequality can occur.

These considerations lead us to consider the relevant question of looking
 for mixtures which do not lead to violation of Bell's inequality, without
 worrying about their specific composition in pure subensembles.
The appropriate formal approach is the following.
Let us consider a given statistical ensemble $E$ of a system of two particles
 and its statistical operator $\rho _{E}(1,2)$ and let us define an
 equivalence relation between ensembles in the following way:
  \begin{equation}
   \label{correlation4}
   [ E^{*}\equiv E ]\Leftrightarrow \left[ \\ \rho_{E^{*}}(1,2)=
   \rho _{E}(1,2)\right] .
  \end{equation}
Our problem can now be reformulated in the following way: given a certain
 statistical operator $\rho (1,2)$ and considering the equivalence class of
 the statistical ensembles having it as its statistical operator, does this
 class contain at least one ensemble which is a statistical union of
 subensembles each of which is associated to a pure and factorized state?
It is obvious that if one can answer in the affirmative to such a question,
 then one can guarantee that the considered statistical operator cannot lead
 to a violation of Bell's inequality. Some relevant investigations in this
 direction have appeared recently~\cite{horo1,per1, Teu} but the general
problem
 is extremely difficult and far from having found a satisfactory solution.


\section{Conclusions}

In this paper we have reviewed the peculiar features displayed by entangled
 quantum states, by analyzing separately the cases of two or many
 distinguishable or identical quantum systems.
We have given the appropriate definitions for the various
 cases of interest and we have derived the  necessary and sufficient conditions
 which must be satisfied in order that a system can be considered entangled or
 non-entangled.
The analysis has been quite exhaustive and, we hope, it has clarified some of
 the subtle questions about this extremely relevant trait of quantum mechanics.


\vspace{0.5cm}

\section*{Acknowledgments}

We would like to thank Prof. D.D\"urr, Prof. G.Calucci and Dr. D.Mauro
 for many helpful discussions as well as the referee of the Journal of
Statistical Physics for useful remarks and stimulating suggestions.



\end{document}